\documentclass[fleqn,usenatbib]{mnras}
\usepackage{newtxtext,newtxmath}
\usepackage[T1]{fontenc}

\usepackage{graphicx}	
\usepackage{amsmath}	
\usepackage{booktabs}
\usepackage{tabularx}
\usepackage{array}
\usepackage{comment}
\usepackage{tikz}
\usepackage{float}
\usepackage{xcolor}

\newif\ifshowcomments
\showcommentstrue       

\usetikzlibrary{patterns, shapes.geometric}
\usetikzlibrary{positioning,arrows.meta,calc}

\newcommand{\Normal}{\mathcal{N}}

\newcommand{\C}{\mathbf{C}}

\newcommand{\bone}{\mathbf{1}}
\newcommand{\by}{\mathbf{y}}

\newcommand{\DRW}{\mathrm{DRW}}
\newcommand{\keyidea}{\textbf{Key idea:}~}

\title[QUELLS]{QUELLS: Detecting Strong-Lensing Echoes in Unresolved Quasar Light Curves}

\author[J. Laguna-Miralles et al.]{
Júlia Laguna-Miralles,$^{1,2}$\thanks{E-mail: jl2268@cam.ac.uk}
Vasily Belokurov,$^{1}$
Miles Cranmer,$^{1,2,3}$
Jacob Price,$^{1}$
Sergey E. Koposov$^{1,4}$
\\
$^{1}$Institute of Astronomy, University of Cambridge, Madingley Road, Cambridge CB3 0HA, United Kingdom\\
$^{2}$Kavli Institute for Cosmology, University of Cambridge, Madingley Road, Cambridge CB3 0HA, United Kingdom\\
$^{3}$Department of Applied Mathematics and Theoretical Physics, University of Cambridge, Wilberforce Road, Cambridge CB3 0WA, United Kingdom \\
$^{4}$Institute for Astronomy, University of Edinburgh, Royal Observatory, Blackford Hill, Edinburgh EH9 3HJ, United Kingdom
}

\date{MNRAS, submitted}

\pubyear{\the\year{}}

\begin{document}
\label{firstpage}
\pagerange{\pageref{firstpage}--\pageref{lastpage}}
\maketitle

\begin{abstract}
Wide-field time-domain surveys monitor millions of quasars, but galaxy-scale quasar lenses often remain unresolved at survey resolution. Their blended light curves can still carry a lensing signature: magnified, time-delayed copies of the same intrinsic variability. We present QUELLS (QUasar Echoes for Lensing in Light curves at Survey scale), a Gaussian-process framework that searches for these echoes by comparing single-source and delayed-mixture damped random walk (DRW) models, fitted jointly across bands at a common delay. Although the blended light curve is not Markovian, the latent DRW is, which allows us to evaluate the delayed-mixture likelihood exactly on irregularly sampled light curves without dense matrix factorisation, making survey-scale searches feasible. We construct a parent sample of \(1\,904\,365\) spectroscopically confirmed SDSS and DESI DR1 quasars with ZTF \(g\) and/or \(r\) light curves, supplemented by Gaia DR3 \(G\). Survey-realistic DRW mocks with 1 per cent false-positive thresholds show that detection and delay recovery improve with short-timescale variability, photometric precision and cadence, and that a common \(g+r\) delay strengthens both. Echoes injected into observed quasar light curves are recovered best at \(20\)--\(60\,{\rm d}\), and most reliably by coherent \(g+r\) fits. However, real unlensed controls exceed the DRW-calibrated thresholds far more often than the nominal rate, demonstrating the need for calibration on observed controls. Injected blends still show an excess above the in-sample observed-control
references, supporting a variability-based search that Gaia
complements by resolving close pairs. QUELLS enables ranking of unresolved lens candidates and their plausible delays for follow-up in ZTF, Gaia DR4 and Rubin/LSST.
\end{abstract}


\section{Introduction}
\label{sec:intro}

Strongly lensed quasars \footnote{We use `quasar' and `AGN'
interchangeably for the unobscured, optically variable type~1 AGN targeted in
this work.} are versatile probes of cosmology, galaxy structure and
quasars themselves. When combined with models of the lens
mass distribution and line-of-sight structure, measured inter-image delays constrain the time-delay distance and provide an independent route to the Hubble constant
\citep{Refsdal_1964,Treu_2016,Millon_2020,Birrer_2024, TDCOSMO_2026}.
Flux-ratio anomalies trace dark-matter substructure
\citep[e.g.][]{Dalal_2002,Gilman_2020}, while microlensing constrains
the size and structure of quasar accretion discs
\citep[e.g.][]{Morgan_2010} and broad-line regions
\citep[e.g.][]{Sluse_2012}. Their population
statistics also constrain cosmological parameters \citep{Oguri_2012}.

However, these applications are limited by sample size: confirmed lensed-quasar samples still comprise only a few hundred systems \citep{Ducourant_2026}, with Euclid now expanding the candidate pool through high-resolution imaging and reporting several hundred lensed AGN candidates in its first major data release \citep{Euclidlenses_2026}. Existing searches draw on combinations of morphology, colour, astrometry and spectroscopy \citep[e.g.][]{Oguri_2006,Lemon_2018}, with method-sensitivity depending on image separation, image contrast and the contribution of the lens galaxy. In seeing-limited surveys, closely separated images can blend into a single photometric source, removing the direct signature of multiple point sources \citep{Oguri_2010,Lemon_2024}. Although higher-resolution imaging can resolve closer pairs, with Euclid progressively expanding this discovery space, its current footprint covers only part of the sky accessible to wide-area monitoring \citep{EuclidDR1_2026}, and time-delayed variability provides complementary evidence to identify candidates and test their lensing nature. For systems that remain blended, spectroscopic searches for superposed galaxy and quasar emission can also miss faint deflectors overwhelmed by the quasar light \citep{More_2016,Lemon_2024}, which is particularly relevant at small separations: for early-type deflectors, smaller separations are associated with lower galaxy luminosities at fixed lens--source geometry
\citep{Inada_2012}.

\begin{figure*}
    \centering
    \includegraphics[width=0.95\textwidth]{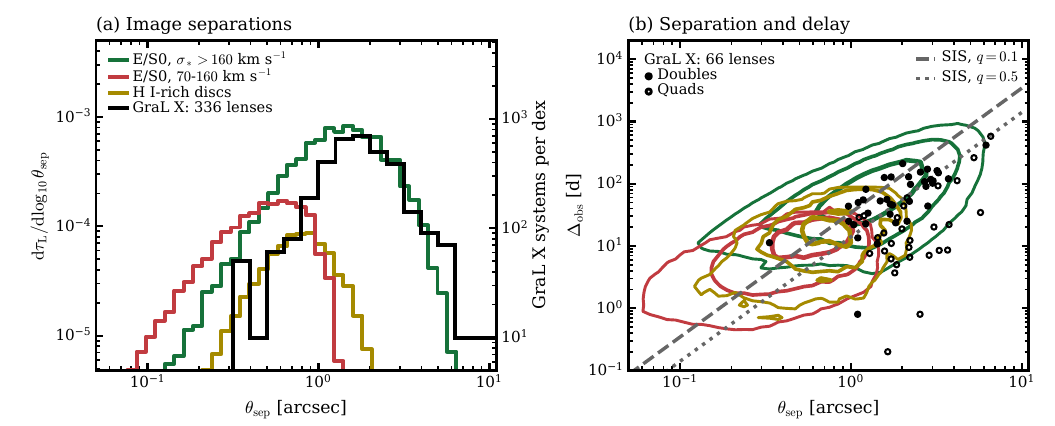}
    \caption{Lensing predictions for \(z=2\) sources and GraL~X lenses
    \citep{Ducourant_2026}; colours apply throughout. (a) Optical depth per dex and catalogue counts (black; right axis), using the maximum image separation per system. The count scale matches the massive E/S0 total over \(2''\)--\(6''\) for shape comparison. (b) Contours mark 1, 10 and 50 per cent of the peak density for each population and require \(q\geq0.1\). GraL X markers show the widest image pair with a measured delay per system, and grey SIS guides use \(d_{\Delta t}=3\,{\rm Gpc}\). \keyidea few lenses are known at small separations despite predicted contributions from several deflector populations, motivating searches sensitive to unresolved images.}
    \label{fig:lens-population-motivation}
\end{figure*}

Figure~\ref{fig:lens-population-motivation} compares the image separations and measured delays of spectroscopically confirmed lensed quasars in the Gaia GraL~X compilation \citep[][hereafter GraL~X]{Ducourant_2026} with illustrative predictions for galaxy populations lensing quasars at \(z=2\). We denote the angular image separation by \(\theta_{\rm sep}\in\mathbb{R}_{>0}\), measured in arcseconds,
and the absolute observer-frame arrival-time difference by \(\Delta_{\rm obs}\in\mathbb{R}_{>0}\), measured in days.
The dimensionless lensing optical depth
\(\tau_{\rm L}\in\mathbb{R}_{\geq0}\) approximates the probability that a randomly placed source is multiply imaged when lensing is rare (\(\tau_{\rm L}\ll1\)), where
\(\mathrm{d}\tau_{\rm L}/\mathrm{d}\log_{10}\theta_{\rm sep}\)
gives the contribution to this probability per dex in image separation.

The calculations combine galaxy abundances with adopted mass profiles
\citep{Choi_2007,Lelli_2016,Zwaan_2010} for foreground lens galaxies at redshift \(z_{\rm l}\in(0,z)\),
comprising elliptical and lenticular (E/S0) systems, split by
line-of-sight stellar velocity dispersion
\(\sigma_*\in\mathbb{R}_{>0}\) in \({\rm km\,s^{-1}}\), and discs
rich in neutral atomic hydrogen (H\,I). For both E/S0 populations, we adopt the singular isothermal sphere (SIS) model, a spherical profile whose total mass density decreases as the inverse square of radius \citep{Turner_1984}, whereas the disc profiles are inferred from rotation curves \citep{Lelli_2016}. For this SIS formulation, the grey guides use the fainter-to-brighter magnification ratio \(q\in(0,1]\) and time-delay distance \(d_{\Delta t}\in\mathbb{R}_{>0}\), measured in Gpc, which encodes the lens--source geometry and cosmology. Further details of the calculations and their assumptions are given in Appendix~\ref{app:lens-population-motivation}.

Panel (a) shows a dearth of known lenses at small separations: the model places about 40 per cent of the optical depth at \(\theta_{\rm sep}<1''\), whereas only 12 per cent of GraL~X systems (39 of 336) have such separations. Because the model includes neither magnification bias nor survey selection, this is a contrast in shape rather than a measure of catalogue incompleteness (Appendix~\ref{app:lens-population-motivation}). Below \(1''\), three deflector populations contribute. Massive E/S0 galaxies reach such small separations only when they lie close to the source, where the lensing geometry is inefficient (median \(z_{\rm l}\simeq1.1\) for \(z=2\), against \(z_{\rm l}\simeq0.6\) for their wider systems). Low-mass E/S0 galaxies dominate below \(\sim0.3''\), and H\,I-rich discs supply about 15 per cent of the optical depth at \(0.3''\)--\(1''\). These last two populations probe the mass distributions of lower-mass galaxies, which are poorly represented among known lenses. Panel (b) shows that compact lenses have observer-frame delays of days to tens of days, a region sparsely represented among catalogued systems with measured delays, and that at fixed separation the delay distinguishes the deflector populations: the large time-delay distances of massive, near-source lenses give delays two to three times longer than those of low-mass early types and discs at the same \(\theta_{\rm sep}\) (medians of 33, 11 and 15\,d at \(0.4''\)--\(0.7''\), for \(q\geq0.1\)). A delay measured from an unresolved light curve, combined with a separation from high-resolution follow-up, therefore indicates the nature of the deflector. The predicted contribution of compact lenses motivates complementary searches for lensing signatures that survive image blending.

Variability offers such an avenue: the combined flux of the blended images contains magnification-weighted, time-delayed copies of the same intrinsic quasar variability
\citep{Pindor_2005,Springer_2021}. For closely separated systems, difference-imaging searches identify spatially extended or multi-peaked variability even when the static morphology is ambiguous
\citep{Kochanek_2006,Lacki_2009,Chao_2020,Dux_2024}. For methods using unresolved total-flux light curves, early work reconstructed the individual components by combining unresolved monitoring with sparse resolved observations \citep{Geiger_1996}, with more recent approaches including entropy-based candidate selection \citep{KroneMartins_2019}, autocorrelation \citep{Shu_2021}, stochastic-process likelihoods \citep{Springer_2021}, fluctuation minimisation \citep{Bag_2022,Bag_2023}, neural networks \citep{Biggio_2022}, and joint analysis of unresolved photometry and time-dependent centroid astrometry \citep{Springer_2021b}. The statistical signature underlying these approaches has been known: the autocorrelation function of the blended light curve is that of the intrinsic variability plus cross terms shifted by \(\pm\Delta_{\rm obs}\), which imprint a periodic modulation on its power spectrum \citep{Geiger_1996}, a result discussed for unresolved quasars with damped random walk (DRW) variability by \citet{Bag_2023}. The same structure arises in photometric reverberation mapping, where a broad-band light curve contains the continuum together with a delayed and scaled emission-line response \citep{Zu_2011,Zu_2016,Chelouche_2013}, and in searches for lensing echoes in gamma-ray light curves of unresolved blazars \citep{Barnacka_2011}. Therefore, probing the compact-lens population in Figure~\ref{fig:lens-population-motivation} motivates variability methods tailored to existing wide-field survey light curves, with Euclid imaging and reconstructions from forthcoming Gaia DR4 scans helping to further characterise the resulting candidates and guide follow-up
\citep{EuclidDR1_2026,Shahaf_2026}.

With this goal, we present QUELLS (QUasar Echoes for Lensing in Light curves at Survey scale), a framework for discovering
lensed quasars through time-delayed echoes of their intrinsic
variability in unresolved survey light curves.
We adopt the likelihood comparison of \citet{Springer_2021}
between a single stochastic light curve and a linear combination
of time-shifted copies, which they formulated for a power-law power spectrum, computing the time-domain covariance by numerical integration and evaluating the likelihood by dense matrix inversion.
To make this comparison practical at survey scale, we model
the intrinsic variability as a damped random walk. Although
the blended light curve is not Markovian, the latent source
process is, allowing fast and exact likelihood evaluation on
irregularly sampled data. We use this likelihood evaluation to identify and refine competing delay modes, calibrate false-positive rates and
recovery efficiency on simulated light curves, and test the portability of this calibration through echo injections into observed quasar light curves. The contribution of QUELLS therefore lies not in the form of the delayed-mixture covariance, which follows from earlier work, but in its exact and efficient evaluation, its calibrated multiband implementation, and its validation on survey-realistic simulations and observed quasar light curves.
We focus on two-image (double) lenses, which dominate predicted
lensed-quasar samples \citep{Oguri_2010}.

The echo search gains an additional consistency check from
observations of the same quasar in several bands, because
the macrolensing delay is independent of wavelength
\citep{Treu_2016}. Intrinsic inter-band lags leave this
common-delay expectation unchanged because they affect both
images equally, although microlensing can introduce day-scale
chromatic shifts \citep{Tie_2018}. For the week-to-month delays
considered here, we therefore follow
\citet[][their section~4.4]{Springer_2021} in fitting a common
delay as a first approximation, with band-specific flux ratios
accommodating differential extinction, microlensing and
lens-galaxy light. We calibrate this multiband implementation
on simulations sharing one intrinsic realisation across bands.

We develop and test QUELLS using the Zwicky Transient Facility (ZTF)
and the ESA Gaia mission. ZTF provides public \(g\)- and \(r\)-band
photometry with a median point-spread function (PSF) full width at half
maximum (FWHM) of about \(2''\), often blending galaxy-scale lenses into
a single catalogue source \citep{Bellm_2019,Masci_2019}.
We build a parent catalogue comprising \(1\,904\,365\) spectroscopically confirmed quasars from the Sloan Digital Sky Survey (SDSS) and the Dark Energy Spectroscopic Instrument first data release (DESI DR1), with ZTF Data Release 24 (DR24) epoch photometry in \(g\), \(r\), or both \citep{Lyke_2020,Karim_2026}. We use DR23 (2018 March--2024 October) for mock calibration and echo-injection tests, and the longer DR24 light curves for application to two known lensed quasars. Gaia DR3 complements these observations with \(G\)-band epoch photometry
over 34 months (2014 July--2017 May) where available \citep{Vallenari_2023,Eyer_2023,Carnerero_2023}, with its scanning law setting the sampling and orientation of the photometric windows, which span roughly \(0.7''\times2.1''\) for faint sources. Depending on separation, flux ratio and scan angle, a sub-arcsecond pair can be resolved, unresolved or blended on only some Gaia transits
\citep{Lemon_2017,Fabricius_2021,Holl_2023}. Therefore, where both images contribute to one Gaia light curve, its independently sampled epochs provide a coherence check on the ZTF delay rather than a standalone detection.

The framework is also designed to scale to forthcoming data releases and larger surveys. Gaia DR4, scheduled for December 2026, will publish epoch photometry for all sources over 66 months, overlapping nearly the first two years of ZTF \citep{Brown_2025}. The Vera C. Rubin Observatory Legacy Survey of Space and Time (LSST) will provide deeper, six-band monitoring over a decade \citep{Ivezic_2019}, with forecasts predicting several thousand lensed quasars, many with detectable variability and measurable delays \citep{Oguri_2010,Yue_2022,Taak_2023,Shajib_2025}.

We describe the QUELLS framework (Section~\ref{sec:framework}) and validate it in two steps (Section~\ref{sec:validation}): survey-realistic DRW mocks show where and how well the search recovers injected lenses, and injections into observed light curves show that it continues to do so in real quasar variability. We then discuss its intended use for survey-scale candidate ranking and delay initialisation, including its application to known lenses (Section~\ref{sec:discussion}), and summarise our results in Section~\ref{sec:conclusions}. The empirical calibration required for deployment and the ranked lens candidates list within the parent sample are deferred to a forthcoming paper analysing the extended ZTF DR24 baseline (Section~\ref{sec:real-calibration}).

\section{The QUELLS framework}
\label{sec:framework}

\begin{figure*}
\centering
\includegraphics[width=\textwidth]{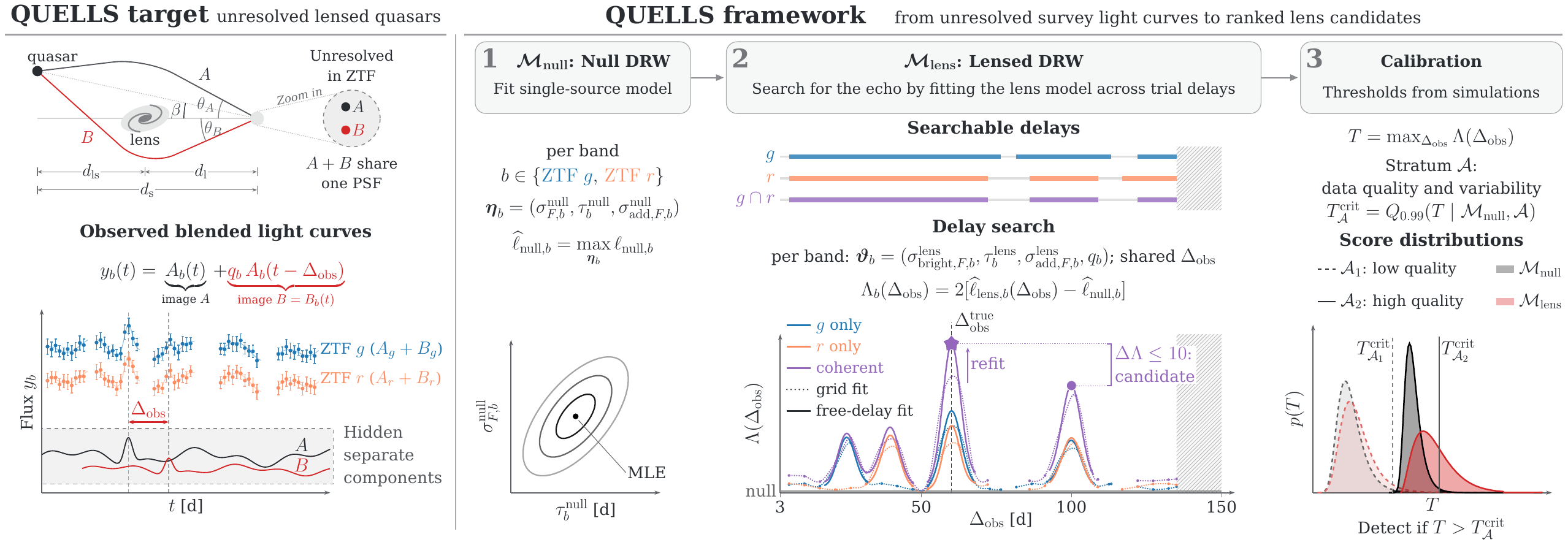}\par
\caption{
QUELLS lens-search and calibration workflow. The QUELLS target is an unresolved lensed quasar: a galaxy-scale lens produces two images whose light is blended in a single ZTF PSF, so each band records one light curve plus a delayed, scaled echo. In the lens sketch, \(d_{\rm l},d_{\rm s},d_{\rm ls}\in\mathbb{R}_{>0}\) are the observer--lens, observer--source and lens--source angular-diameter distances, respectively. (1) Each band is first fitted with a DRW without lensing. (2) The lensed DRW model is fitted on a grid of trial delays, refined with the delay free near each peak, and combined across bands at a common delay. (3) Detection thresholds are set per data-quality stratum from simulated or control unlensed quasars. Model definitions are given in Sections~\ref{sec:drw-source-model}--\ref{sec:lens-model}; fitting, detection statistics and calibration are described in Sections~\ref{sec:likelihood-fitting}--\ref{sec:detection-calibration}. Insets are schematic and use arbitrary scales.}
\label{fig:pipeline_comparison}
\end{figure*}

Figure~\ref{fig:pipeline_comparison} summarises the QUELLS framework. We compare two models for the light curve of a quasar in a single band: under \(\mathcal M_{\rm null}\) it is one stochastic quasar signal plus observational noise, and under \(\mathcal M_{\rm lens}\) the same signal plus a delayed, scaled copy of itself. The latter is what a two-image lens produces when the PSF of the survey blends its images into a single time series. Given the survey-dependent blending discussed in Section~\ref{sec:intro}, we fit each band separately and, where possible, combine them only through a common macrolensing delay assumed to be achromatic.

The search proceeds in four steps. We clean each light curve with
exposure-level quality cuts and convert it to relative linear flux
(Section~\ref{sec:preprocessing}). We then fit a single DRW, a widely used stochastic model of optical quasar variability
\citep{Kelly_2009,MacLeod_2010}, together with an additional white-noise term, so that its maximised likelihood defines the null reference
(Sections~\ref{sec:drw-source-model} and~\ref{sec:null-model}).
Next, we scan the delayed-mixture model (Section~\ref{sec:lens-model}) over
the trial delays supported by the cadence, retaining and refining every local
maximum of the delay profile so that competing solutions are kept
(Section~\ref{sec:likelihood-fitting}). Finally, we combine the per-band
statistics at a common observer-frame delay and turn them into detections with
thresholds calibrated on null samples, either simulations or observed
controls, processed through the same search
(Section~\ref{sec:detection-calibration}). Table~\ref{tab:notation_summary}
summarises the notation.

\subsection{Survey light curves and preprocessing}
\label{sec:preprocessing}

For a single object observed in one photometric band \(b\), let
\(
  \left\{
  t_{j,{\rm obs}},\,m_j,\,s_{m,j}
  \right\}_{j=1}^{n}
\)
denote the retained light-curve measurements, where
\(n\in\mathbb{N}\) is the number of retained epochs,
\(t_{j,{\rm obs}}\in\mathbb{R}\) are the observer-frame observation times,
\(m_j\in\mathbb{R}\) are the apparent magnitudes, and
\(s_{m,j}\in\mathbb{R}_{>0}\) are their associated uncertainties, with epochs ordered such that
\(
  t_{1,{\rm obs}}\leq t_{2,{\rm obs}}\leq\cdots\leq t_{n,{\rm obs}}.
\)
We omit the index \(b\) on these and all derived single-band quantities, restoring it only where bands are compared or combined.

We use the ZTF PSF-fit light curves \citep{Masci_2019}, excluding the
more sparsely sampled \(i\) band, which adds little coherence gain in
our mocks. For sources in overlapping fields, we select the light curve
with the most epochs to retain a single field calibration.
We retain epochs with \(\texttt{catflags}=0\)
\citep{Masci_2019,ZSDS_Supplement} and a \(5\sigma\) limiting magnitude
\(\texttt{limitmag}\) at least one magnitude fainter than the median
magnitude of the light curve.
This depth requirement excludes shallow exposures affected by bright
sky, including moonlight, poor transparency or poor seeing
\citep[e.g.][]{Guidry_2021}, but does not explicitly model residual
Moon-dependent photometric bias
(discussed in Sections~\ref{sec:real-calibration} and~\ref{sec:known-lenses}).
We apply the same criteria to null and lensed mocks, avoiding PSF-shape
cuts (\texttt{chi}, \texttt{sharp}) that could reject genuine blends and sigma-clipping that selects directly on measured flux excursions. We require finite measurements with positive fluxes and uncertainties in both surveys. For Gaia, we retain \(G\)-band epochs with \(\texttt{variability\_flag\_g\_reject}=0\) and exclude \(G_{\rm BP}\) and \(G_{\rm RP}\) because their photometry is too noisy at the apparent
magnitudes of most quasars in our sample \citep{Vallenari_2023}.

We require a known spectroscopic redshift \(z\in\mathbb{R}_{\geq0}\) for each fitted object. Source-process times, damping timescales,
delays and variability rates are defined in its rest frame throughout this work, unless marked with the subscript \(\mathrm{obs}\). For a reference epoch
\(t_{0,{\rm obs}}\in\mathbb{R}\),
\begin{equation}
t_j\coloneq\frac{t_{j,{\rm obs}}-t_{0,{\rm obs}}}{1+z},
\qquad
\Delta\coloneq\frac{\Delta_{\rm obs}}{1+z},
\label{eq:rest-frame-time}
\end{equation}
where \(t_j\in\mathbb{R}_{\geq0}\), and \(\Delta,\Delta_{\rm obs}\in\mathbb{R}_{>0}\) are the rest- and observer-frame lensing delays. We take \(t_{0,{\rm obs}}=t_{1,{\rm obs}}\), which is arbitrary because the covariance is stationary, and report recovered delays as \(\Delta_{\rm obs}\).

We model all light curves in dimensionless relative linear flux. Writing \(\gamma\coloneq0.4\ln10\), we take \(s_{m,j}\) as the reported magnitude uncertainty for ZTF and \(s_{m,j}=s_{F,j}/(\gamma F_j)\) for Gaia \(G\), whose archive provides the epoch flux \(F_j\) and its uncertainty \(s_{F,j}\). For each object and band, we define \(m_{\rm ref}\coloneq\operatorname{median}_{j}(m_j)\in\mathbb{R}\) and transform
\begin{equation}
  y_j\coloneq10^{-0.4(m_j-m_{\rm ref})},
  \qquad
  s_j\coloneq\gamma\,y_j\,s_{m,j},
  \label{eq:magnitude-to-flux}
\end{equation}
with \(y_j,s_j\in\mathbb{R}_{>0}\), so that the flux scale is centred near unity and \(s_j\) follows from first-order error propagation. We write \(\mathbf{y}\coloneq(y_1,\ldots,y_n)^{\mathsf T}\) and \(\mathbf{s}\coloneq(s_1,\ldots,s_n)^{\mathsf T}\), both in \(\mathbb{R}_{>0}^{\,n}\), and summarise the single-epoch photometric precision using the signal-to-noise (S/N) ratio
\begin{equation}
  ({\rm S/N})_{\rm epoch}
  \coloneq
  \left[
    \gamma\,\operatorname{median}_{j}(s_{m,j})
  \right]^{-1}
  \in\mathbb{R}_{>0},
  \label{eq:epoch-snr}
\end{equation}
computed from the reported catalogue magnitude uncertainties. Positivity applies to the retained catalogue measurements, not to the modelled fluxes.

\subsection{Stochastic model for intrinsic quasar variability}
\label{sec:drw-source-model}

Three processes shape an unresolved lensed light curve: the intrinsic variability of the quasar, the summation of the fluxes of its images, and the photometric measurement noise. The first is described most naturally in logarithmic flux, the other two in linear flux. Intrinsic quasar variability is thought to be multiplicative: fluctuations in the
accretion flow modulate the emitted flux by fractional amounts, which
produces the linear rms--flux relation and lognormal flux distributions
observed in accreting systems \citep{Uttley_2005}, and is the reason DRW
models in logarithmic flux or magnitude are the most widely used
description of optical quasar variability
\citep{Kelly_2009,MacLeod_2010,Kozlowski_2010,Kelly_2014}. Lensing, by
contrast, is additive: the fluxes of the unresolved images sum
linearly, so the lensed light curve is a linear operation on the source
flux and not on its logarithm. Photometric noise is also additive in flux: it arises from photon counting from source, sky and detector photons and is approximately Gaussian in counts \citep{Howell_2006,Masci_2019}, whereas
catalogue magnitude errors are Gaussian only in the high-S/N limit
\citep{Lupton_1999,Sesar_2007}.

We therefore fit in relative linear flux, where lensing and measurement
noise are described exactly and only the source model is approximate.
Modelling the source as a Gaussian process in flux keeps the
delayed mixture Gaussian, giving a likelihood available in
closed form. We evaluate it exactly by exploiting the Markov
structure of the latent DRW, with a cost that grows linearly
with the number of epochs for a fixed number of retained
latent states (Section~\ref{sec:likelihood-fitting}). This is what makes survey-scale fitting tractable. The standard log-flux or magnitude DRW has no such property: its flux is lognormal, the delayed sum of two
lognormal processes is not Gaussian, and its likelihood has no closed
form. Because the two are distinct stochastic models rather than
reparametrisations of one another, fitting in linear flux is an
approximation, and we measure its cost instead of assuming it away. The
calibration mocks are generated with the physically motivated model, a
magnitude DRW exponentiated to flux with the images and Gaussian noise
added in flux (Section~\ref{sec:calibration-grid}), and fitted with the
linear-flux model. Additionally, we compare the formulations directly on observed AGN light curves (Appendix~\ref{sec:four-model-DRW-comparison}).

Therefore, for a single DRW component we define
\[
  y(t)\coloneq\mu_F+x_F(t),
  \qquad
  x_F(t)\sim\mathrm{GP}\!\left[0,k_F(t,t')\right],
\]
where \(\mu_F\in\mathbb{R}\) is a constant mean and
\(x_F:\mathbb{R}\rightarrow\mathbb{R}\) is a zero-mean Gaussian process with
covariance \(k_F:\mathbb{R}\times\mathbb{R}\rightarrow\mathbb{R}\).
For the fitted process, we adopt the DRW covariance
\begin{equation}
  k_{\DRW,F}(t,t';\sigma_F,\tau)
  \coloneq
  \sigma_F^2
  \exp\!\left(-\frac{|t-t'|}{\tau}\right),
  \label{eq:drw-kernel}
\end{equation}
with
\(\sigma_F,\tau\in\mathbb{R}_{>0}\).
Here \(\sigma_F\) is the stationary rms amplitude and \(\tau\) the damping
timescale, such that
\(k_{\DRW,F}(t,t;\sigma_F,\tau)=\sigma_F^2\). 

Writing \(\delta t\coloneq t-t'\), the corresponding second-order
structure function is
\begin{equation}
  \mathrm{SF}_{\DRW,F}^2(\delta t)
  =
  2\sigma_F^2
  \left[
    1-\exp\!\left(-\frac{|\delta t|}{\tau}\right)
  \right],
  \qquad
  \delta t\in\mathbb{R}.
  \label{eq:drw-sf}
\end{equation}
For \(|\delta t|\ll\tau\),
\begin{equation}
  \mathrm{SF}_{\DRW,F}^2(\delta t)
  \simeq
  2D|\delta t|,
  \qquad
  D\coloneq\frac{\sigma_F^2}{\tau},
  \qquad
  D\in\mathbb{R}_{>0},
  \label{eq:D-definition}
\end{equation}
where \(D\) is the short-lag variability rate, in relative-flux squared
per rest-frame day.

Cosmological time dilation gives \(\tau_{\rm obs}=(1+z)\tau \) and \(D_{\rm obs}=\frac{D}{1+z},\)
with \(\tau_{\rm obs},D_{\rm obs}\in\mathbb{R}_{>0}\). Parameters fitted under the two hypotheses carry superscripts
\(\mathrm{null}\) and \(\mathrm{lens}\), respectively.
  
\subsection{Single-source null model}
\label{sec:null-model}

Under \(\mathcal{M}_{\rm null}\), the observed relative-flux light curve is
modelled as a single DRW process plus independent Gaussian noise. Writing
\(\bone\coloneq(1,\ldots,1)^\top\), we have
\begin{equation}
  \by\mid\mathcal{M}_{\rm null}
  \sim
  \Normal\!\left(
    \mu_F^{\rm null}\bone,\,
    \C_F^{\rm null}
  \right),
  \label{eq:null-generative-distribution}
\end{equation}
with
\begin{equation}
  \left(\C_F^{\rm null}\right)_{jj'}
  =
  k_{\DRW,F}\!\left(
    t_j,t_{j'};
    \sigma_F^{\rm null},
    \tau^{\rm null}
  \right)
  +
  \delta_{jj'}
  \left[
    s_j^2+
    \left(\sigma_{\rm add,F}^{\rm null}\right)^2
  \right],
  \label{eq:null-covariance}
\end{equation}
where \(\mu_F^{\rm null}\in\mathbb{R} \), \(\sigma_F^{\rm null}\), \(\tau^{\rm null}\in\mathbb{R}_{>0}\),\(\sigma_{\rm add,F}^{\rm null}\in\mathbb{R}_{\geq0}\)
and \(\delta_{jj'}\) is the Kronecker delta.
The fitted term \(\sigma_{\rm add,F}^{\rm null}\) is a constant,
epoch-independent white-noise amplitude in relative flux, added in
quadrature to the uncertainties \(s_j\) and optimised in
\(\ln\sigma_{\rm add,F}\). It absorbs excess scatter that is uncorrelated between epochs and not described by the pipeline uncertainties, such as underestimated photometric noise, epoch-to-epoch calibration or PSF errors, and intrinsic variability unresolved by the sampling \citep{Zu_2013,Kozlowski_2016,Suberlak_2021,Arevalo_2026}. Treating the additional noise as constant in flux rather than magnitude
matters mainly for strongly variable sources, and the flux prescription
adopted here is preferred for about three quarters of the ZTF light
curves in our comparison sample
(Appendix~\ref{sec:four-model-DRW-comparison}).

For fixed covariance parameters, the mean is determined analytically,
\(
  \widehat{\mu}_F^{\rm null}
  =
  \frac{
    \bone^\top
    \left(\C_F^{\rm null}\right)^{-1}\by
  }{
    \bone^\top
    \left(\C_F^{\rm null}\right)^{-1}\bone
  }.
\)

\subsection{Delayed-mixture lens model}
\label{sec:lens-model}

Under \(\mathcal{M}_{\rm lens}\), the unresolved light curve contains a DRW
component and a delayed, scaled copy of the same stochastic process. We order
the two components by mean flux and define the fainter-to-brighter flux ratio
\begin{equation}
  q \coloneq
  \frac{\bar F_{\rm faint}}{\bar F_{\rm bright}},
  \qquad
  0<q\leq1,
  \label{eq:lens-flux-ratio}
\end{equation}
where \(\bar F_{\rm bright},\bar F_{\rm faint}\in\mathbb{R}_{>0}\) are the mean fluxes of the brighter and fainter images. The upper bound \(q=1\) follows from ordering the components by flux,
and physically \(q\) can take any value in \((0,1]\). However, we restrict the search to \(0.1 \le q \le 1\) to avoid the weak-secondary regime, in which the delayed component becomes increasingly difficult to distinguish from the null at the survey signal-to-noise ratio considered here. Let
\(x_{{\rm bright},F}^{\rm lens}(t)\) be a zero-mean DRW process with
\(\sigma_{{\rm bright},F}^{\rm lens}>0\) and
\(\tau^{\rm lens}>0\). The stochastic unresolved signal is
\[
  x_F^{\rm lens}(t)
  \coloneq
  x_{{\rm bright},F}^{\rm lens}(t)
  +
  q\,x_{{\rm bright},F}^{\rm lens}(t-\Delta).
\]
The two terms correspond to images A and B in
Figure~\ref{fig:pipeline_comparison}.
The corresponding covariance is
\begin{equation}
\begin{aligned}
  k_F^{\rm lens}(\delta t)
  &=
  \left(\sigma_{{\rm bright},F}^{\rm lens}\right)^2
  \Bigg[
    (1+q^2)
    \exp\!\left(-\frac{|\delta t|}{\tau^{\rm lens}}\right)
    \\
  &\qquad
    +q\exp\!\left(-\frac{|\delta t+\Delta|}{\tau^{\rm lens}}\right)
    +q\exp\!\left(-\frac{|\delta t-\Delta|}{\tau^{\rm lens}}\right)
  \Bigg].
  \label{eq:lens-kernel}
\end{aligned}
\end{equation}
The first term contains the two autocovariances and the remaining terms the cross-covariances. This three-term structure is not specific to the DRW: \citet{Geiger_1996} derived it, together with the corresponding power spectrum (Eq.~\ref{eq:psd-lens}), for a blended two-image lens with arbitrary intrinsic variability, and \citet{Bag_2023} discussed it for unresolved quasars with DRW variability. Equation~\eqref{eq:lens-kernel} is also the zero-width limit of the top-hat transfer-function model used for single-band photometric reverberation mapping \citep{Zu_2011,Zu_2016}. Reversing \(\Delta\) interchanges the cross-covariance terms, giving \(k_F^{\rm lens}(\delta t;\Delta)=k_F^{\rm lens}(\delta t;-\Delta)\) and the same unresolved-flux likelihood for either arrival order under this stationary Gaussian model, even for unequal image brightnesses. Only the delay magnitude is therefore identifiable, making the assignment of the unshifted component to the brighter image a convention. At the observed epochs,
\begin{equation}
  \left(\C_F^{\rm lens}\right)_{jj'}
  =
  k_F^{\rm lens}(t_j-t_{j'})
  +
  \delta_{jj'}
  \left[
    s_j^2+
    \left(\sigma_{\rm add,F}^{\rm lens}\right)^2
  \right],
  \label{eq:lens-covariance}
\end{equation}
with
\(\sigma_{\rm add,F}^{\rm lens}\in\mathbb{R}_{\geq0}\), and
\begin{equation}
  \by\mid\mathcal{M}_{\rm lens}
  \sim
  \Normal\!\left(
    \mu_F^{\rm lens}\bone,\,
    \C_F^{\rm lens}
  \right).
  \label{eq:lens-generative-distribution}
\end{equation}
Here \(\mu_F^{\rm lens}\in\mathbb{R}\) is the mean total unresolved
flux, determined analytically as in the null model, while
\(\sigma_{{\rm bright},F}^{\rm lens}\) is the DRW amplitude of the
brighter-image contribution.

\begin{figure*}
    \centering
    \includegraphics[width=\textwidth]{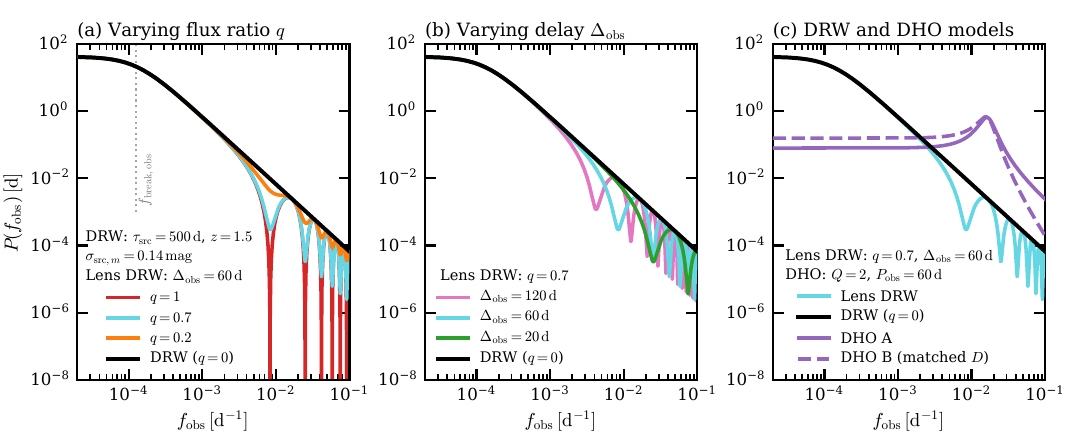}
      \caption{Analytic power spectra of unlensed and unresolved two-image lensed DRWs, and two illustrative DHOs. The lensed models use \(\sigma_{{\rm bright},F}^{\rm lens}=\sigma_{{\rm src},F}/(1+q)\), keeping the total mean flux fixed.
      (a) Flux-ratio dependence of the lensed kernel comb structure; the dotted line marks the DRW break
      \(f_{\rm break,obs}=(2\pi\tau_{\rm obs})^{-1}\).
      (b) Delay dependence, showing the \(1/\Delta_{\rm obs}\) spacing
      of the interference pattern.
      (c) Comparison with two DHOs, both normalised to the stationary variance of the unlensed DRW. DHO B additionally matches its short-lag variability rate \(D_{\rm obs}=\sigma_{{\rm src},F}^{2}/\tau_{\rm obs}\), and hence its asymptotic high-frequency PSD normalisation. The driving-noise perturbation timescales are
      \(\tau_{\rm perturb,obs}=9.55\,\mathrm{d}\) for A and
      \(1.19\,\mathrm{d}\) for B. Here \(Q\) denotes the quality factor
      and \(P_{\rm obs}\) the observer-frame undamped natural period,
      chosen to equal the lens delay \citep[see][]{Yu_2022}. \keyidea lensing imprints a comb with frequency spacing \(1/\Delta_{\rm obs}\) on the whole DRW spectrum, which a single DHO does not reproduce, providing a characteristic signature beyond DHO-like departures from a DRW.}
    \label{fig:drw-lensed-dho-psd}
\end{figure*}

The same kernel can be read in the frequency domain, where its signature
is explicit. A DRW with amplitude \(\sigma_F\) and observer-frame timescale \(\tau_{\rm obs}\) has the power spectral density (PSD)
\(P_{\rm DRW}(f_{\rm obs})=2\sigma_F^2\tau_{\rm obs}/[1+(2\pi f_{\rm obs}\tau_{\rm obs})^2]\),
and the delayed copy multiplies it by
\(|1+q\,e^{-2\pi i f_{\rm obs}\Delta_{\rm obs}}|^2\), giving the PSD \(P_{\rm lens}\) of the unresolved lensed signal:
\begin{equation}
  P_{\rm lens}(f_{\rm obs})
  =
  P_{\rm DRW}(f_{\rm obs})
  \left[1+q^2+2q\cos(2\pi f_{\rm obs}\Delta_{\rm obs})\right],
  \label{eq:psd-lens}
\end{equation}
with \(P_{\rm DRW}\) evaluated at \(\sigma_{{\rm bright},F}^{\rm lens}\) and \(\tau_{\rm obs}^{\rm lens}=(1+z)\tau^{\rm lens}\). Lensing in our model thus imprints a comb on the DRW PSD with frequency
spacing \(1/\Delta_{\rm obs}\) and a modulation factor ranging from
\((1-q)^2\) to \((1+q)^2\), reaching zero for equal-brightness images.

Figure~\ref{fig:drw-lensed-dho-psd} shows this comb for typical quasar variability parameters and compares it with a widely used extension of the DRW, the damped harmonic oscillator
\citep[DHO, or CARMA(2,1) process;][]{Kelly_2014,Kasliwal_2017,Moreno_2019,Yu_2022,Yu_2025}. A DHO can produce a single resonance of finite width, whereas lensing modulates the whole spectrum at fixed frequency spacing \(1/\Delta_{\rm obs}\), so DHO-like departures from a DRW in real quasars are not expected to mimic the full spectral signature of a delayed echo mixture. Equivalently, in the time domain the lensed DRW covariance has a kink at lag \(\Delta\), where its slope jumps by \(-2q(\sigma_{{\rm bright},F}^{\rm lens})^2/\tau^{\rm lens}\), whereas the covariance of a DHO, or of any finite-order CARMA process, is smooth at all non-zero lags.

In QUELLS, we work in the time domain, exploiting the DRW Markov structure for efficient, exact Gaussian likelihood evaluation at irregular epochs. The same model can also be fitted in Fourier space by using the joint covariance of the coefficients, which accounts for the observing window and retains delay information despite the blurred spectral comb. Appendix~\ref{sec:fourier-sampling} illustrates this connection with closely agreeing time- and Fourier-domain delay estimates for one realisation.

\subsection{Likelihood evaluation and numerical fitting}
\label{sec:likelihood-fitting}

Both hypotheses describe the light curve as Gaussian with a constant
mean and differ only in their covariance
(Eqs~\ref{eq:null-covariance} and~\ref{eq:lens-covariance}). With the
mean determined analytically, we define the log likelihood
in band \(b\) for either model
\(\mathcal{M}\in\{\mathcal{M}_{\rm null},\mathcal{M}_{\rm lens}\}\) as
\begin{equation}
\begin{aligned}
  \ell_{\mathcal M,b}
    \coloneq
  -\frac{1}{2}\Big[
    &(\by-\widehat{\mu}_F^{\mathcal M}\bone)^{\mathsf T}
    (\C_F^{\mathcal M})^{-1}
    (\by-\widehat{\mu}_F^{\mathcal M}\bone)
    \\
    &+\ln\det\C_F^{\mathcal M}
    +n\ln(2\pi)
  \Big],
\end{aligned}
\label{eq:relative-flux-profile-likelihood}
\end{equation}
where
\(\widehat{\mu}_F^{\mathcal M}\in\mathbb{R}\) is profiled analytically as with the null model, and \(\widehat{\ell}_{\mathcal M,b}\in\mathbb{R}\) denotes the log likelihood at the maximum-likelihood estimate of the free parameters. The delayed-model likelihood is evaluated exactly by augmenting the latent DRW process at the observed and delay-shifted epochs, conditioning
sequentially and discarding latent states after their last use. Although the blended light curve is not Markovian, the latent
DRW is, and each observation depends on only two source states,
at \(t_j\) and \(t_j-\Delta\). This chronological conditioning avoids inversions of DRW transition variances, including at coincident or nearly coincident epochs, and is used throughout the production fits.

The resulting likelihood agrees with a direct dense-matrix
calculation and requires \(\mathcal{O}(nw^2)\) operations,
where \(w\) is the maximum number of retained latent states.
The cost grows linearly with \(n\) at fixed \(w\), while denser
sampling and longer trial delays generally increase \(w\).
Fast exact methods for an ordinary DRW
\citep{Rybicki_1995,Foreman-Mackey_2017} do not apply directly
to the blend, whose covariance has a kink at non-zero lag
\(\Delta\) (Section~\ref{sec:lens-model}) that no fixed sum
of exponential kernel terms reproduces.

The null model is fitted first by bounded multistart optimisation
\citep[L-BFGS-B, with a Powell fallback;][]{Byrd_1995,Powell_1964}, and its best solution initialises the lens search. We scan observer-frame delays on a \(1\,{\rm d}\) grid over
\(3\,{\rm d}\leq\Delta_{\rm obs}\leq
\min(150\,{\rm d},\,0.45B_{\rm obs})\), where \(B_{\rm obs}\coloneq t_{n,{\rm obs}}-t_{1,{\rm obs}}\). These limits define the calibrated search domain rather than physical bounds on lensing delays; the upper limit covers the delays expected for pairs that blend in ZTF, as shown in Section~\ref{sec:calibration-grid}, and the final refinement in \(\Delta_{\rm obs}\) is continuous.

Within this range we exclude delays that would be constrained mainly by
extrapolation across seasonal gaps. Let \(\{\widetilde t_{\ell,{\rm obs}}\}_{\ell=1}^{n_{\rm ep}}\) denote the distinct observer-frame epochs, where \(n_{\rm ep}\in\mathbb{N}\) is their number and
\(\widetilde t_{\ell,{\rm obs}}\in\mathbb{R}\). We split these epochs into observing blocks at gaps longer than \(90\,{\rm d}\) and extend each block by \(7\,{\rm d}\) at both ends to obtain intervals \(W_k\subset\mathbb{R}\). For a candidate delay, the fractions of epochs whose shifted counterpart falls inside a block are
\begin{equation}
\begin{aligned}
  S_{\pm}(\Delta_{\rm obs})
  &\coloneq
  \frac{1}{n_{\rm ep}}
  \sum_{\ell=1}^{n_{\rm ep}}
  \mathbb{I}\!\left[
    \widetilde t_{\ell,{\rm obs}}\pm\Delta_{\rm obs}
    \in\bigcup_k W_k
  \right],\\
  S_{\rm sym}(\Delta_{\rm obs})
  &\coloneq
  \max\!\left[S_-(\Delta_{\rm obs}),S_+(\Delta_{\rm obs})\right],
\end{aligned}
\label{eq:delay-support-fraction}
\end{equation}
with \(S_\pm,S_{\rm sym}\in[0,1]\), where \(\mathbb{I}[\cdot]\in\{0,1\}\) is the
indicator function. We retain delays with \(S_{\rm sym}\geq0.60\), which
defines the supported domain \(\Omega_{b,{\rm obs}}\subset\mathbb{R}_{>0}\) in
band \(b\). Tests on high-S/N ZTF-\(r\) mocks give comparable detection
and delay recovery when requiring \(S_{\rm sym}\geq0.50\)
or \(S_{\rm sym}\geq0.60\), with a small recovery loss
for \(S_{\rm sym}\geq0.70\). Therefore, we adopt the \(0.60\) support threshold as a compromise between recovery and computational cost. Restricting the search in this way, together with the upper limit of \(0.45B_{\rm obs}\), also guards against the spurious modes near the edges of the observing window that Gaussian-process time-delay likelihoods develop because delay estimation is intrinsically extrapolative \citep{Kroupa_2026a}.
  
At each supported grid delay, we maximise the lens likelihood over the
band-specific nuisance parameters
\begin{equation}
  \boldsymbol{\vartheta}_b
  \coloneq
  \left(
    \sigma_{{\rm bright},F,b}^{\rm lens},
    \tau_b^{\rm lens},
    \sigma_{{\rm add},F,b}^{\rm lens},
    q_b
  \right), \quad
  \boldsymbol{\vartheta}_b
  \in
  \mathbb{R}_{>0}^{2}
  \times\mathbb{R}_{\geq0}
  \times[0.1,1]
  \label{eq:lens-nuisance-vector}
\end{equation}
within the numerical bounds
\(10^{-5}\leq\sigma_{{\rm bright},F}^{\rm lens}\leq10\),
\(10^{-4}\,{\rm d}\leq\tau^{\rm lens}\leq10^{6}\,{\rm d}\) and
\(10^{-8}\leq\sigma_{{\rm add},F}^{\rm lens}\leq1\); the same limits apply to the null model. We use broad numerical bounds to reduce boundary effects on the maximum-likelihood estimates, with small positive lower bounds because the amplitudes are optimised in logarithmic form. This gives the profile likelihood
\begin{equation}
  \widehat{\ell}_{{\rm lens},b}(\Delta_{\rm obs})
  \coloneq
  \max_{\boldsymbol{\vartheta}_b}
  \ell_{{\rm lens},b}
  (\Delta_{\rm obs},\boldsymbol{\vartheta}_b).
  \label{eq:band-profile-likelihood}
\end{equation}

We maximise the likelihood with bounded L-BFGS-B at each
grid delay, traversing each supported interval in increasing
and decreasing delay order. Each traversal uses the preceding solution as an initial guess, and we retain the higher likelihood at each delay. Peaks within 10 log-likelihood units of the best grid point first receive a local \(0.25\,{\rm d}\) scan, and we then refine every local maximum continuously in
\(\Delta_{\rm obs}\) and \(\boldsymbol{\vartheta}_b\)
with bounded SLSQP \citep{Kraft_1988}. For each band we retain the null fit, delay profile, refined lens solutions, likelihoods and quality flags, and the subsequent common-delay
search reuses these results and performs additional band-specific fits only where needed.

A complete two-band search, including the individual \(g\) and \(r\)
fits and their coherent combination, has an estimated mean cost of
approximately \(1\,{\rm CPU\,min}\) per paired DR23-era entry.
Accounting for the retained epoch counts and cadence of the parent
sample, and fitting only qualifying bands, gives an estimated total
ZTF fitting cost of about \(7\times10^3\)~CPU~h
(Section~\ref{sec:observing-regimes}). Because each source can be fitted independently, the search can be parallelised, making a survey-wide application computationally feasible. We defer this application to a subsequent paper analysing ZTF DR24 (Section~\ref{sec:real-calibration}).

\subsection{Detection statistics and calibration}
\label{sec:detection-calibration}

For band \(b\), let
\(\widehat{\ell}_{{\rm null},b}\in\mathbb{R}\) denote the
maximum of Eq.~\eqref{eq:relative-flux-profile-likelihood}
over \((\sigma_{F,b}^{\rm null},\tau_b^{\rm null},
\sigma_{{\rm add},F,b}^{\rm null})\), and
\(\widehat{\ell}_{{\rm lens},b}(\Delta_{\rm obs})
\in\mathbb{R}\) the maximum over
\(\boldsymbol{\vartheta}_b\) at fixed delay
(Eq.~\eqref{eq:band-profile-likelihood}), with both means determined analytically (see Section~\ref{sec:likelihood-fitting}). For an evaluated delay
\(\Delta_{\rm obs}\in\Omega_{b,{\rm obs}}\), we define
\begin{equation}
  \Lambda_b(\Delta_{\rm obs})
  \coloneq
  2\left[
    \widehat{\ell}_{{\rm lens},b}(\Delta_{\rm obs})
    -
    \widehat{\ell}_{{\rm null},b}
  \right].
  \label{eq:band-delay-statistic}
\end{equation}

Let \(\mathcal E_b\subset\Omega_{b,{\rm obs}}\) denote the
finite set of candidate observer-frame delays returned by
the single-band grid search and continuous refinement
(Section~\ref{sec:likelihood-fitting}). The single-band detection statistic is
\begin{equation}
  T_b
  \coloneq
  \max_{\Delta_{\rm obs}\in\mathcal E_b}
  \Lambda_b(\Delta_{\rm obs}),
  \qquad
  T_b\in\mathbb{R},
  \label{eq:single-band-detection-statistic}
\end{equation}
where larger values indicate greater improvement under the delayed-mixture model.

For a familiar scale we also report the Bayesian information criterion \citep[BIC;][]{Schwarz_1978} difference
\begin{equation}
  \Delta\mathrm{BIC}_b
  \coloneq
  \mathrm{BIC}_{\mathcal{M}_{\rm null},b}
  -\mathrm{BIC}_{\mathcal{M}_{\rm lens},b}
  =
  T_b-2\ln n_b,
  \label{eq:delta-bic-definition}
\end{equation}
where \(\mathrm{BIC}_{\mathcal M,b}\coloneq k_{\mathcal M}\ln n_b-2\widehat{\ell}_{\mathcal M,b}\), with \(\widehat{\ell}_{{\rm lens},b}\) evaluated at the delay that attains \(T_b\), and \(n_b\in\mathbb N\) is the number of fitted epochs in band \(b\) while \(k_{\rm null}=4\) and \(k_{\rm lens}=6\) count the fitted parameters including the analytical mean; the lens model adds \(q\) and \(\Delta_{\rm obs}\). We use \(\Delta\mathrm{BIC}_b\in\mathbb R\) only descriptively. Its sign does not fix a false-positive rate, and neither likelihood-ratio \(\chi^2\) asymptotics \citep{Wilks_1938,Protassov_2002} nor the BIC approximation to the evidence \citep{Drton_2013} apply, because the null is reached only in singular limits (\(q\rightarrow0\) or \(\Delta_{\rm obs}\rightarrow0\)). Detection thresholds must therefore be calibrated with null simulations or observed control light curves processed through the same analysis pipeline as the data.

Calibrating through the full search also accounts for the look-elsewhere
effect. Neighbouring trial delays are strongly correlated: for representative
dense ZTF mocks the profile correlation length is \(7\)--\(9\,{\rm d}\), giving
roughly ten effective trials across the \(3\)--\(150\,{\rm d}\) range, and the
null thresholds include this multiplicity automatically.

We calibrate thresholds within observable strata \(\mathcal A\), defined
by survey quality and, where used, the fitted null-model observer-frame
short-lag variability rate
\begin{equation}
  \widehat D_{b,{\rm obs}}
  \coloneq
  \frac{
    (\widehat\sigma_{F,b}^{\rm null})^2
  }{
    (1+z)\widehat\tau_b^{\rm null}
  }
  =
  \frac{
    (\widehat\sigma_{F,b}^{\rm null})^2
  }{
    \widehat\tau_{b,{\rm obs}}^{\rm null}
  },
  \qquad
  \widehat D_{b,{\rm obs}}\in\mathbb{R}_{>0}.
  \label{eq:fitted-observer-D}
\end{equation}
We calibrate in \(\widehat D_{\rm obs}\) rather than in amplitude and damping time separately
because finite survey baselines need not sample the DRW turnover. In this
regime, different \((\sigma_F,\tau)\) pairs can produce nearly indistinguishable
short-lag variability, while the likelihood primarily constrains their
combination \(D=\sigma_F^2/\tau\) \citep{Laguna-Miralles_2026}. Consistently,
\citet{Brewer_2026} show that short-term volatility is substantially better
constrained by finite quasar light curves than the damping timescale itself.
Thus \(\widehat D_{\rm obs}\) provides a more robust observable coordinate
for calibration and directly quantifies variability on the timescales to
which the delay search is most sensitive.

For any calibrated statistic \(T\), the threshold corresponding to a target
false-positive rate (FPR) \(\alpha\in(0,1)\) is
\begin{equation}
  T_{\mathcal A}^{\rm crit}(\alpha)
  \coloneq
  Q_{1-\alpha}
  \left(
    T\mid\mathcal M_{\rm null},\mathcal A
  \right),
  \label{eq:empirical-detection-threshold}
\end{equation}
where \(Q_{1-\alpha}\) is the empirical null quantile. We estimate
thresholds at \(\alpha=0.01\)
from one null sample and verify their achieved FPR on an independent held-out sample.

Because the lensing delay is common to all bands while flux ratios need not be (Section~\ref{sec:intro}), each band keeps its own \(q_b\) and nuisance parameters, and we combine the per-band profile-likelihood-ratio curves at a common delay. Their sum is a working-independence composite likelihood ratio \citep{Varin_2011}: it ignores the cross-band correlation in its form but not in its calibration. ZTF \(g\) and \(r\) sample the same intrinsic variability, making the \(\Lambda_b\) strongly correlated under both hypotheses. For this reason, we construct DRW mocks that share one latent source realisation across bands, allowing their null thresholds to account for this correlation.

For a band set \(\mathcal B\in\{\{g,r\},\{g,r,G\}\}\), we search for a
common delay that maximises the summed per-band likelihood improvement. We restrict this search to delays supported in both ZTF bands,
\(\Omega_{\rm coh}\coloneq\Omega_{g,{\rm obs}}\cap\Omega_{r,{\rm obs}}\subset\mathbb{R}_{>0}\), but Gaia does not restrict this domain further: at each common delay we fit the Gaia light curve directly, rather than interpolating a Gaia profile across delays that Gaia alone does not support. At each trial delay we add the per-band improvements \(\Lambda_b\), and the coherent statistic is the largest such sum,
\begin{equation}
  T_{\rm coh}^{\mathcal B}
  \coloneq
  \max_{\Delta_{\rm obs}\in\mathcal E_{\rm coh,\mathcal B}}
  \sum_{b\in\mathcal B}
  \Lambda_b(\Delta_{\rm obs}),
  \qquad T_{\rm coh}^{\mathcal B}\in\mathbb{R},
  \label{eq:coherent-statistic}
\end{equation}
where \(\mathcal E_{{\rm coh},\mathcal B}\subseteq\Omega_{\rm coh}\) is the set
of common delays at which we evaluate the sum. We seed it from maxima of the
summed profile and from the retained single-band solutions, and refine
continuously every basin within 10 summed log-likelihood units of the best seed, refitting the nuisance parameters of each band at any new delay.

For descriptive consistency with the single-band score, let
\(N_{\mathcal B}\coloneq\sum_{b\in\mathcal B}n_b\) be the total number of fitted
measurements. For \(B\coloneq|\mathcal B|\) bands, the null model has \(4B\)
parameters, while the lens model adds one \(q_b\) per band and one shared
delay, giving \(5B+1\). Using the total measurement count \(N_{\mathcal B}\) to set the penalty, we define the descriptive BIC-like score
\begin{equation}
\Delta\mathrm{BIC}_{\rm coh}^{\mathcal B}
\coloneq
T_{\rm coh}^{\mathcal B}
-(B+1)\ln N_{\mathcal B}.
\label{eq:coherent-delta-bic-definition}
\end{equation}
The penalty is therefore \(3\ln(n_g+n_r)\) for \(g+r\) and
\(4\ln(n_g+n_r+n_G)\) for \(g+r+G\); the band counts need not be equal or the observations simultaneous.

We calibrate empirical null thresholds separately for each
statistic and band set, recording searches with empty
supported domains as unscoreable. The coherent \(\Delta\mathrm{BIC}\) is descriptive, not a
composite-likelihood information criterion or Bayes-factor
approximation \citep{Varin_2011}, and its \(N_{\mathcal B}\)-dependent penalty means a fixed \(\Delta\mathrm{BIC}_{\rm coh}\) threshold implies different \(T_{\rm coh}\) thresholds across sources.

\begin{figure*}
    \centering
    \includegraphics[width=\textwidth]{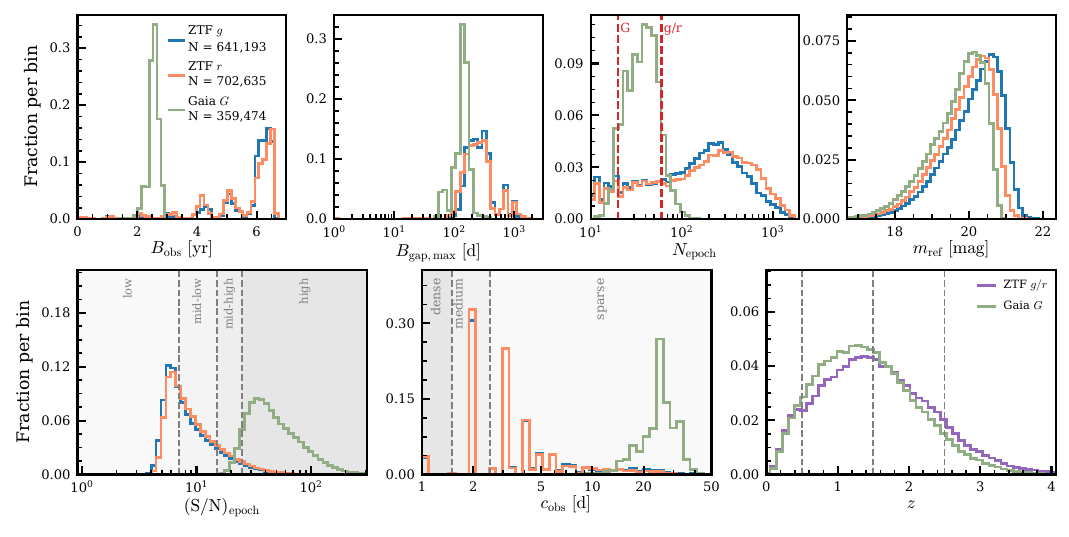}
    \caption{Distributions of observer-frame baseline ($B_{\rm obs}$), maximum gap ($B_{\rm gap,max}$), retained epoch count ($N_{\rm epoch}$), and reference magnitude ($m_{\rm ref}$) (top row); per-epoch relative-flux S/N ratio ($({\rm S/N})_{\rm epoch}$), median intra-season night cadence ($c_{\rm obs}$), and spectroscopic redshift ($z$) (bottom row). ZTF ($g$/$r$) curves with at least ten epochs retained after the depth cut, and Gaia ($G$) curves with at least ten retained epochs and a qualifying ZTF counterpart, are overlaid, with fractions normalised within each displayed range and cadence shown logarithmically. Red dashed lines mark calibration epoch requirements; grey guides show the observing regimes and source redshifts adopted for mock calibration (Table~\ref{tab:mock-calibration-grid}). \keyidea the parent sample spans broad cadence and S/N ranges, requiring different calibration regimes.}
    \label{fig:multiband_survey_distributions}
\end{figure*}

\section{Validation experiments}
\label{sec:validation}

We validate QUELLS in two steps. In
Section~\ref{sec:mock-calibration}, survey-realistic mocks with known source and lens parameters show how the false-positive rate, detection efficiency and delay recovery depend on the source, lens and observing conditions. These mocks have intrinsic
variability following a DRW, albeit in log-flux rather than in the linear flux
assumed by the fit, so they test the flux approximation of the detector
but not departures from the DRW itself. In
Section~\ref{sec:empirical-injection} we inject delayed copies
into observed AGN light curves, whose variability and systematics are
those of real quasars, to test whether the search remains effective
beyond the DRW family.

\subsection{Target AGN sample and observing regimes}
\label{sec:observing-regimes}

Figure~\ref{fig:multiband_survey_distributions} summarises the parent
sample of Section~\ref{sec:intro}, using DR23-era ZTF photometry and
the epoch-quality criteria of Section~\ref{sec:preprocessing}. We count each photometric identifier once per band, requiring at least ten retained epochs. For Gaia, the nearest match within \(2''\) must also have qualifying ZTF photometry and a valid spectroscopic redshift, giving \(359\,474\) \(G\) curves. Known lenses remain included even when separate Gaia images share a blended ZTF curve; in the ZTF redshift panel, these images are grouped into one entry per verified lens system (known lenses in our sample are discussed in Section~\ref{sec:known-lenses}).

Table~\ref{tab:mock-calibration-grid} summarises the source, lens and
survey parameters used for the controlled DRW calibration.
We stratify the ZTF light curves by
\(({\rm S/N})_{\rm epoch}\), defined in Eq.~\eqref{eq:epoch-snr}, and the
median observer-frame intra-season cadence
\(c_{\rm obs}\in\mathbb{R}_{>0}\), measured in days after grouping same-night observations
and splitting at gaps longer than \(90\,{\rm d}\). Four S/N bins and three cadence bins (bottom rows of Table~\ref{tab:mock-calibration-grid}) define 12 observing regimes, where for paired \(g+r\) calibration, we require at least 60
retained epochs per band and matching regime assignments. For
mocks including Gaia \(G\), we require at least
20 retained Gaia epochs, allowing for its sparser
scanning-law cadence. After the quality cuts, \(536\,482\) catalogue entries have at least 60 retained epochs in \(g\), \(r\), or both. Of these, \(428\,388\) satisfy the two-band epoch requirement and \(275\,142\) also have matching regime assignments (Table~\ref{tab:regime-counts}). For these ZTF catalogue entries with matching \(g\)- and \(r\)-band regimes, we find \(166\,884\) distinct Gaia \(G\) light curves satisfying the Gaia epoch cut. The regime-matching requirement applies only to the coherent \(g+r\) calibration; all sources meeting the ZTF epoch requirement remain eligible for separately calibrated single-band searches.

\subsection{Survey-realistic DRW mock calibration}
\label{sec:mock-calibration}

The controlled mocks preserve the sampling and heteroscedastic
uncertainties of real survey light curves while providing known
source and lens parameters. Following the motivation in
Section~\ref{sec:drw-source-model}, we generate log-flux DRW
variability with additive Gaussian flux noise and analyse it
with the Gaussian-flux detector, corresponding to Models~2 and~1,
respectively, in Appendix~\ref{sec:four-model-DRW-comparison}.
This tests the effect of the intrinsic-variability
approximation of the detector on false positives, detection efficiency and
delay recovery.

\subsubsection{Mock construction and calibration grid}
\label{sec:calibration-grid}

Each calibration cell represents one combination of the parameter
values listed in Table~\ref{tab:mock-calibration-grid}. From each regime we randomly select 50 eligible AGN templates, indexed by \(i\in\mathbb{N}_{>0}\), and reuse their observer-frame epochs \(t_{b,i,j,{\rm obs}}\) and
relative-flux uncertainties \(s_{b,i,j}\) across parameter combinations in both the null and lensed ensembles, so that differences reflect the injected parameters at fixed sampling and reported noise.

The redshifts span the bulk of the parent sample, while the damping timescales and amplitudes cover representative quasar-variability regimes \citep{MacLeod_2010,Suberlak_2021}, and the lens parameters (flux ratio $q$ and observer-frame delays $\Delta_{\rm obs}$) cover different galaxy-scale double-lens configurations (Section~\ref{sec:observing-regimes}). 

\begin{figure}
\centering
\includegraphics[width=\columnwidth]{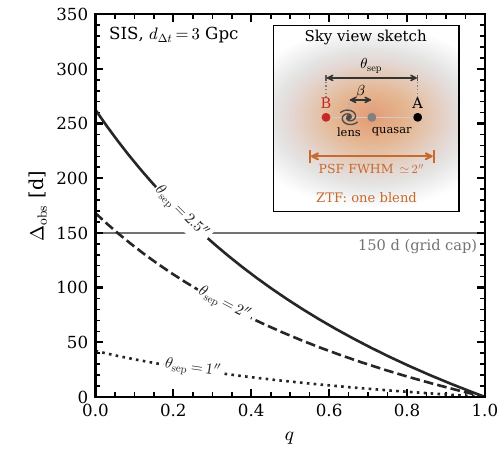}
\caption{
SIS flux-ratio--delay relation (Eq.~\eqref{eq:sis-delay}) at the reference distance; the grey line marks the \(150\,{\rm d}\) search limit.
Inset: sky geometry in the notation of Figure~\ref{fig:pipeline_comparison}, showing
images A (black) and B (red), the deflector galaxy and the unlensed source direction
(grey). Orange shading illustrates blending in ZTF, with the arrow indicating a typical PSF FWHM of \(2''\).
\keyidea at fixed separation (driven by the maximum separation not resolved by ZTF) and distance (the distribution of redshifts in our quasar sample), a brighter secondary image implies a shorter SIS delay, motivating our calibration and delay search limits.
}
\label{fig:sis-delay-paper}
\end{figure}

The grid varies $q$ and $\Delta_{\rm obs}$ independently, but in real lenses both are set by the same geometry. For an SIS lens with Einstein radius
\(\theta_{\rm E}\in\mathbb{R}_{>0}\), a source at angular
offset \(\beta\in(0,\theta_{\rm E})\) produces two images
on opposite sides of the deflector, at angular distances
\(\theta_{\rm E}+\beta\) and \(\theta_{\rm E}-\beta\)
from its centre.
Their separation is
\(\theta_{\rm sep}=2\theta_{\rm E}\in\mathbb{R}_{>0}\),
and their flux ratio is
\(q=(\theta_{\rm E}-\beta)/(\theta_{\rm E}+\beta)\)
(Eq.~\eqref{eq:lens-flux-ratio}), with the corresponding observer-frame delay
\begin{equation}
\Delta_{\rm obs}=\frac{d_{\Delta t}}{2c}\,\theta_{\rm sep}^{2}
  \frac{1-q}{1+q}\simeq42\,{\rm d}
  \left(\frac{\theta_{\rm sep}}{1''}\right)^{2}
  \frac{1-q}{1+q}
  \left(\frac{d_{\Delta t}}{3\,{\rm Gpc}}\right),
\label{eq:sis-delay}
\end{equation}
where \(c\) is the speed of light and
\(\theta_{\rm sep}\) is in radians in the first equality.
Using the notation of Figure~\ref{fig:pipeline_comparison}, the time-delay distance is \(d_{\Delta t}\coloneq
(1+z_{\rm l})d_{\rm l}d_{\rm s}/d_{\rm ls}
\in\mathbb{R}_{>0}\), where \(z_{\rm l}\in\mathbb{R}_{>0}\) is the lens redshift. We adopt \(d_{\Delta t}=3\,{\rm Gpc}\) as a reference value,
representative of a galaxy lens at \(z_{\rm l}=0.5\) and a quasar
source at \(z=2\). This source redshift lies near the middle
of our sample distribution (Figure~\ref{fig:multiband_survey_distributions}), with other choices of
\(d_{\Delta t}\) rescaling the delay linearly.
Figure~\ref{fig:sis-delay-paper} shows this relation for
\(\theta_{\rm sep}\in\{1'',2'',2.5''\}\).

With a typical PSF FWHM of \(\simeq2''\), ZTF can blend
pairs out to \(\simeq2.5''\), depending on seeing, S/N
and image contrast. Within this blended regime, brighter secondaries produce stronger echoes (Eq.~\eqref{eq:lens-kernel}) and, at fixed separation and distance, shorter delays.
At the reference \(d_{\Delta t}\),
\(\theta_{\rm sep}\leq2''\) gives
\(\Delta_{\rm obs}<150\,{\rm d}\) throughout the searched
range \(q\geq0.1\), while at \(2.5''\) this holds for
\(q\gtrsim0.28\).
Together, these considerations motivate the
\(150\,{\rm d}\) search limit. Larger \(d_{\Delta t}\) or fainter secondaries permit longer delays, but seasonal gaps increasingly limit the sampled overlap with the delayed copy
(Section~\ref{sec:likelihood-fitting}).

The independent grid of Table~\ref{tab:mock-calibration-grid}
therefore measures how detection efficiency changes with
each parameter while holding the others fixed.
It is a controlled experiment rather than a lens-population
prior; we return to this distinction in
Section~\ref{sec:operating-region}.

Crossing these parameters with the
12 observing regimes defines the conditional detection efficiency: the probability of selecting an injected lens at the corresponding null-calibrated threshold,
\begin{equation} P_{\rm det} = P_{\rm det}\!\left( z,\tau_{\rm src},\sigma_{{\rm src},m}, q,\Delta_{\rm obs}, ({\rm S/N})_{\rm epoch},c_{\rm obs} \right) \in[0,1]. \label{eq:calibrated-selection-function} \end{equation}

\begin{table}
\centering
\footnotesize
\setlength{\tabcolsep}{3.5pt}
\begin{tabular}{@{}llp{4.8cm}c@{}}
\toprule
Class & Coordinate & Grid values or definition & Levels \\
\midrule

Source &
\(z\) &
\(\{0.5,1.5,2.5\}\) &
3 \\

Source &
\(\tau_{{\rm src}}\) &
\(\{50,100,300,500,700,1000\}\ {\rm d}\) &
6 \\

Source &
\(\sigma_{{\rm src},m}\) &
\(\{0.08,0.14,0.20\}\ {\rm mag}\) (\(r\) band) &
3 \\

Lens &
\(q\) &
\(\{0.2,0.4,0.7,1.0\}\) &
4 \\

Lens &
\(\Delta_{\rm obs}\) &
\(\{10,20,30,40,60,90,120\}\ {\rm d}\) &
7 \\

Survey &
\(({\rm S/N})_{\rm epoch}\)  &
low (\(\leq7\)), mid-low (\(7,15]\),\\
& &
mid-high (\(15,25\)), high (\(\geq25\)) &
4 \\

Survey &
\(c_{\rm obs}\) &
dense (\(\leq1.5\,{\rm d}\)),\\
& &
medium (\((1.5,2.5]\,{\rm d}\)),\\
& &
sparse (\(>2.5\,{\rm d}\)) &
3 \\

\bottomrule
\end{tabular}
 \caption{Mock calibration grid, with \(18\,144\) lensed cells across 12 observing regimes. The \(r\)-band magnitude amplitudes define controlled weak, reference and strong variability scenarios.}
\label{tab:mock-calibration-grid}
\end{table}

\begin{table}
\centering
\begin{tabular}{lrrr}
\hline
S/N regime & sparse & medium & dense\\
\hline
high      & 8\,394  & 11\,530 & 1\,674\\
mid-high  & 16\,834 & 19\,314 & 2\,189\\
mid-low   & 83\,257 & 73\,125 & 4\,943\\
low       & 36\,985 & 16\,674 & 223\\
\hline
\end{tabular}
\caption{Number of ZTF catalogue entries in the 12 observing regimes
satisfying the paired \(g+r\) requirements after quality cuts
(\(275\,142\) entries in total).}
\label{tab:regime-counts}
\end{table}

Each grid redshift is assigned independently of the redshift of the template source,
and its epochs are converted to rest-frame time with
Eq.~\eqref{eq:rest-frame-time}, so sampling and uncertainties stay fixed while
the rest-frame cadence changes with \(z\).

For each simulated system and ensemble
\(e\in\{{\rm null},{\rm lens}\}\), we draw an independent realisation
\(x_{{\rm src},m}^{e}(t)\) of a stationary zero-mean DRW in \(r\)-band
magnitude, with covariance
\(\sigma_{{\rm src},m}^{2}\exp(-|t-t'|/\tau_{\rm src})\) in rest-frame
time. All bands share this realisation and its damping timescale and differ only in
amplitude, through the factors
\begin{equation}
a_r=1,\qquad a_g=1.17,\qquad a_G=1.03,
\qquad a_b\in\mathbb{R}_{>0},
\label{eq:band-amplitude-factors}
\end{equation}
since the observed wavelength dependence of the damping timescale is weak
compared with the population spread in \(\tau\) \citep{MacLeod_2010,Yu_2026}. \footnote{The relation \(SF_\infty\propto\lambda^{-0.479}\) of
\citet{MacLeod_2010} gives \(a_g\simeq1.15\) at the ZTF effective wavelengths,
whereas more recent measurements find steeper blue enhancements, such as
\(\sigma_{\rm DHO}\propto\lambda^{-0.73\pm0.07}\) \citep{Yu_2022} and a
timescale-dependent dependence in ZTF \(g\) and \(r\) \citep{Patel_2025}; we
adopt the conservative (higher variability improves recovery) value \(a_g=1.17\). For Gaia, \(a_G=1.03\) follows from
integrating the MacLeod scaling over the \(G\) passband for
\(f_\nu\propto\nu^{-0.5}\).}

The intrinsic relative-flux
fluctuation in band \(b\in\{g,r,G\}\) is then
\begin{equation}
x_{{\rm src},F,b}^{e}(t)
\coloneq
\exp\!\left[
\gamma a_b x_{{\rm src},m}^{e}(t)
-\frac{(\gamma a_b\sigma_{{\rm src},m})^2}{2}
\right]-1 \in (-1,\infty),
\label{eq:mock-source-parameters}
\end{equation}
where the final term ensures \(\langle 1 + x_{{\rm src},F,b}^{e} \rangle = 1\).

\begin{figure*}
    \centering
    \includegraphics[width=\textwidth]{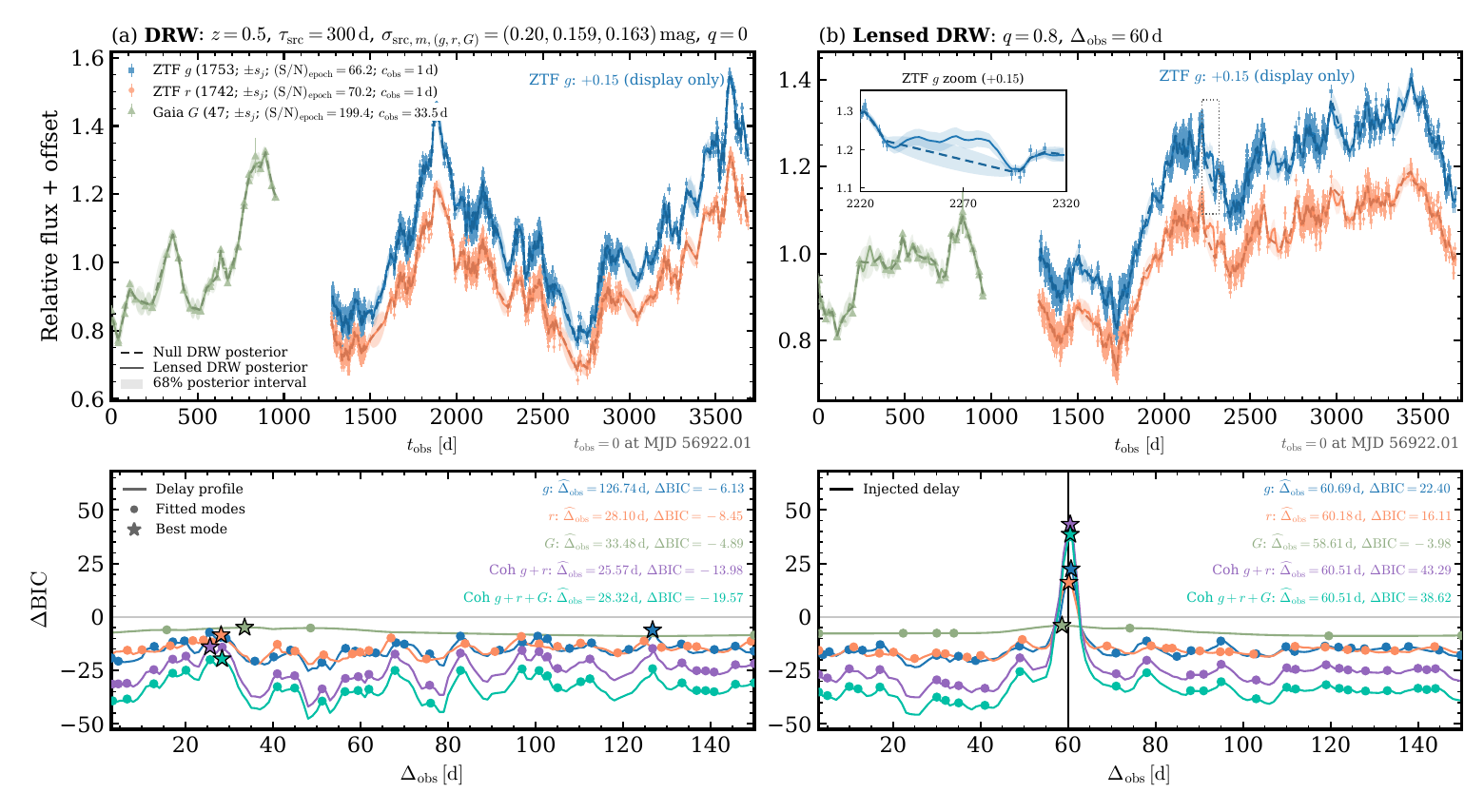}
    \caption{Example DRW mock light curves and delay searches at ZTF
    \(g,r\) and Gaia \(G\) epochs, using the sampling and photometric
    uncertainties of SDSS J170717.75+453610.5.
    (a) Unlensed control; (b) lensed mock, sharing the source parameters
    listed in (a).
    Top: mock observations with error bars \(\pm s_j\), null (dashed)
    and delayed-mixture (solid) posterior means, and shaded 68 per cent
    posterior intervals. Bottom: individual-band and coherent searches, annotated with
    maximum-likelihood delays and descriptive \(\Delta\mathrm{BIC}\)
    values. 
    \keyidea individual ZTF bands constrain the injected delay, and
    coherent fitting reinforces their shared peak, with no corresponding
    feature in the unlensed control.}
    \label{fig:mock_singleband_example_fits}
\end{figure*}

At the rest-frame epochs \(t_{b,i,j}\) of
template \(i\) in band \(b\), the null and lensed mock measurements are
\begin{align}
y^{\rm null}_{b,i,j}
&\coloneq
1+x_{{\rm src},F,b}^{\rm null}(t_{b,i,j})
+\epsilon^{\rm null}_{b,i,j},
\label{eq:calibration-null-mock}
\\
y^{\rm lens}_{b,i,j}
&\coloneq
1+\frac{
x_{{\rm src},F,b}^{\rm lens}(t_{b,i,j})
+q\,x_{{\rm src},F,b}^{\rm lens}(t_{b,i,j}-\Delta)
}{1+q}
+\epsilon^{\rm lens}_{b,i,j},
\label{eq:calibration-lens-mock}
\end{align}
in relative flux, where division by \(1+q\) preserves unit mean flux. We characterise the injected source variability by its rest-frame \(r\)-band rate \(D_m\coloneq\sigma_{{\rm src},m}^2/\tau_{\rm src}
\in\mathbb{R}_{>0}\), measured in
\({\rm mag}^2\,{\rm d}^{-1}\).

Figure~\ref{fig:mock_singleband_example_fits} compares null and
lensed realisations in the high-S/N, dense-cadence ZTF regime,
holding the sampling and photometric uncertainties fixed.
Under these conditions, the lensed realisation exhibits a common
peak near the injected delay, whereas the unlensed control shows
no corresponding feature.
While the resulting \(\Delta\mathrm{BIC}\) profiles illustrate
the separation between these two realisations, detection
significance must be established through separate null
calibration for each band and band combination
(Section~\ref{sec:detection-calibration}).

Noise is added once, in relative flux, after the two image contributions
have been combined, as for an unresolved measurement. For each template
epoch it includes the reported uncertainty \(s_{b,i,j}\) and the fitted
white-noise excess, \(\epsilon^e_{b,i,j}\sim\mathcal N\big(0,\,s_{b,i,j}^{2}+(\widehat{\sigma}_{{\rm add},F,b,i}^{\rm null})^2\big)\).
The draws are independent across epochs, bands and ensembles and of the
source process. We hold the epoch variance fixed as the simulated flux
varies and treat the fitted excess as a constant additive term in flux;
Appendix~\ref{sec:four-model-DRW-comparison} compares the latter choice
with a magnitude-constant prescription on observed light curves.

The \(g+r+G\) mocks use Gaia epoch photometry where available
and assume that both images contribute to one unresolved
\(G\)-band light curve. The full grid
contains \(907\,200\) lensed and \(162\,000\) null mocks. Within each
observing regime, the 50 templates are split into two fixed folds; nulls
from one fold calibrate thresholds applied to the other, with all
realisations of a template kept together.

\subsubsection{Detection performance, delay recovery and multiband gains}
\label{sec:mock-detection-performance}

\begin{figure*}
    \centering
    \includegraphics[width=1\linewidth]{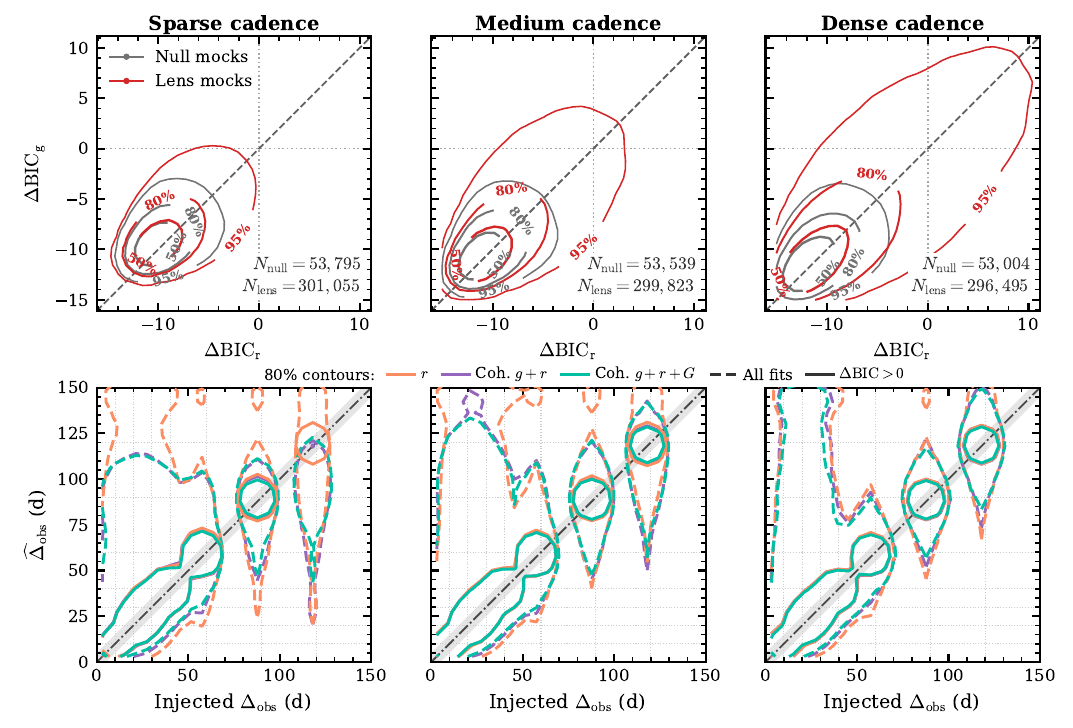}
    \caption{Model separation and delay recovery across observing cadences.
    Top: joint distributions of the single-band scores for null and lensed
    mocks, with contours enclosing 50, 80 and 95 per cent of each sample.
    Bottom: recovered versus injected delays, with contours enclosing 80 per cent of the smoothed density, normalised separately for all usable lensed fits (dashed) and the $\Delta\mathrm{BIC}>0$ subset (solid); the grey band marks agreement within $\pm5$\,d. Upper panels require usable scores
    in both ZTF bands, whereas lower samples are selected separately for
    each band combination and require a finite recovered delay. Among usable lensed fits in the lower panels, the percentages with $\Delta\mathrm{BIC}>0$, in the order $r$, $g+r$, $g+r+G$, are 2.3, 2.7 and 1.8 per cent for sparse cadence; 4.7, 6.1 and 4.2 per cent for medium cadence;
    and 10.7, 12.7 and 10.6 per cent for dense cadence. \keyidea denser sampling separates null and lensed scores more clearly and improves delay recovery, with coherent fits performing best.}
    \label{fig:p6d_band_score_delay}
\end{figure*}

Figure~\ref{fig:p6d_band_score_delay} shows how cadence affects detection and
delay recovery in the DRW mocks. In the top row, the null distributions are concentrated
at negative \(\Delta\mathrm{BIC}\), whereas the lensed
distributions extend to increasingly positive values
with denser sampling, improving discrimination in each band. The two scores are correlated along the diagonal because both
bands sample the same intrinsic variability.

In the bottom row, coherent \(g+r\) and \(g+r+G\) fits recover the injected delay better than a single band (illustrated with ZTF \(r\)) at every cadence. This is because combining bands helps reduce the influence of independent photometric noise, although substantial off-diagonal structure remains at sparse cadence. Restricting to \(\Delta\mathrm{BIC}>0\) tightens recovery
further, indicating that stronger model preference is associated with more reliable delays. These recovery results are obtained with log-flux DRW mocks, whereas the detector fits a linear-flux DRW
(Section~\ref{sec:drw-source-model}). Successful recovery under this deliberate model mismatch supports the linear-flux approximation as an effective approach to delay searches at these variability amplitudes.

\begin{figure}
    \centering
    \includegraphics[width=1\columnwidth]{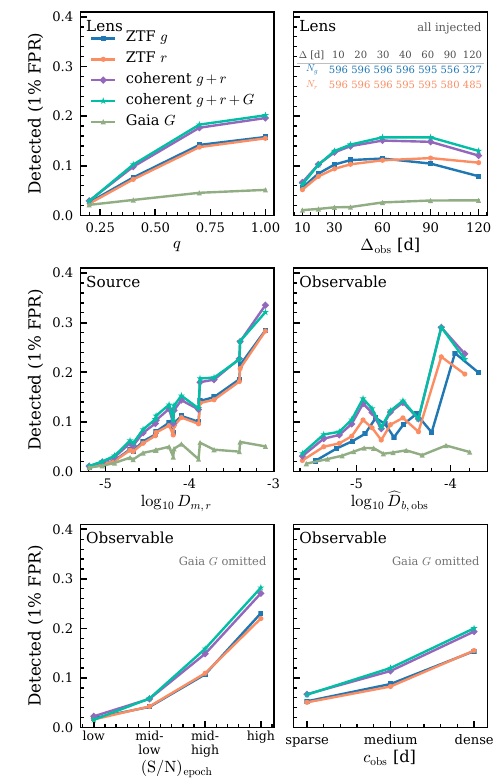}
    \caption{Detection fractions at a nominal 1 per cent DRW-calibrated FPR for individual bands and coherent combinations, shown against lens parameters (top row), injected source variability (middle left) and observable quantities (middle right and bottom row). The delay panel includes all injected systems, counting unsupported or unusable cases as non-detections, where \(N_g\) and \(N_r\) give the numbers of templates supporting each delay (all other panels condition on temporal support). The injected variability is expressed through \(D_{m,r}=D_m\), the rest-frame \(r\)-band magnitude-space rate, and in the middle-right panel, each single-band curve uses the rate measured in that band, while both coherent curves use the \(r\)-band rate. The other two observable panels use ZTF S/N and cadence, so the Gaia \(G\)-only curve is omitted there. Both channels involving Gaia, \(G\) and \(g+r+G\), have lower-bound detection fractions because the low-S/N, sparse-cadence regime contains too few null mocks to calibrate their thresholds. \keyidea strong short-timescale variability, comparable image fluxes, high S/N and dense sampling favour detection, with sensitivity being highest at delays of tens of days, and coherent fitting providing the highest detection efficiency.}
    \label{fig:p6d-sanity-checks}
\end{figure}

We reiterate that a positive $\Delta\mathrm{BIC}$ does not set a FPR (Section~\ref{sec:detection-calibration}). For this reason, Figure~\ref{fig:p6d-sanity-checks} compares detection fractions at a nominal 1 per cent rate, using null-calibrated thresholds for each likelihood-ratio statistic and observing regime. At fixed observing conditions, detection depends on the lens parameters: more equal image fluxes (higher brightness ratio \(q\)) strengthen the delayed echo contribution, whereas small \(q\)
leaves only a weak perturbation of the primary light curve. Efficiency rises with delay to a broad plateau at 20--90\,d, since short-delay blends can resemble a single variable source, while at longer delays seasonal gaps remove the epochs whose delayed counterparts would be observed. At 120\,d this loss is larger in \(g\) than in \(r\), because fewer \(g\) templates keep enough overlap (327 against 485 of 596), so the \(r\) and coherent curves decline less. Coherent fitting is the most efficient channel throughout, while Gaia \(G\) alone is much less sensitive, and adding it to \(g+r\) changes the efficiency only modestly.

For the quasar source parameters, detection favours shorter
damping times and larger variability amplitudes. We summarise
both trends with the injected rest-frame \(r\)-band variability
rate \(D_m\), where larger \(D_m\) supplies stronger
short-timescale structure, making the delayed correlation easier
to distinguish from measurement noise.
Redshift enters only through time dilation
(Eq.~\eqref{eq:rest-frame-time}), so we do not show it separately.
The fitted null-model rate \(\widehat D_{b,{\rm obs}}\), which is
measurable for real sources, shows a noisier trend because each
bin mixes S/N and cadence, which also control detection. Efficiency also increases steadily with per-epoch S/N and with cadence: higher S/N preserves the correlation signature against noise, and denser sampling resolves variability on the timescales of the delay. These dependences motivate calibrating thresholds in cells of S/N, cadence and \(\widehat D_{b,{\rm obs}}\), all measurable for observed sources.

\begin{figure*}
    \centering
    \includegraphics[width=1\linewidth]{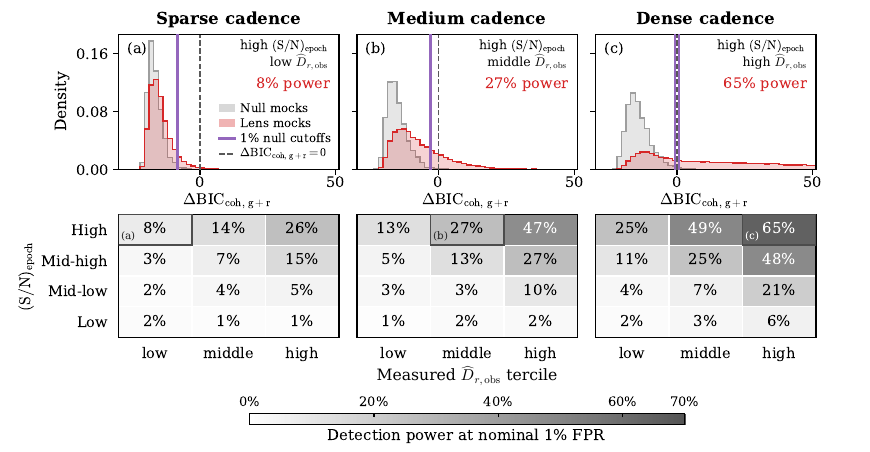}
      \caption{Observable-space calibration of coherent $g+r$ lens selection.
      Columns show sparse, medium and dense cadences. Top: density-normalised
      $\Delta\mathrm{BIC}_{\rm coh,\,g+r}$ distributions of null (grey) and lensed
      (red) mocks in the outlined cells below. The two purple lines mark the
      99th-percentile null cutoffs estimated separately from the two template
      folds, each targeting a nominal 1 per cent FPR; dashed lines
      mark zero. Bottom: detection fractions in 36 cells of
      $({\rm S/N})_{\rm epoch}$, cadence and measured $\widehat D_{r,{\rm obs}}$ tercile.
      In each cell, tercile boundaries and cutoffs learned from one fold
      are applied to the other, and vice versa; the percentage is the total
      number of detected lensed mocks divided by the total number of usable
      lensed mocks across both folds. \keyidea detection power depends strongly on the observable cell, from about 1 per cent at low S/N to 65 per cent in dense, high-S/N, high-\(\widehat D_{r,{\rm obs}}\) cells.}
    \label{fig:p6d-observable-calibration}
\end{figure*}

Figure~\ref{fig:p6d-observable-calibration} shows coherent $g+r$
DRW calibration and selection in this observable space. Null mocks determine the
$\Delta\mathrm{BIC}_{\rm coh,\,g+r}$ thresholds within each observing
regime and null-defined $\widehat D_{r,{\rm obs}}$ tercile, while lensed mocks measure
the corresponding efficiencies. Assigning an observed source to a
calibration cell therefore requires no knowledge of its lens
parameters, which are unknown for real sources. The portability of this DRW calibration to observed data is
discussed in Section~\ref{sec:real-calibration}. The all-injected fractions differ from the ones in Figure~\ref{fig:p6d-sanity-checks} because we now condition on usable fits within the fitted-$\widehat D_{r,{\rm obs}}$
window, without an injected-delay-support cut. Fixing S/N and cadence removes the mixing that blurs the \(\widehat D_{r,{\rm obs}}\) trend in Figure~\ref{fig:p6d-sanity-checks}: at mid-high and high S/N, efficiency rises with the \(\widehat D_{r,{\rm obs}}\) tercile at every cadence, up to 65 per cent in the dense, high-S/N, upper-tercile cell, where the null and lensed \(\Delta\)BIC distributions separate most. However, the contribution of these high-efficiency regimes
to a survey-selected lens sample also depends on how many
catalogue sources occupy them (Table~\ref{tab:regime-counts})
and on the abundance and properties of the underlying lens
population (Figure~\ref{fig:sis-delay-paper};
Section~\ref{sec:discussion}).

\subsection{Empirical injection--recovery with observed AGN light curves}
\label{sec:empirical-injection}

We now extend the echo-injection tests to observed quasar
light curves. We add delayed copies of each observed light curve to itself and use the original light curve, on the same retained epochs, as an echo-free control (this control is not an independently confirmed unlensed source). To separate the effect of real variability from that of the injection procedure, we process a matched DRW realisation on the same epochs in exactly the same way.

\subsubsection{Real AGN light curves and injection procedure}

We use the ZTF $g$ and $r$ light curves of the dense-cadence templates from Section~\ref{sec:calibration-grid}. We restrict the experiment to this cadence because the delayed image must be reconstructed from the observed data: at medium cadence the reconstruction relies mainly on interpolation across gaps between observing nights, and the outcome would then test the construction rather than the method. For the same reason, we require at least 300 epochs and no gap longer than 700\,d in both bands, so that enough observations survive the delay mask to constrain a delayed correlation. Since this mask retains only epochs whose delayed counterparts can be reconstructed and removes up to
half of the observations at the longest delays, we further require at least 200 retained epochs per template, band and injected delay. Of the 150 dense-cadence templates in the three higher S/N groups, six fail these cuts and three are catalogued blazars, identified through positional cross-matches against SIMBAD \citep{Wenger_2000} and Roma-BZCAT \citep{Massaro_2015}. Three further sources fail the cleaned-data requirements and one lacks sufficient echo support, leaving 137 sources: 45 mid-low, 45 mid-high and 47 high S/N, spanning \(0.12<z<3.41\).

\begin{figure*}
    \centering
    \includegraphics[width=\textwidth]{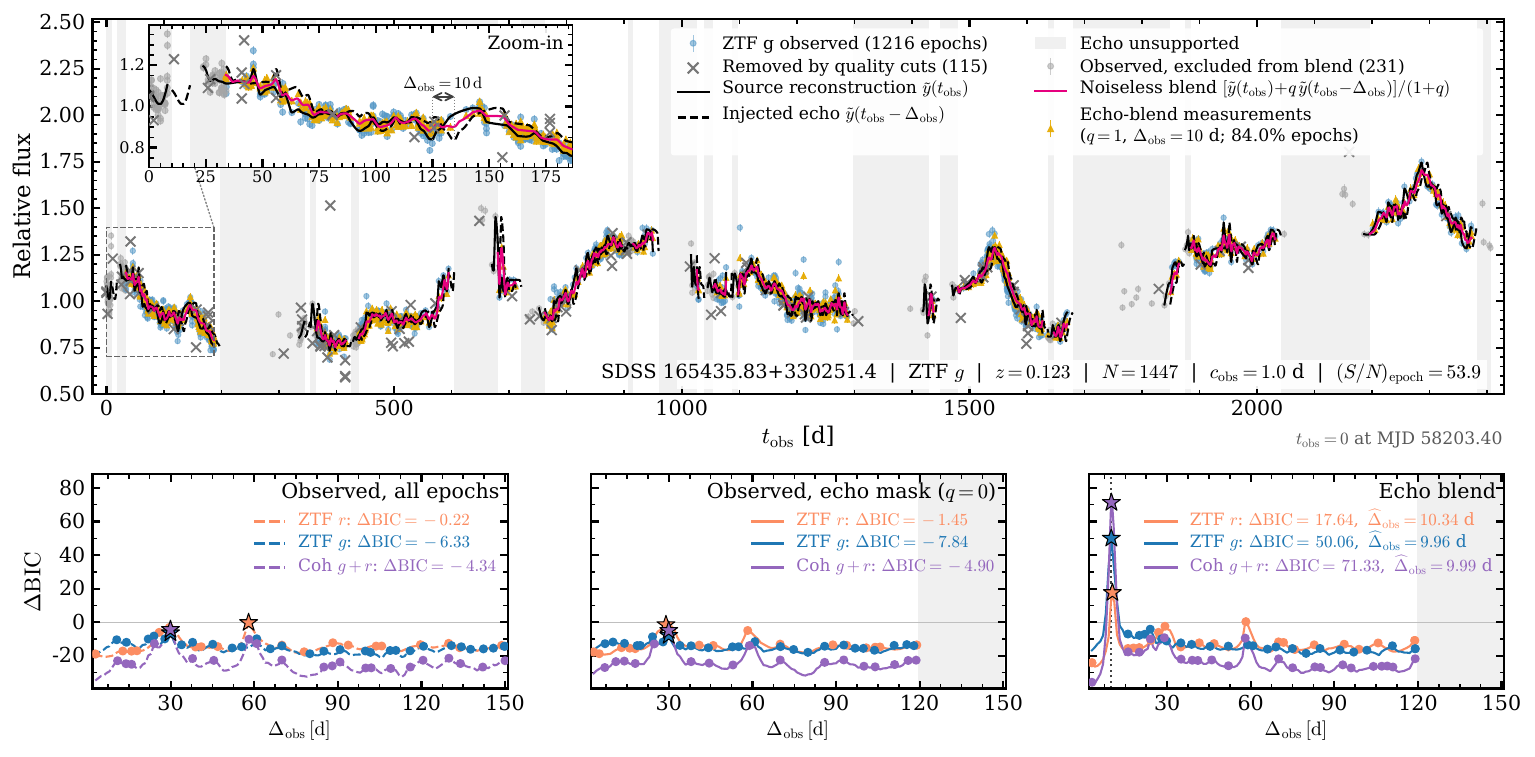}
    \caption{Construction and recovery of an echo injected into
    an observed quasar light curve.
    Top: the reconstructed variability supplies a delayed copy
    that is combined with the observations to form the blend.
    Grey shading marks epochs without a reconstructed delayed
    counterpart, and crosses indicate quality rejections.
    Bottom: single-band and coherent searches compare all cleaned
    observations (left), the matched echo-free control (middle)
    and the injected blend (right), with grey shading marking
    delays outside all displayed search domains.
    \keyidea the lensed DRW search from QUELLS recovers the injected delay from observed quasar variability, without assuming a DRW
    source process in the injection construction.}
    \label{fig:empirical_echo_blend}
\end{figure*}

The measured curve supplies the primary image; for the delayed copy we construct a smoothed, interpolated relative-flux curve
\(\tilde{y}_{b,i}(t_{\rm obs})\). Figure~\ref{fig:empirical_echo_blend} illustrates this construction for one
source, which we use as a running example: the reconstruction follows
the observed variability, and its copy, shifted by the injected delay,
is added to form the blend. To build it, we
combine observations within 0.5 observer-frame days with inverse-variance
weights and split the result at gaps longer than 10~d, keeping segments
with at least five such groups.

Within each segment we apply inverse-variance-weighted Gaussian smoothing of width \(h\in\mathbb{R}_{>0}\), truncated at $3h$, followed by piecewise cubic Hermite interpolation \citep{FritschButland_1984}, without extrapolation or interpolation across segment boundaries. We use $h=1$\,d at $\Delta_{\rm obs}=10$\,d and otherwise \(h=\max(2\,{\rm d},\,1.5\,c_{b,{\rm obs}})\), where \(c_{b,{\rm obs}}\) is the median intra-season cadence (Section~\ref{sec:observing-regimes}) of the cleaned light curve; the latter widths are 2--3\,d in this suite. We note how smoothing suppresses the short-timescale variability
that helps constrain the delay, while leaving residual correlated
measurement noise in the reconstructed echo. This loss of useful
structure makes our recovery tests more demanding, particularly at
\(\Delta_{\rm obs}=10\)\,d, where recovery relies strongly on
short-timescale variability.

As in the mocks (Eq.~\ref{eq:calibration-lens-mock}), the injected measurements are
\begin{equation}
    y^{\rm lens}_{b,i,j}
    \coloneq
    \frac{1}{1+q}\Bigl[
        y_{b,i,j}
        + q\,\tilde{y}_{b,i}\!\bigl(t_{b,i,j,{\rm obs}}-\Delta_{\rm obs}\bigr)
    \Bigr]
    + \epsilon^{\rm lens}_{b,i,j},
    \label{eq:empirical-blend}
\end{equation}
with the delay applied in the observer frame and the factor \((1+q)^{-1}\) keeping the mean flux equal to that of the observed curve.

Scaling the observed curve by \((1+q)^{-1}\) also scales its measurement noise,
to variance \(s_{b,i,j}^2/(1+q)^2\). We therefore add fresh Gaussian noise,
\begin{equation}
    \epsilon^{\rm lens}_{b,i,j}
    \sim
    \mathcal N\!\Bigl(0,\,
        s_{b,i,j}^{2}\bigl[1-(1+q)^{-2}\bigr]\Bigr),
    \label{eq:empirical-fresh-noise}
\end{equation}
so that the white-noise variance of each injected epoch returns to the declared
\(s_{b,i,j}^2\), which is also what the detector receives. Two differences from
an ideal blend remain: the reconstructed echo carries some smoothed, correlated
noise from the observed curve, and any unmodelled excess scatter of the primary
is scaled by \((1+q)^{-1}\) like the rest of the curve. 

The echo-free control retains the same masked observations
without adding an echo or fresh noise.
In Figure~\ref{fig:empirical_echo_blend}, it shows no strong
delayed-mixture preference, whereas the blend peaks at the
injected 10-d delay. For comparison, matched DRW light curves use the same cleaned epochs, the fitted \(r\)-band null-model timescale and amplitude, and the template redshift. Equation~\eqref{eq:band-amplitude-factors} sets the other band amplitudes, while measurement noise combines the declared photometric uncertainties and fitted excess scatter. Observed and simulated curves undergo the same reconstruction,
masking and noise rebalancing, with shared fresh-noise draws. This holds the injection procedure fixed, although the comparison cannot distinguish intrinsic departures from a DRW from photometric systematics.

\subsubsection{Experiment, scoring and recovery}
\label{sec:empirical-recovery}

\begin{figure*}
    \centering
    \includegraphics[width=1\linewidth]{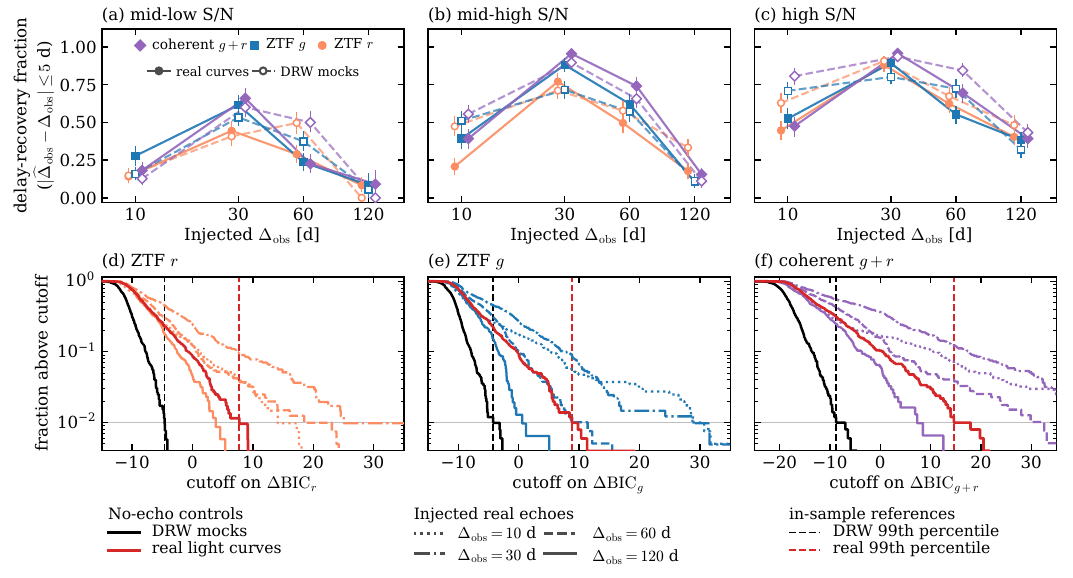}
    \caption{Delay recovery and score distributions for 137 dense-cadence templates: 45 mid-low, 45 mid-high and 47 high S/N. (a)--(c) Delay recovery for observed light curves and matched DRW mocks, shown separately by S/N and pooling the three injected flux ratios. Recovery is the fraction of usable blends whose preferred delay lies within \(5\,{\rm d}\) of the injected value, with error bars showing the central 68 per cent source-bootstrap intervals. (d)--(f) Fractions of usable fits exceeding each \(\Delta\mathrm{BIC}\) cutoff, comparing observed and DRW controls with injected observed blends across the three S/N regimes. Dashed vertical lines mark the 99th percentiles of the displayed control samples; because these references are estimated from the same samples, they do not establish independently validated false-positive thresholds. \keyidea delay recovery remains effective in observed light curves, but excess control scores relative to matched DRW realisations require empirical calibration of survey thresholds.
}
\label{fig:empirical_echo_suite}
\end{figure*}

We inject delays \(\Delta_{\rm obs}\in\{10,30,60,120\}\,{\rm d}\)
with flux ratios \(q\in\{0.4,0.7,1\}\) into both ZTF bands,
using one noise realisation per combination. Each delay defines a mask of epochs with a reconstructable delayed counterpart, and on each usable mask, we retain one echo-free control and construct three blends, one per flux ratio. This gives up to four controls and twelve blends per template and band, separately for the observed curves and matched DRW simulations (cases with too few retained epochs are omitted). All fits use the production search of
Section~\ref{sec:likelihood-fitting}, without access to the
injected parameters. Figure~\ref{fig:empirical_echo_suite} shows recovery separately
by S/N and pools the three S/N regimes for the score distributions.

We distinguish delay recovery from detection.
Recovery requires the preferred delay to lie within
\(5\,{\rm d}\) of the injected value, irrespective of
detection significance. The recovery fractions use available, usable blends, with uncertainty ranges estimated from \(2\,000\) bootstrap resamples of sources within each S/N regime, keeping all injections from a given source together.

Panels (a)--(c) show the strongest recovery at \(30\,{\rm d}\).
For real blends, coherent \(g+r\) recovery reaches 66,
96 and 96 per cent in the mid-low, mid-high and high-S/N
groups, respectively.
At \(60\,{\rm d}\), the corresponding fractions are 23,
74 and 70 per cent. Because \(30\,{\rm d}\) lies near the synodic period,
we repeat the recovery comparison after excluding
sources whose matched control already prefers a delay
within \(5\,{\rm d}\) of \(30\,{\rm d}\).
Coherent recovery remains 63, 94 and 92 per cent in
the same S/N order, showing that recovery is not
restricted to sources with a pre-existing control
preference near the injected delay.
Recovery generally deteriorates at the shortest and longest
tested delays, with differences between real variability
and the matched DRW realisations.

Panels (d)--(f) compare score exceedance rather than correct
delay recovery.
At the displayed DRW 99th-percentile references, approximately
22, 23 and 32 per cent of real controls exceed the cutoff
in \(g\), \(r\) and coherent \(g+r\), respectively.
Using the real-control 99th percentiles instead retains
approximately 9, 9 and 12 per cent of the real
\(30\,{\rm d}\) blends.
These conditional, in-sample comparisons demonstrate the
difference between real and DRW control distributions;
they do not establish survey false-positive rates or
completeness.
Survey deployment therefore requires independent, matched
observed controls and explicit treatment of lunar photometric
systematics (Section~\ref{sec:real-calibration}).

\section{Discussion}
\label{sec:discussion}

\subsection{Intended use and operating region}
\label{sec:operating-region}

QUELLS ranks unresolved sources by their evidence for delayed variability
and retains competing delay solutions to guide high-resolution imaging,
spectroscopy and resolved monitoring. Confirmation of lensing and a
definitive delay measurement require follow-up.

The calibration of Section~\ref{sec:mock-calibration} identifies where this
ranking is most informative. Detection efficiency increases with per-epoch S/N, cadence, flux ratio and short-lag variability rate, and peaks at observer-frame delays of approximately 20--90~d
(Fig.~\ref{fig:p6d-sanity-checks}): below \(\sim20\)~d the two images are hard to distinguish from a single process with the available light curves, and beyond \(\sim120\)~d seasonal gaps
reduce their sampled overlap. The empirical injections likewise show that recovery
depends on injected delay and photometric quality
(Section~\ref{sec:empirical-recovery}). The most favourable targets therefore
combine well-sampled week-to-month variability with relatively high
\(\widehat D_{\rm obs}\).

This operating region overlaps the compact-lens population motivating QUELLS (Figure~\ref{fig:lens-population-motivation}). In particular, the SIS relation (Eq.~\ref{eq:sis-delay}; Figure~\ref{fig:sis-delay-paper}) shows that, for the reference \(d_{\Delta t}\), systems with
\(\theta_{\rm sep}\lesssim2.5''\) can produce delays of 20--90\,d while remaining blended in ZTF, depending on the flux ratio. At fixed separation and time-delay distance, however, \(q\) and \(\Delta_{\rm obs}\) are coupled by the lens geometry, whereas the calibration grid of
Table~\ref{tab:mock-calibration-grid} varies them independently. The grid therefore over-represents the nearly equal-flux, long-delay configurations to which the detector is most sensitive. Estimating a population-averaged efficiency requires accounting for the joint distribution of these properties among unresolved lenses \citep[e.g.][]{Oguri_2010}, a calculation we defer to the DR24 application. The same model indicates which compact deflectors fall within this operating region. At \(\theta_{\rm sep}<1''\), the delays of 20--90\,d to which QUELLS is most sensitive are typical of massive E/S0 lenses lying close to the source, whereas low-mass early types and discs at the same separations have typical delays of about 3--30\,d (16th--84th percentiles), largely at the short end of our sensitivity (Figure~\ref{fig:p6d-sanity-checks}). Reaching these lower-mass deflectors therefore requires improved sensitivity to delays of order 10\,d, through denser or more precise monitoring than most current survey light curves provide.

\subsection{Real-data calibration and candidate purity}
\label{sec:real-calibration}

In a survey application, detection thresholds must be calibrated
against unlensed controls matched in per-epoch S/N,
intra-season cadence and fitted \(\widehat D_{\rm obs}\),
so that they have comparable variability power on the
timescales of the delay search. This is motivated by Figure~\ref{fig:empirical_echo_suite}, where we saw how the real and matched-DRW score distributions remain clearly distinct after the adopted quality cuts, despite the matched observing windows. An additional analysis of high-S/N controls found similar
delay-profile correlation lengths (\(10\)--\(12\,{\rm d}\))
for real and matched-DRW light curves, with median counts
of 13--14 local peaks per profile over the searched
\(3\)--\(150\,{\rm d}\) delay range, suggesting that the excess
scores of the real controls reflect higher peaks rather
than more peaks. The mismatch can reflect correlated photometric errors
and intrinsic departures from the DRW spectrum
\citep{Mushotzky_2011,Kelly_2014,Kasliwal_2015},
which the present comparison cannot disentangle, given that
the additional white-noise term cannot absorb arbitrary
correlations between epochs, since it models only
uncorrelated scatter.
One remaining relevant systematic is the documented Moon-dependent
bias in ZTF PSF photometry
\citep[Sect.~13.6, note~8]{ZSDS_Supplement}, present even after applying the quality cuts. Although the control-conditioned test in
Section~\ref{sec:empirical-recovery} shows that
\(30\,{\rm d}\) recovery persists beyond pre-existing delay
preferences, it does not isolate the lunar contribution to the excess control scores. Quantifying and treating this contribution remains necessary for the DR24 survey calibration, particularly near synodic
multiples such as the \(30\)- and \(60\,{\rm d}\) injections.

The importance of calibrating detection significance follows
from the rarity of lenses: with a lens fraction of order
\(10^{-3}\) \citep{Oguri_2010}, even a modest false-positive
rate can leave contaminants dominating the selection.
For illustration, assuming 70 per cent completeness,
false-positive rates of 1 and 30 per cent imply roughly
one lens per fifteen and per four hundred selected sources,
respectively; these are not measured survey yields.
Therefore, the predominantly unlensed parent sample provides
a large comparison population from which thresholds can be
learned within the intended observing strata and tested
on independent sources, while injections on matched
observing windows can quantify completeness at the achieved
control-exceedance rate.
Independent indicators, including Gaia astrometric or
multi-peak flags, colours and morphology, can further enrich
the selected sample, although false-positive rates and
completeness must then be established for the combined
selection as well as the adopted survey, source model
and noise treatment.

Building on these results, a forthcoming paper will present
a ranked catalogue of lens candidates from the full QUELLS
search on ZTF DR24, with empirical calibration on matched
observed controls and treatment of the lunar systematic.
The simulation calibration and empirical injection suite
presented here use DR23, whereas DR24 extends the public-survey
baseline by approximately one year, providing additional
epochs to constrain delayed signals and distinguish
competing delay solutions.
\subsection{Role of Gaia}
\label{sec:gaia-role}

Gaia enters QUELLS as an independently sampled test of a common delay
(Eq.~\ref{eq:coherent-statistic}): its sparse scanning-law cadence limits its
standalone sensitivity, but its different observing window can suppress ZTF
sampling aliases. This requires both images to contribute to the same Gaia
epoch-photometry source, which depends on separation, contrast and scan angle
\citep{Lemon_2017,Fabricius_2021,Holl_2023} and must be checked object by
object. The SIS relation (Figure~\ref{fig:sis-delay-paper}) helps explain which
systems benefit: pairs blended in Gaia tend to have short delays,
where ZTF recovery is weakest, whereas systems with delays of tens
of days can be wide enough for Gaia to resolve, as for the known lens
HS~2209+1914 (Section~\ref{sec:known-lenses}). Pairs in between may be
blended on some transits and resolved on others, producing a time-varying effective flux ratio that the constant-\(q_{b}\) model does not describe. Gaia therefore contributes primarily as a resolver, through astrometric and multi-peak flags that identify close pairs independently of their variability, with the coherence channel as a secondary check. Gaia DR4 should strengthen both roles through its 66-month photometric baseline, epoch astrometry and image products.

\subsection{Application to known lenses}
\label{sec:known-lenses}

\begin{figure}
    \centering
    \includegraphics[width=0.85\columnwidth]{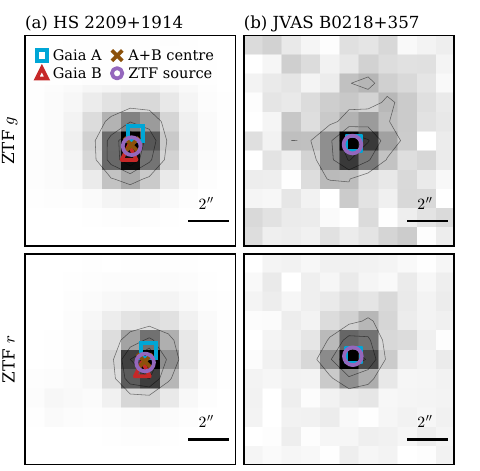}
    \caption{ZTF \(g\) (top) and \(r\) (bottom) reference images (\(10''\) fields; \(1.01''\) pixels) of the known galaxy-scale double lenses in Section~\ref{sec:known-lenses}. Markers locate Gaia images A/B, their \(G\)-flux-weighted centre and the selected ZTF source. Gaia resolves the \(1.04''\) pair in HS~2209+1914, whose ZTF source lies \(0.06''\) from the flux-weighted centre.  For JVAS~B0218+357 (\(0.34''\)), Gaia records a single source, labelled A, within \(0.1''\) of the ZTF position. \keyidea both systems provide the blended ZTF light curves targeted by QUELLS, while Gaia resolves HS~2209+1914 owing to its finer resolution (Section~\ref{sec:gaia-role}).}
    \label{fig:known-lens-cutouts}
        
\end{figure}

\begin{figure*}
    \centering
    \includegraphics[width=\textwidth]{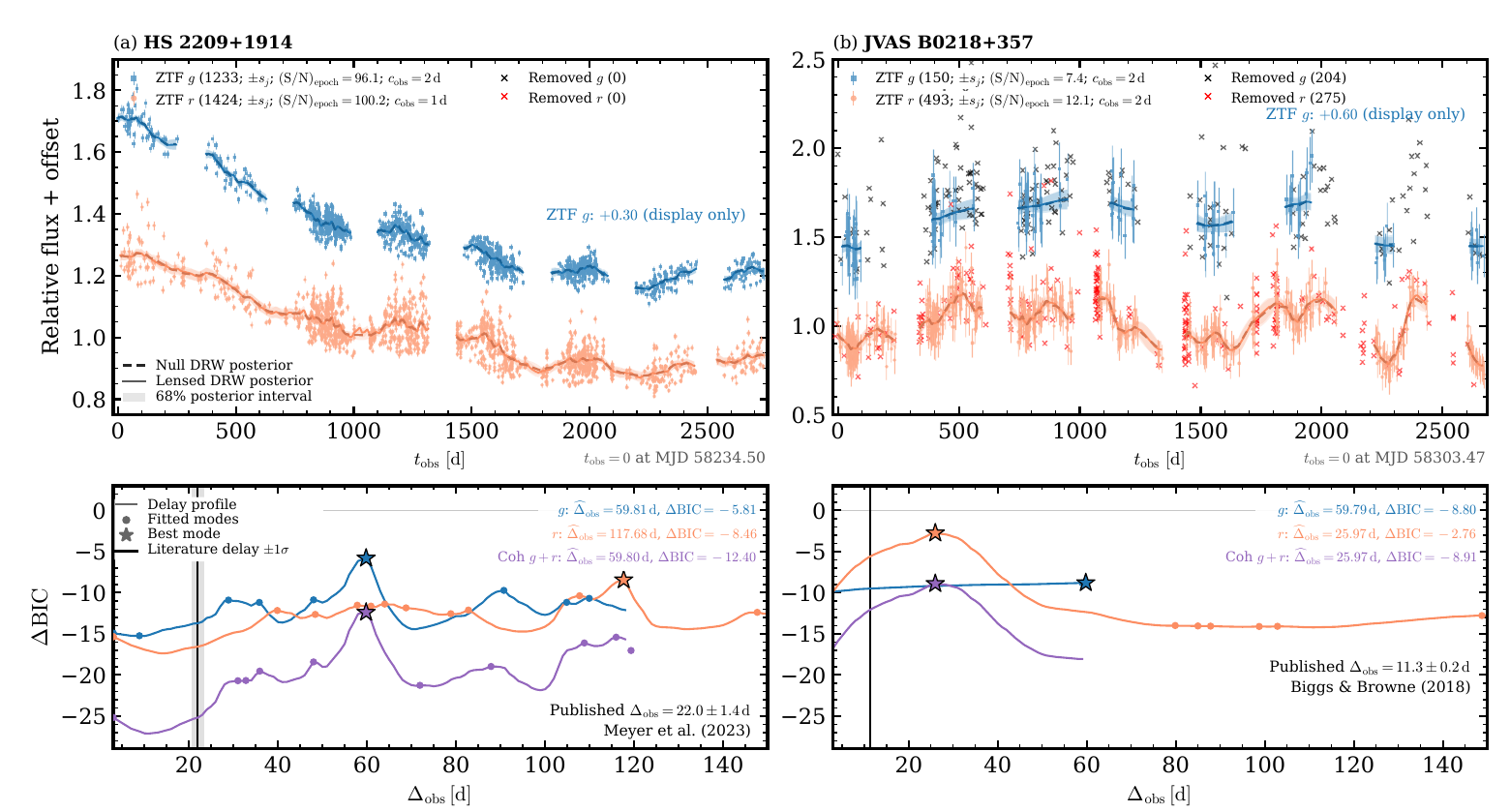}
    \caption{ZTF DR24 light curves and delay searches for HS~2209+1914 \citep{Meyer_2023} and JVAS~B0218+357 \citep{Biggs_Browne_2018}, whose photometry blends both images of each lens (Figure~\ref{fig:known-lens-cutouts}). Top: relative flux in \(g\) and \(r\) after the epoch cuts of Section~\ref{sec:preprocessing}, with crosses marking epochs removed by the quality cuts. Curves show the null and lensed DRW posterior means, with shading indicating 68 per cent posterior intervals, both omitted across gaps longer than 90\,d. Bottom: \(\Delta\mathrm{BIC}\) profiles over 3--150\,d for single-band and coherent searches, with temporal support restricting the \(g\) and coherent \(g+r\) searches of B0218+357 to 3--59.8\,d. Published delays are used only for comparison, and no
    false-positive calibration is applied.
    \keyidea neither published delay is recovered as the preferred solution, illustrating the challenges posed by weak short-timescale variability and short delays with faint, unequal image contributions.}
    \label{fig:literature_delay_lenses}
\end{figure*}

Of the \(\sim300\) known lensed quasars, few probe the regime QUELLS targets:
most are resolved by ZTF or are quadruply imaged, and only a small subset have
published delays. 
We therefore screened the known-delay double lenses of the Gaia GraL compilation with separations below \(3''\), focusing on compact systems potentially blended in ZTF (Figure~\ref{fig:lens-population-motivation}). We further required usable ZTF \(g\) and \(r\) photometry, a published delay within
3--150~d, evidence that both images contribute to the ZTF catalogue flux
(Figure~\ref{fig:known-lens-cutouts}), and no documented strong differential microlensing. This last criterion excludes SDSS~J1339+1310, whose image B is
strongly microlensed \citep{Fian_2024}. Only HS~2209+1914 and JVAS~B0218+357 pass these cuts, and we show
their light curves and delay searches in Figure~\ref{fig:literature_delay_lenses}. We note that neither system provides a usable Gaia coherence
channel: HS~2209+1914 is resolved by Gaia, as anticipated in Section~\ref{sec:gaia-role}, with epoch photometry available only for image B, while B0218+357 is unresolved but has no Gaia epoch photometry. For these DR24 examples, the DR23 calibration provides only
qualitative guidance; significance requires calibration matched to their sampling and noise, including the different \(g,r\) cadence strata of HS~2209+1914.

HS~2209+1914 illustrates the importance of source variability despite excellent photometry. Both bands lie in the high-S/N regime, with medium cadence in \(g\) and dense cadence in \(r\). Although ZTF records a verified blend of similarly bright images, the light curve is dominated by a smooth \(\sim30\) per cent decline over seven years, with little week-to-month
structure. The weak short-timescale fractional variability is qualitatively consistent with its high intrinsic continuum luminosity,
\(\lambda L_\lambda(3000\,\text{\AA})\simeq
5.1\times10^{45}\,{\rm erg\,s^{-1}}\) \citep{Ding_2021}, given the observed anticorrelation between quasar luminosity and fractional variability \citep{MacLeod_2010}. Since our mock calibration shows reduced sensitivity when variability structure on the delay timescale is weak (Figure~\ref{fig:p6d-sanity-checks}), the absence of a lens-model preference in this system is consistent with that expectation. The delay profiles also contain structure near multiples of the synodic period despite the quality cuts, consistent with a possible residual lunar contribution
and motivating explicit modelling of this systematic in
the DR24 calibration (Section~\ref{sec:real-calibration}).

B0218+357 provides a complementary test at a short published delay.
Its \(0.34''\) pair is unresolved by Gaia at \(G\simeq20\).
After \({\tt catflags}=0\) selection, the one-magnitude depth cut
removes 204 of 354 \(g\) epochs (58 per cent) and 275 of 768
\(r\) epochs (36 per cent), leaving 150 and 493 epochs,
respectively. Both retained bands lie in the mid-low-S/N, medium-cadence
regime (Table~\ref{tab:mock-calibration-grid};
Figure~\ref{fig:literature_delay_lenses}).
The cleaned \(r\)-band and coherent \(g+r\) profiles show a broad
enhancement over short delays that encompasses the published
11.3-d delay \citep{Biggs_Browne_2018,Cheung_2014}.
Their maxima occur at 26.0~d, with
\(\Delta{\rm BIC}_r=-2.76\) and
\(\Delta{\rm BIC}_{{\rm coh},g+r}=-8.91\), respectively. However, the small and chromatic optical flux ratio, extinction exceeding
one magnitude towards image A \citep{Falco_1999}, and dilution
by the \(z=0.68\) lens galaxy make the short-delay signal
challenging to isolate. The broad profile structure motivates
denser, more precise photometry and simulations at the observed
epochs and uncertainties to assess sensitivity to the published delay.

Together, these systems illustrate the different requirements
of spatially selected lens catalogues and an unresolved-variability
search. QUELLS can provide a complementary test when both images
contribute stably to the measured flux, with the published delays
providing an independent reference for the two examples considered
here.

\subsection{Model assumptions and extensions}

\paragraph*{Intrinsic variability.}
Successful delay recovery in log-flux DRW mocks and echoes injected
into observed light curves supports the linear-flux DRW approximation
under the tested conditions (Sections~\ref{sec:calibration-grid}
and~\ref{sec:empirical-recovery}). Departures from the assumed variability and noise models can nevertheless affect model preference,
requiring calibration on observed controls
(Section~\ref{sec:real-calibration}).
The QUELLS framework could be adapted to test alternative
stochastic source models, such as DHO or higher-order CARMA
processes. For resolved lenses, \citet{Kroupa_2026b} find that a smoother Mat\'ern-3/2 kernel is preferred over the DRW for two of three quasars, and that non-stationary trends can be significant; their joint modelling of slow drifts offers a template for extending QUELLS.

\paragraph*{Intrinsic reverberation.}
Broad emission lines falling in a photometric band respond to the
continuum with delays of tens to hundreds of days, adding a scaled,
delayed and smoothed copy of the continuum variability to the band
light curve \citep{Zu_2016}. In the narrow-response limit this has
the same covariance as a lensing echo
(Section~\ref{sec:lens-model}), making reverberation a possible
physical source of the excess scores in observed controls
(Section~\ref{sec:real-calibration}). Because the line contribution
and its delay depend on band and redshift, multiband coherence can
help distinguish reverberation from lensing but does not establish
a gravitational origin on its own. The control excess persists at
higher redshifts, while the low-redshift sources show no clear
additional \(r\)-band enhancement. Low-redshift line reverberation
alone therefore cannot explain the overall excess, although
reverberation from other lines or continuum responses may contribute.

\paragraph*{Image multiplicity and dilution.}
QUELLS targets double lenses: the model assumes two images, one delay and a
constant flux ratio. Quads whose delays are closely spaced may still be
approximately described by two components, but in general a quad requires
further delayed terms, which the kernel can accommodate at the cost of extra
delay and flux-ratio parameters; we do not calibrate this case here. Constant light from the lens galaxy or other contaminants can also dilute the relative variability, lowering \(\widehat D_{\rm obs}\) and hence sensitivity without changing the physical delay.

\paragraph*{Microlensing.}
Microlensing by stars in the lens galaxy changes the magnification of each image separately, often on observed timescales of months to years \citep{Mosquera_2011}. In a blended light curve, this introduces image-specific variability with no shared delayed counterpart and makes
the effective flux ratio evolve, complicating echo recovery. The magnification also depends on the size of the emitting region, which is generally larger at longer wavelengths, making the image flux ratios chromatic. Our band-specific \(q_b\) values accommodate a fixed difference between bands, but not its evolution in time. By changing the relative
contributions of disc regions with different light-travel delays, microlensing can also shift measured delays by days
\citep{Tie_2018}. These image-specific distortions motivated excluding
SDSS~J1339+1310, whose image B is strongly microlensed, from the known-lens test (Section~\ref{sec:known-lenses}). To accommodate such variability, resolved delay measurements fit a separate slowly varying component for each image \citep[e.g.][]{Tewes_2013}, suggesting a possible extension of QUELLS with analogous components at the cost of additional parameters.

\paragraph*{Photometric errors.}
The additional white-noise term, with amplitude \(\sigma_{{\rm add},F}\), accommodates uncorrelated scatter beyond the reported uncertainties, including underestimated measurement noise and variability not captured by the fitted DRW \citep{Sesar_2007,Zu_2013,Kozlowski_2016,Suberlak_2021,Arevalo_2026}.
It is therefore a phenomenological nuisance term rather than
a direct estimate of instrumental noise. Appendix~\ref{sec:four-model-DRW-comparison}
also tests its assumed brightness dependence: a flux-constant excess is
preferred to a magnitude-constant prescription for about three quarters
of the weighted ZTF-\(r\) sample, although neither description can absorb
correlated systematics. This distinction matters for lens searches because correlated errors
can produce structured delay profiles rather than merely broaden the
photometric scatter. Moon-dependent photometric bias and
seeing-dependent host contamination are possible contributors,
motivating independent photometric checks and empirical control
calibration, particularly for near-synodic candidate delays and
extended sources.

\paragraph*{Seasonal sampling.}
The cadence-support rule (Eq.~\eqref{eq:delay-support-fraction}) excludes delays with insufficient observing-window overlap, while the calibration mocks preserve the actual observing times of real templates,
allowing seasonal windows and gaps to affect both the null-score distributions and the measured recovery efficiencies. For the most promising candidates on real data searches, targeted simulations and echo injections at their measured epochs and uncertainties can further assess delay recovery and competing solutions, complementing the population-level empirical calibration.

\section{Conclusions}
\label{sec:conclusions}

We have presented QUELLS, a Gaussian-process framework that searches
for strong-lensing echoes in unresolved quasar light curves.
Its natural targets are compact, quasar-dominated double lenses,
including systems with 20--90~d delays whose image separations remain
unresolved by ZTF and whose faint lens galaxies leave little trace
in fibre spectra. QUELLS complements morphological and spectroscopic
searches for these unresolved systems by building on the delayed-mixture
covariance of earlier work \citep{Geiger_1996,Springer_2021,Bag_2023}.
It combines exact, efficient likelihood evaluation for a DRW source
with calibration for survey-scale searches.

Three practical lessons follow. First, multi-band coherence
is the most effective addition to single-band fitting because the echo is shared by all bands, although correlated intrinsic variability and shared photometric systematics mean that coherence alone does not guarantee a lens (Section~\ref{sec:empirical-recovery}). 
Second, echo injections demonstrate delay recovery in observed
light curves, extending the validation beyond DRW mocks.
However, matched controls exceed the mock-calibrated thresholds more often than expected, motivating empirical calibration and explicit
treatment of Moon-dependent photometric systematics before survey
deployment. Third, Gaia contributes primarily as a resolver of close pairs. Its comparatively sparse cadence favours detection at longer
delays, but, for comparable flux ratios and lensing geometry,
these correspond to wider pairs that Gaia is more likely to resolve.
More compact pairs that remain blended in Gaia tend to have shorter
delays, where Gaia-alone sensitivity is poor and ZTF recovery is also
weaker, limiting the overlap in which Gaia can provide a photometric
coherence channel.

Overall, QUELLS makes survey-scale searches computationally feasible
through independent source fits that can be parallelised.
The estimated ZTF fitting cost is about \(7\times10^3\)~CPU~h for
the \(536\,482\) eligible catalogue entries in the DR23-era reference
sample. We will apply this framework to ZTF DR24 with empirical
recalibration and treatment of the lunar systematic, using Gaia DR4
astrometry, high-resolution Euclid imaging and spectroscopy to
assess the resulting lens candidates and guide follow-up.
Improved sensitivity to shorter delays would also broaden future
searches with Rubin/LSST to compact lenses produced by lower-mass
deflectors.

\section*{Acknowledgements}

JLM acknowledges support from the ``la Caixa'' Foundation (ID 100010434; fellowship code LCF\slash BQ\slash EU24\slash 12060091), the Wolfson College PhD Studentship for Women in the Physical Sciences, and the Cambridge Centre for Doctoral Training in Data Intensive Science. VB is grateful for the support from the Leverhulme Trust Research Project Grant RPG-2021-205 ‘The Faint Universe Made Visible with Machine Learning’. MC is grateful for support from the Isaac Newton Trust and the Schmidt Sciences AI2050 Early Career Fellowship. SEK acknowledges support from the Science \& Technology Facilities Council (STFC) grant ST/Y001001/1.

GPT-6.1 Sol (OpenAI) and Claude Opus 5.5 (Anthropic) assisted with
text revision, code development, and consistency checks. The authors reviewed all outputs and take full responsibility for the presented work.

Based on observations obtained with the Samuel Oschin 48-inch
Telescope at Palomar Observatory as part of the Zwicky
Transient Facility project. ZTF is supported by the National
Science Foundation under grants AST-1440341 and AST-2034437
and a collaboration including current partners Caltech, IPAC,
the Oskar Klein Center at Stockholm University, the University
of Maryland, the University of California, Berkeley, the
University of Wisconsin at Milwaukee, the University of
Warwick, Ruhr University, Cornell University, Northwestern
University and Drexel University. Operations are conducted
by COO, IPAC and UW.

This work has made use of data from the European Space Agency
(ESA) mission \textit{Gaia}
(\url{https://www.cosmos.esa.int/gaia}), processed by the
\textit{Gaia} Data Processing and Analysis Consortium
(DPAC, \url{https://www.cosmos.esa.int/web/gaia/dpac/consortium}).
Funding for the DPAC has been provided by national
institutions, in particular the institutions participating
in the \textit{Gaia} Multilateral Agreement.

This research has made use of the NASA/IPAC Infrared Science
Archive, which is funded by the National Aeronautics and Space
Administration and operated by the California Institute
of Technology.

\section*{Data Availability}

The survey light curves and quasar catalogues used here are
publicly available through
\href{https://irsa.ipac.caltech.edu/Missions/ztf.html}{IRSA},
the \href{https://gea.esac.esa.int/archive/}{Gaia Archive},
\href{https://www.sdss4.org/dr16/algorithms/qso_catalog/}{SDSS}
and \href{https://data.desi.lbl.gov/doc/access/}{DESI}.
Processed data, simulations, calibration results and analysis
code are available from the corresponding author on reasonable
request, with a public release of the QUELLS code planned
alongside a forthcoming paper presenting the ranked
lens-candidate catalogue.

\bibliographystyle{mnras}
\bibliography{quells} 

@ARTICLE{Kelly_2009,
       author = {{Kelly}, Brandon C. and {Bechtold}, Jill and {Siemiginowska}, Aneta},
        title = "{Are the Variations in Quasar Optical Flux Driven by Thermal Fluctuations?}",
      journal = {\apj},
         year = 2009,
        month = jun,
       volume = {698},
       number = {1},
        pages = {895-910},
          doi = {10.1088/0004-637X/698/1/895},
archivePrefix = {arXiv},
       eprint = {0903.5315},
 primaryClass = {astro-ph.CO},
       adsurl = {https://ui.adsabs.harvard.edu/abs/2009ApJ...698..895K}
}

@article{Refsdal_1964,
    author = "Refsdal, Sjur",
    title = "{On the Possibility of Determining Hubble's Parameter and the Masses of Galaxies from the Gravitational Lens Effect}",
    doi = "10.1093/mnras/128.4.307",
    journal = "Mon. Not. Roy. Astron. Soc.",
    volume = "128",
    number = "4",
    pages = "307--310",
    year = "1964"
}

@ARTICLE{Birrer_2024,
       author = {{Birrer}, S. and {Millon}, M. and {Sluse}, D. and {Shajib}, A.~J. and {Courbin}, F. and {Erickson}, S. and {Koopmans}, L.~V.~E. and {Suyu}, S.~H. and {Treu}, T.},
        title = "{Time-Delay Cosmography: Measuring the Hubble Constant and Other Cosmological Parameters with Strong Gravitational Lensing}",
      journal = {\ssr},
         year = 2024,
        month = aug,
       volume = {220},
       number = {5},
          eid = {48},
        pages = {48},
          doi = {10.1007/s11214-024-01079-w},
archivePrefix = {arXiv},
       eprint = {2210.10833},
 primaryClass = {astro-ph.CO},
       adsurl = {https://ui.adsabs.harvard.edu/abs/2024SSRv..220...48B}
}

@ARTICLE{Dalal_2002,
       author = {{Dalal}, N. and {Kochanek}, C.~S.},
        title = "{Direct Detection of Cold Dark Matter Substructure}",
      journal = {\apj},
         year = 2002,
        month = jun,
       volume = {572},
       number = {1},
        pages = {25-33},
          doi = {10.1086/340303},
archivePrefix = {arXiv},
       eprint = {astro-ph/0111456},
 primaryClass = {astro-ph},
       adsurl = {https://ui.adsabs.harvard.edu/abs/2002ApJ...572...25D}
}

@ARTICLE{Morgan_2010,
       author = {{Morgan}, Christopher W. and {Kochanek}, C.~S. and {Morgan}, Nicholas D. and {Falco}, Emilio E.},
        title = "{The Quasar Accretion Disk Size-Black Hole Mass Relation}",
      journal = {\apj},
         year = 2010,
        month = apr,
       volume = {712},
       number = {2},
        pages = {1129-1136},
          doi = {10.1088/0004-637X/712/2/1129},
archivePrefix = {arXiv},
       eprint = {1002.4160},
 primaryClass = {astro-ph.CO},
       adsurl = {https://ui.adsabs.harvard.edu/abs/2010ApJ...712.1129M}
}

@ARTICLE{Shu_2021,
       author = {{Shu}, Yiping and {Belokurov}, Vasily and {Evans}, N. Wyn},
        title = "{Discovering strongly lensed QSOs from unresolved light curves}",
      journal = {\mnras},
         year = 2021,
        month = apr,
       volume = {502},
       number = {2},
        pages = {2912-2921},
          doi = {10.1093/mnras/stab241},
archivePrefix = {arXiv},
       eprint = {2011.04667},
 primaryClass = {astro-ph.IM},
       adsurl = {https://ui.adsabs.harvard.edu/abs/2021MNRAS.502.2912S}
}

@ARTICLE{Springer_2021,
       author = {{Springer}, Ofer M. and {Ofek}, Eran O.},
        title = "{Measuring time delays - I. Using a flux time series that is a linear combination of time-shifted light curves}",
      journal = {\mnras},
         year = 2021,
        month = sep,
       volume = {506},
       number = {1},
        pages = {864-876},
          doi = {10.1093/mnras/stab1600},
archivePrefix = {arXiv},
       eprint = {2101.11017},
 primaryClass = {astro-ph.IM},
       adsurl = {https://ui.adsabs.harvard.edu/abs/2021MNRAS.506..864S}
}

@ARTICLE{Springer_2021b,
       author = {{Springer}, Ofer M. and {Ofek}, Eran O.},
        title = "{Measuring time delays - II. Using observations of the unresolved flux and astrometry}",
      journal = {\mnras},
         year = 2021,
        month = dec,
       volume = {508},
       number = {3},
        pages = {3166-3180},
          doi = {10.1093/mnras/stab2432},
archivePrefix = {arXiv},
       eprint = {2101.11024},
 primaryClass = {astro-ph.IM},
       adsurl = {https://ui.adsabs.harvard.edu/abs/2021MNRAS.508.3166S}
}

@ARTICLE{Bag_2022,
       author = {{Bag}, Satadru and {Shafieloo}, Arman and {Liao}, Kai and {Treu}, Tommaso},
        title = "{Identifying Lensed Quasars and Measuring Their Time Delays from Unresolved Light Curves}",
      journal = {\apj},
         year = 2022,
        month = mar,
       volume = {927},
       number = {2},
          eid = {191},
        pages = {191},
          doi = {10.3847/1538-4357/ac51cb},
archivePrefix = {arXiv},
       eprint = {2110.15315},
 primaryClass = {astro-ph.CO},
       adsurl = {https://ui.adsabs.harvard.edu/abs/2022ApJ...927..191B}
}

@ARTICLE{Bag_2023,
       author = {{Bag}, Satadru and {Sohn}, Wuhyun and {Shafieloo}, Arman and {Liao}, Kai},
        title = "{Harnessing Unresolved Lensed Quasars: The Mathematical Foundation of the Fluctuation Curve}",
      journal = {\apj},
         year = 2023,
        month = jun,
       volume = {950},
       number = {2},
          eid = {158},
        pages = {158},
          doi = {10.3847/1538-4357/accea3},
archivePrefix = {arXiv},
       eprint = {2209.11078},
 primaryClass = {astro-ph.CO},
       adsurl = {https://ui.adsabs.harvard.edu/abs/2023ApJ...950..158B}
}

@ARTICLE{Biggio_2022,
       author = {{Biggio}, L. and {Domi}, A. and {Tosi}, S. and {Vernardos}, G. and {Ricci}, D. and {Paganin}, L. and {Bracco}, G.},
        title = "{Time-delay estimation in unresolved lensed quasars}",
      journal = {\mnras},
         year = 2022,
        month = oct,
       volume = {515},
       number = {4},
        pages = {5665-5672},
          doi = {10.1093/mnras/stac2034},
archivePrefix = {arXiv},
       eprint = {2110.01012},
 primaryClass = {astro-ph.IM},
       adsurl = {https://ui.adsabs.harvard.edu/abs/2022MNRAS.515.5665B}
}

@ARTICLE{Taak_2023,
       author = {{Taak}, Yoon Chan and {Treu}, Tommaso},
        title = "{Strong lensed QSOs with variability detectable by LSST: How many are there?}",
      journal = {\mnras},
         year = 2023,
        month = oct,
       volume = {524},
       number = {4},
        pages = {5446-5453},
          doi = {10.1093/mnras/stad2201},
archivePrefix = {arXiv},
       eprint = {2304.02784},
 primaryClass = {astro-ph.GA},
       adsurl = {https://ui.adsabs.harvard.edu/abs/2023MNRAS.524.5446T}
}

@ARTICLE{Lemon_2024,
       author = {{Lemon}, Cameron and {Courbin}, Fr{\'e}d{\'e}ric and {More}, Anupreeta and {Schechter}, Paul and {Ca{\~n}ameras}, Raoul and {Delchambre}, Ludovic and {Leung}, Calvin and {Shu}, Yiping and {Spiniello}, Chiara and {Hezaveh}, Yashar and et al.},
        title = "{Searching for Strong Gravitational Lenses}",
      journal = {\ssr},
         year = 2024,
        month = feb,
       volume = {220},
       number = {2},
          eid = {23},
        pages = {23},
          doi = {10.1007/s11214-024-01042-9},
archivePrefix = {arXiv},
       eprint = {2310.13466},
 primaryClass = {astro-ph.GA},
       adsurl = {https://ui.adsabs.harvard.edu/abs/2024SSRv..220...23L}
}

@ARTICLE{Shajib_2025,
       author = {{Shajib}, Anowar J. and {Smith}, Graham P. and {Birrer}, Simon and {Verma}, Aprajita and {Arendse}, Nikki and {Collett}, Thomas and {Daylan}, Tansu and {Serjeant}, Stephen and {LSST Strong Lensing Science Collaboration}},
        title = "{Strong gravitational lenses from the Vera C. Rubin Observatory}",
      journal = {Philosophical Transactions of the Royal Society of London Series A},
         year = 2025,
        month = may,
       volume = {383},
       number = {2295},
          eid = {20240117},
        pages = {20240117},
          doi = {10.1098/rsta.2024.0117},
archivePrefix = {arXiv},
       eprint = {2406.08919},
 primaryClass = {astro-ph.CO},
       adsurl = {https://ui.adsabs.harvard.edu/abs/2025RSPTA.38340117S}
}

@ARTICLE{Millon_2020,
       author = {{Millon}, M. and {Courbin}, F. and {Bonvin}, V. and {Paic}, E. and {Meylan}, G. and {Tewes}, M. and {Sluse}, D. and {Magain}, P. and {Chan}, J.~H.~H. and {Galan}, A. and et al.},
        title = "{COSMOGRAIL. XIX. Time delays in 18 strongly lensed quasars from 15 years of optical monitoring}",
      journal = {\aap},
         year = 2020,
        month = aug,
       volume = {640},
          eid = {A105},
        pages = {A105},
          doi = {10.1051/0004-6361/202037740},
archivePrefix = {arXiv},
       eprint = {2002.05736},
 primaryClass = {astro-ph.CO},
       adsurl = {https://ui.adsabs.harvard.edu/abs/2020A&A...640A.105M}
}

@ARTICLE{Bellm_2019,
       author = {{Bellm}, Eric C. and {Kulkarni}, Shrinivas R. and {Graham}, Matthew J. and {Dekany}, Richard and {Smith}, Roger M. and {Riddle}, Reed and {Masci}, Frank J. and {Helou}, George and {Prince}, Thomas A. and {Adams}, Scott M. and {Barbarino}, C. and {Barlow}, Tom and {Bauer}, James and {Beck}, Ron and {Belicki}, Justin and {Biswas}, Rahul and {Blagorodnova}, Nadejda and {Bodewits}, Dennis and {Bolin}, Bryce and {Brinnel}, Valery and {Brooke}, Tim and {Bue}, Brian and {Bulla}, Mattia and {Burruss}, Rick and {Cenko}, S. Bradley and {Chang}, Chan-Kao and {Connolly}, Andrew and {Coughlin}, Michael and {Cromer}, John and {Cunningham}, Virginia and {De}, Kishalay and {Delacroix}, Alex and {Desai}, Vandana and {Duev}, Dmitry A. and {Eadie}, Gwendolyn and {Farnham}, Tony L. and {Feeney}, Michael and {Feindt}, Ulrich and {Flynn}, David and {Franckowiak}, Anna and {Frederick}, S. and {Fremling}, C. and {Gal-Yam}, Avishay and {Gezari}, Suvi and {Giomi}, Matteo and {Goldstein}, Daniel A. and {Golkhou}, V. Zach and {Goobar}, Ariel and {Groom}, Steven and {Hacopians}, Eugean and {Hale}, David and {Henning}, John and {Ho}, Anna Y.~Q. and {Hover}, David and {Howell}, Justin and {Hung}, Tiara and {Huppenkothen}, Daniela and {Imel}, David and {Ip}, Wing-Huen and {Ivezi{\'c}}, {\v{Z}}eljko and {Jackson}, Edward and {Jones}, Lynne and {Juric}, Mario and {Kasliwal}, Mansi M. and {Kaspi}, S. and {Kaye}, Stephen and {Kelley}, Michael S.~P. and {Kowalski}, Marek and {Kramer}, Emily and {Kupfer}, Thomas and {Landry}, Walter and {Laher}, Russ R. and {Lee}, Chien-De and {Lin}, Hsing Wen and {Lin}, Zhong-Yi and {Lunnan}, Ragnhild and {Giomi}, Matteo and {Mahabal}, Ashish and {Mao}, Peter and {Miller}, Adam A. and {Monkewitz}, Serge and {Murphy}, Patrick and {Ngeow}, Chow-Choong and {Nordin}, Jakob and {Nugent}, Peter and {Ofek}, Eran and {Patterson}, Maria T. and {Penprase}, Bryan and {Porter}, Michael and {Rauch}, Ludwig and {Rebbapragada}, Umaa and {Reiley}, Dan and {Rigault}, Mickael and {Rodriguez}, Hector and {van Roestel}, Jan and {Rusholme}, Ben and {van Santen}, Jakob and {Schulze}, S. and {Shupe}, David L. and {Singer}, Leo P. and {Soumagnac}, Maayane T. and {Stein}, Robert and {Surace}, Jason and {Sollerman}, Jesper and {Szkody}, Paula and {Taddia}, F. and {Terek}, Scott and {Van Sistine}, Angela and {van Velzen}, Sjoert and {Vestrand}, W. Thomas and {Walters}, Richard and {Ward}, Charlotte and {Ye}, Quan-Zhi and {Yu}, Po-Chieh and {Yan}, Lin and {Zolkower}, Jeffry},
        title = "{The Zwicky Transient Facility: System Overview, Performance, and First Results}",
      journal = {\pasp},
         year = 2019,
        month = jan,
       volume = {131},
       number = {995},
        pages = {018002},
          doi = {10.1088/1538-3873/aaecbe},
archivePrefix = {arXiv},
       eprint = {1902.01932},
 primaryClass = {astro-ph.IM},
       adsurl = {https://ui.adsabs.harvard.edu/abs/2019PASP..131a8002B}
}

@ARTICLE{Masci_2019,
       author = {{Masci}, Frank J. and {Laher}, Russ R. and {Rusholme}, Ben and {Shupe}, David L. and {Groom}, Steven and {Surace}, Jason and {Jackson}, Edward and {Monkewitz}, Serge and {Beck}, Ron and {Flynn}, David and {Terek}, Scott and {Landry}, Walter and {Hacopians}, Eugean and {Desai}, Vandana and {Howell}, Justin and {Brooke}, Tim and {Imel}, David and {Wachter}, Stefanie and {Ye}, Quan-Zhi and {Lin}, Hsing-Wen and {Cenko}, S. Bradley and {Cunningham}, Virginia and {Rebbapragada}, Umaa and {Bue}, Brian and {Miller}, Adam A. and {Mahabal}, Ashish and {Bellm}, Eric C. and {Patterson}, Maria T. and {Juri{\'c}}, Mario and {Golkhou}, V. Zach and {Ofek}, Eran O. and {Walters}, Richard and {Graham}, Matthew and {Kasliwal}, Mansi M. and {Dekany}, Richard G. and {Kupfer}, Thomas and {Burdge}, Kevin and {Cannella}, Christopher B. and {Barlow}, Tom and {Van Sistine}, Angela and {Giomi}, Matteo and {Fremling}, Christoffer and {Blagorodnova}, Nadejda and {Levitan}, David and {Riddle}, Reed and {Smith}, Roger M. and {Helou}, George and {Prince}, Thomas A. and {Kulkarni}, Shrinivas R.},
        title = "{The Zwicky Transient Facility: Data Processing, Products, and Archive}",
      journal = {\pasp},
         year = 2019,
        month = jan,
       volume = {131},
       number = {995},
        pages = {018003},
          doi = {10.1088/1538-3873/aae8ac},
archivePrefix = {arXiv},
       eprint = {1902.01872},
 primaryClass = {astro-ph.IM},
       adsurl = {https://ui.adsabs.harvard.edu/abs/2019PASP..131a8003M}
}

@ARTICLE{Vallenari_2023,
       author = {{Gaia Collaboration} and {Vallenari}, A. and {Brown}, A.~G.~A. and {Prusti}, T. and {de Bruijne}, J.~H.~J. and {Arenou}, F. and {Babusiaux}, C. and {Biermann}, M. and {Creevey}, O.~L. and {Ducourant}, C. and et al.},
        title = "{Gaia Data Release 3. Summary of the content and survey properties}",
      journal = {\aap},
         year = 2023,
        month = jun,
       volume = {674},
          eid = {A1},
        pages = {A1},
          doi = {10.1051/0004-6361/202243940},
archivePrefix = {arXiv},
       eprint = {2208.00211},
 primaryClass = {astro-ph.GA},
       adsurl = {https://ui.adsabs.harvard.edu/abs/2023A&A...674A...1G}
}

@ARTICLE{Eyer_2023,
       author = {{Eyer}, L. and {Audard}, M. and {Holl}, B. and {Rimoldini}, L. and {Carnerero}, M.~I. and {Clementini}, G. and {De Ridder}, J. and {Distefano}, E. and {Evans}, D.~W. and {Gavras}, P. and et al.},
        title = "{Gaia Data Release 3. Summary of the variability processing and analysis}",
      journal = {\aap},
         year = 2023,
        month = jun,
       volume = {674},
          eid = {A13},
        pages = {A13},
          doi = {10.1051/0004-6361/202244242},
archivePrefix = {arXiv},
       eprint = {2206.06416},
 primaryClass = {astro-ph.SR},
       adsurl = {https://ui.adsabs.harvard.edu/abs/2023A&A...674A..13E}
}

@ARTICLE{Lyke_2020,
       author = {{Lyke}, Brad W. and {Higley}, Alexandra N. and {McLane}, J.~N. and {Schurhammer}, Danielle P. and {Myers}, Adam D. and {Ross}, Ashley J. and {Dawson}, Kyle and {Chabanier}, Sol{\`e}ne and {Martini}, Paul and {Busca}, Nicol{\'a}s G. and {Mas des Bourboux}, H{\'e}lion du and {Salvato}, Mara and {Streblyanska}, Alina and {Zarrouk}, Pauline and {Burtin}, Etienne and {Anderson}, Scott F. and {Bautista}, Julian and {Bizyaev}, Dmitry and {Brandt}, W.~N. and {Brinkmann}, Jonathan and {Brownstein}, Joel R. and {Comparat}, Johan and {Green}, Paul and {de la Macorra}, Axel and {Mu{\~n}oz Guti{\'e}rrez}, Andrea and {Hou}, Jiamin and {Newman}, Jeffrey A. and {Palanque-Delabrouille}, Nathalie and {P{\^a}ris}, Isabelle and {Percival}, Will J. and {Petitjean}, Patrick and {Rich}, James and {Rossi}, Graziano and {Schneider}, Donald P. and {Smith}, Alexander and {Vivek}, M. and {Weaver}, Benjamin Alan},
        title = "{The Sloan Digital Sky Survey Quasar Catalog: Sixteenth Data Release}",
      journal = {\apjs},
         year = 2020,
        month = sep,
       volume = {250},
       number = {1},
          eid = {8},
        pages = {8},
          doi = {10.3847/1538-4365/aba623},
archivePrefix = {arXiv},
       eprint = {2007.09001},
 primaryClass = {astro-ph.GA},
       adsurl = {https://ui.adsabs.harvard.edu/abs/2020ApJS..250....8L}
}

@ARTICLE{Karim_2026,
       author = {{DESI Collaboration} and {Abdul Karim}, M. and {Adame}, A.~G. and {Aguado}, D. and {Aguilar}, J. and {Ahlen}, S. and {Alam}, S. and {Aldering}, G. and {Alexander}, D.~M. and {Alfarsy}, R. and {Allen}, L. and {Allende Prieto}, C. and {Alves}, O. and {Anand}, A. and {Andrade}, U. and {Armengaud}, E. and {Avila}, S. and {Aviles}, A. and {Awan}, H. and {Bailey}, S. and {Baleato Lizancos}, A. and {Ballester}, O. and {Bault}, A. and {Bautista}, J. and {Bean}, R. and {Behera}, J. and {BenZvi}, S. and {Beraldo e Silva}, L. and {Bermejo-Climent}, J.~R. and {Beutler}, F. and {Bianchi}, D. and {Blake}, C. and {Blum}, R. and {Bolton}, A.~S. and {Bonici}, M. and {Brieden}, S. and {Brodzeller}, A. and {Brooks}, D. and {Buckley-Geer}, E. and {Burtin}, E. and {Bystr{\"o}m}, A. and {Canning}, R. and {Carnero Rosell}, A. and {Carr}, A. and {Carrilho}, P. and {Casas}, L. and {Castander}, F.~J. and {Cereskaite}, R. and {Cervantes-Cota}, J.~L. and {Chaussidon}, E. and {Chaves-Montero}, J. and {Chen}, S. and {Chen}, X. and {Circosta}, C. and {Claybaugh}, T. and {Cole}, S. and {Cooper}, A.~P. and {Cousinou}, M.-C. and {Cuceu}, A. and {Davis}, T.~M. and {Dawson}, K.~S. and {de Belsunce}, R. and {de la Cruz}, R. and {de la Macorra}, A. and {de Mattia}, A. and {Deiosso}, N. and {Della Costa}, J. and {Demina}, R. and {Demirbozan}, U. and {DeRose}, J. and {Dey}, A. and {Dey}, B. and {Ding}, J. and {Ding}, Z. and {Doel}, P. and {Douglass}, K. and {Dowicz}, M. and {Ebina}, H. and {Edelstein}, J. and {Eisenstein}, D.~J. and {Elbers}, W. and {Emas}, N. and {Escoffier}, S. and {Fagrelius}, P. and {Fan}, X. and {Fanning}, K. and {Favole}, G. and {Fawcett}, V.~A. and {Fern{\'a}ndez-Garc{\'\i}a}, E. and {Ferraro}, S. and {Findlay}, N. and {Font-Ribera}, A. and {Forero-Romero}, J.~E. and {Forero-S{\'a}nchez}, D. and {Frenk}, C.~S. and {G{\"a}nsicke}, B.~T. and {Galbany}, L. and {Garc{\'\i}a-Bellido}, J. and {Garcia-Quintero}, C. and {Garrison}, L.~H. and {Gazta{\~n}aga}, E. and {Gil-Mar{\'\i}n}, H. and {Gloudemans}, A. and {Gnedin}, O.~Y. and {Gontcho A Gontcho}, S. and {Gonzalez}, D. and {Gonzalez-Morales}, A.~X. and {Gonzalez-Perez}, V. and {Gordon}, C. and {Graur}, O. and {Green}, D. and {Gruen}, D. and {Gsponer}, R. and {Guandalin}, C. and {Gutierrez}, G. and {Guy}, J. and {Hahn}, C. and {Han}, J.~J. and {Han}, J. and {He}, S. and {Herrera-Alcantar}, H.~K. and {Heydenreich}, S. and {Honscheid}, K. and {Hou}, J. and {Howlett}, C. and {Huterer}, D. and {Ir{\v{s}}i{\v{c}}}, V. and {Ishak}, M. and {Jacques}, A. and {Jiang}, L. and {Jimenez}, J. and {Jing}, Y.~P. and {Joachimi}, B. and {Joudaki}, S. and {Joyce}, R. and {Jullo}, E. and {Juneau}, S. and {Kara{\c{c}}ayl{\i}}, N.~G. and {Karim}, T. and {Kehoe}, R. and {Kent}, S. and {Khederlarian}, A. and {Kirkby}, D. and {Kisner}, T. and {Kitaura}, F.-S. and {Kizhuprakkat}, N. and {Kong}, H. and {Koposov}, S.~E. and {Kremin}, A. and {Krolewski}, A. and {Lahav}, O. and {Lai}, Y. and {Lamman}, C. and {Lan}, T.-W. and {Landriau}, M. and {Lang}, D. and {Lange}, J.~U. and {Lasker}, J. and {Le Goff}, J.~M. and {Le Guillou}, L. and {Leauthaud}, A. and {Levi}, M.~E. and {Li}, S. and {Li}, T.~S. and {Liu}, W. and {Lodha}, K. and {Lokken}, M. and {Luo}, Y. and {Magneville}, C. and {Manera}, M. and {Manser}, C.~J. and {Margala}, D. and {Martini}, P. and {Maus}, M. and {McCullough}, J. and {McDonald}, P. and {Medina}, G.~E. and {Medina-Varela}, L. and {Meisner}, A. and {Mena-Fern{\'a}ndez}, J. and {Menegas}, A. and {Meneses-Rizo}, J. and {Mezcua}, M. and {Miquel}, R. and {Montero-Camacho}, P. and {Moon}, J. and {Moustakas}, J. and {Mu{\~n}oz-Guti{\'e}rrez}, A. and {Mu noz-Santos}, D. and {Myers}, A.~D. and {Myles}, J. and {Nadathur}, S. and {Najita}, J. and {Napolitano}, L. and {Newman}, J.~A. and {Nikakhtar}, F. and {Nikutta}, R. and {Niz}, G. and {Noriega}, H.~E. and {Nugent}, P.},
        title = "{Data Release 1 of the Dark Energy Spectroscopic Instrument}",
      journal = {\aj},
         year = 2026,
        month = may,
       volume = {171},
       number = {5},
          eid = {285},
        pages = {285},
          doi = {10.3847/1538-3881/ae4c43},
archivePrefix = {arXiv},
       eprint = {2503.14745},
 primaryClass = {astro-ph.CO},
       adsurl = {https://ui.adsabs.harvard.edu/abs/2026AJ....171..285D}
}

@ARTICLE{Ivezic_2019,
       author = {{Ivezi{\'c}}, {\v{Z}}eljko and {Kahn}, Steven M. and {Tyson}, J. Anthony and {Abel}, Bob and {Acosta}, Emily and {Allsman}, Robyn and {Alonso}, David and {AlSayyad}, Yusra and {Anderson}, Scott F. and {Andrew}, John and {Angel}, James Roger P. and {Angeli}, George Z. and {Ansari}, Reza and {Antilogus}, Pierre and {Araujo}, Constanza and {Armstrong}, Robert and {Arndt}, Kirk T. and {Astier}, Pierre and {Aubourg}, {\'E}ric and {Auza}, Nicole and {Axelrod}, Tim S. and {Bard}, Deborah J. and {Barr}, Jeff D. and {Barrau}, Aurelian and {Bartlett}, James G. and {Bauer}, Amanda E. and {Bauman}, Brian J. and {Baumont}, Sylvain and {Bechtol}, Ellen and {Bechtol}, Keith and {Becker}, Andrew C. and {Becla}, Jacek and {Beldica}, Cristina and {Bellavia}, Steve and {Bianco}, Federica B. and {Biswas}, Rahul and {Blanc}, Guillaume and {Blazek}, Jonathan and {Blandford}, Roger D. and {Bloom}, Josh S. and {Bogart}, Joanne and {Bond}, Tim W. and {Booth}, Michael T. and {Borgland}, Anders W. and {Borne}, Kirk and {Bosch}, James F. and {Boutigny}, Dominique and {Brackett}, Craig A. and {Bradshaw}, Andrew and {Brandt}, William Nielsen and {Brown}, Michael E. and {Bullock}, James S. and {Burchat}, Patricia and {Burke}, David L. and {Cagnoli}, Gianpietro and {Calabrese}, Daniel and {Callahan}, Shawn and {Callen}, Alice L. and {Carlin}, Jeffrey L. and {Carlson}, Erin L. and {Chandrasekharan}, Srinivasan and {Charles-Emerson}, Glenaver and {Chesley}, Steve and {Cheu}, Elliott C. and {Chiang}, Hsin-Fang and {Chiang}, James and {Chirino}, Carol and {Chow}, Derek and {Ciardi}, David R. and {Claver}, Charles F. and {Cohen-Tanugi}, Johann and {Cockrum}, Joseph J. and {Coles}, Rebecca and {Connolly}, Andrew J. and {Cook}, Kem H. and {Cooray}, Asantha and {Covey}, Kevin R. and {Cribbs}, Chris and {Cui}, Wei and {Cutri}, Roc and {Daly}, Philip N. and {Daniel}, Scott F. and {Daruich}, Felipe and {Daubard}, Guillaume and {Daues}, Greg and {Dawson}, William and {Delgado}, Francisco and {Dellapenna}, Alfred and {de Peyster}, Robert and {de Val-Borro}, Miguel and {Digel}, Seth W. and {Doherty}, Peter and {Dubois}, Richard and {Dubois-Felsmann}, Gregory P. and {Durech}, Josef and {Economou}, Frossie and {Eifler}, Tim and {Eracleous}, Michael and {Emmons}, Benjamin L. and {Fausti Neto}, Angelo and {Ferguson}, Henry and {Figueroa}, Enrique and {Fisher-Levine}, Merlin and {Focke}, Warren and {Foss}, Michael D. and {Frank}, James and {Freemon}, Michael D. and {Gangler}, Emmanuel and {Gawiser}, Eric and {Geary}, John C. and {Gee}, Perry and {Geha}, Marla and {Gessner}, Charles J.~B. and {Gibson}, Robert R. and {Gilmore}, D. Kirk and {Glanzman}, Thomas and {Glick}, William and {Goldina}, Tatiana and {Goldstein}, Daniel A. and {Goodenow}, Iain and {Graham}, Melissa L. and {Gressler}, William J. and {Gris}, Philippe and {Guy}, Leanne P. and {Guyonnet}, Augustin and {Haller}, Gunther and {Harris}, Ron and {Hascall}, Patrick A. and {Haupt}, Justine and {Hernandez}, Fabio and {Herrmann}, Sven and {Hileman}, Edward and {Hoblitt}, Joshua and {Hodgson}, John A. and {Hogan}, Craig and {Howard}, James D. and {Huang}, Dajun and {Huffer}, Michael E. and {Ingraham}, Patrick and {Innes}, Walter R. and {Jacoby}, Suzanne H. and {Jain}, Bhuvnesh and {Jammes}, Fabrice and {Jee}, M. James and {Jenness}, Tim and {Jernigan}, Garrett and {Jevremovi{\'c}}, Darko and {Johns}, Kenneth and {Johnson}, Anthony S. and {Johnson}, Margaret W.~G. and {Jones}, R. Lynne and {Juramy-Gilles}, Claire and {Juri{\'c}}, Mario and {Kalirai}, Jason S. and {Kallivayalil}, Nitya J. and {Kalmbach}, Bryce and {Kantor}, Jeffrey P. and {Karst}, Pierre and {Kasliwal}, Mansi M. and {Kelly}, Heather and {Kessler}, Richard and {Kinnison}, Veronica and {Kirkby}, David and {Knox}, Lloyd and {Kotov}, Ivan V. and {Krabbendam}, Victor L. and {Krughoff}, K. Simon and {Kub{\'a}nek}, Petr and {Kuczewski}, John and {Kulkarni}, Shri and {Ku}, John and {Kurita}, Nadine R. and {Lage}, Craig S. and {Lambert}, Ron and {Lange}, Travis and {Langton}, J. Brian and {Le Guillou}, Laurent and {Levine}, Deborah and {Liang}, Ming and {Lim}, Kian-Tat and {Lintott}, Chris J. and {Long}, Kevin E. and {Lopez}, Margaux and {Lotz}, Paul J. and {Lupton}, Robert H. and {Lust}, Nate B. and {MacArthur}, Lauren A. and {Mahabal}, Ashish and {Mandelbaum}, Rachel and {Markiewicz}, Thomas W. and {Marsh}, Darren S. and {Marshall}, Philip J. and {Marshall}, Stuart and {May}, Morgan and {McKercher}, Robert and {McQueen}, Michelle and {Meyers}, Joshua and {Migliore}, Myriam and {Miller}, Michelle and {Mills}, David J.},
        title = "{LSST: From Science Drivers to Reference Design and Anticipated Data Products}",
      journal = {\apj},
         year = 2019,
        month = mar,
       volume = {873},
       number = {2},
          eid = {111},
        pages = {111},
          doi = {10.3847/1538-4357/ab042c},
archivePrefix = {arXiv},
       eprint = {0805.2366},
 primaryClass = {astro-ph},
       adsurl = {https://ui.adsabs.harvard.edu/abs/2019ApJ...873..111I}
}

@ARTICLE{Oguri_2010,
       author = {{Oguri}, Masamune and {Marshall}, Philip J.},
        title = "{Gravitationally lensed quasars and supernovae in future wide-field optical imaging surveys}",
      journal = {\mnras},
         year = 2010,
        month = jul,
       volume = {405},
       number = {4},
        pages = {2579-2593},
          doi = {10.1111/j.1365-2966.2010.16639.x},
archivePrefix = {arXiv},
       eprint = {1001.2037},
 primaryClass = {astro-ph.CO},
       adsurl = {https://ui.adsabs.harvard.edu/abs/2010MNRAS.405.2579O}
}

@ARTICLE{MacLeod_2010,
       author = {{MacLeod}, C.~L. and {Ivezi{\'c}}, {\v{Z}}. and {Kochanek}, C.~S. and {Koz{\l}owski}, S. and {Kelly}, B. and {Bullock}, E. and {Kimball}, A. and {Sesar}, B. and {Westman}, D. and {Brooks}, K. and {Gibson}, R. and {Becker}, A.~C. and {de Vries}, W.~H.},
        title = "{Modeling the Time Variability of SDSS Stripe 82 Quasars as a Damped Random Walk}",
      journal = {\apj},
         year = 2010,
        month = oct,
       volume = {721},
       number = {2},
        pages = {1014-1033},
          doi = {10.1088/0004-637X/721/2/1014},
archivePrefix = {arXiv},
       eprint = {1004.0276},
 primaryClass = {astro-ph.CO},
       adsurl = {https://ui.adsabs.harvard.edu/abs/2010ApJ...721.1014M}
}

@ARTICLE{Kochanek_2006,
       author = {{Kochanek}, C.~S. and {Mochejska}, B. and {Morgan}, N.~D. and {Stanek}, K.~Z.},
        title = "{A Simple Method to Find All Lensed Quasars}",
      journal = {\apjl},
         year = 2006,
        month = feb,
       volume = {637},
       number = {2},
        pages = {L73-L76},
          doi = {10.1086/500559},
archivePrefix = {arXiv},
       eprint = {astro-ph/0512031},
 primaryClass = {astro-ph},
       adsurl = {https://ui.adsabs.harvard.edu/abs/2006ApJ...637L..73K}
}

@ARTICLE{Lacki_2009,
       author = {{Lacki}, Brian C. and {Kochanek}, Christopher S. and {Stanek}, Krzysztof Z. and {Inada}, Naohisa and {Oguri}, Masamune},
        title = "{Difference Imaging of Lensed Quasar Candidates in the Sloan Digital Sky Survey Supernova Survey Region}",
      journal = {\apj},
         year = 2009,
        month = jun,
       volume = {698},
       number = {1},
        pages = {428-438},
          doi = {10.1088/0004-637X/698/1/428},
archivePrefix = {arXiv},
       eprint = {0801.3432},
 primaryClass = {astro-ph},
       adsurl = {https://ui.adsabs.harvard.edu/abs/2009ApJ...698..428L}
}

@ARTICLE{Chao_2020,
       author = {{Chao}, Dani C.-Y. and {Chan}, James H.-H. and {Suyu}, Sherry H. and {Yasuda}, Naoki and {More}, Anupreeta and {Oguri}, Masamune and {Morokuma}, Tomoki and {Jaelani}, Anton T.},
        title = "{Lensed quasar search via time variability with the HSC transient survey}",
      journal = {\aap},
         year = 2020,
        month = aug,
       volume = {640},
          eid = {A88},
        pages = {A88},
          doi = {10.1051/0004-6361/201936806},
archivePrefix = {arXiv},
       eprint = {1910.01140},
 primaryClass = {astro-ph.GA},
       adsurl = {https://ui.adsabs.harvard.edu/abs/2020A&A...640A..88C}
}

@ARTICLE{Dux_2024,
       author = {{Dux}, Fr{\'e}d{\'e}ric and {Lemon}, Cameron and {Courbin}, Fr{\'e}d{\'e}ric and {Neira}, Favio and {Anguita}, Timo and {Galan}, Aymeric and {Kim}, Sam and {Hempel}, Maren and {Hempel}, Angela and {Lachaume}, R{\'e}gis},
        title = "{Nine lensed quasars and quasar pairs discovered through spatially extended variability in Pan-STARRS}",
      journal = {\aap},
         year = 2024,
        month = feb,
       volume = {682},
          eid = {A47},
        pages = {A47},
          doi = {10.1051/0004-6361/202347598},
archivePrefix = {arXiv},
       eprint = {2307.13729},
 primaryClass = {astro-ph.GA},
       adsurl = {https://ui.adsabs.harvard.edu/abs/2024A&A...682A..47D}
}

@ARTICLE{Yue_2022,
       author = {{Yue}, Minghao and {Fan}, Xiaohui and {Yang}, Jinyi and {Wang}, Feige},
        title = "{A Mock Catalog of Gravitationally-lensed Quasars for the LSST Survey}",
      journal = {\aj},
         year = 2022,
        month = mar,
       volume = {163},
       number = {3},
          eid = {139},
        pages = {139},
          doi = {10.3847/1538-3881/ac4cb0},
archivePrefix = {arXiv},
       eprint = {2201.06761},
 primaryClass = {astro-ph.GA},
       adsurl = {https://ui.adsabs.harvard.edu/abs/2022AJ....163..139Y}
}

@ARTICLE{Treu_2016,
       author = {{Treu}, Tommaso and {Marshall}, Philip J.},
        title = "{Time delay cosmography}",
      journal = {\aapr},
         year = 2016,
        month = dec,
       volume = {24},
       number = {1},
          eid = {11},
        pages = {11},
          doi = {10.1007/s00159-016-0096-8},
archivePrefix = {arXiv},
       eprint = {1605.05333},
 primaryClass = {astro-ph.CO},
       adsurl = {https://ui.adsabs.harvard.edu/abs/2016A&ARv..24...11T}
}

@ARTICLE{Fabricius_2021,
       author = {{Fabricius}, C. and {Luri}, X. and {Arenou}, F. and {Babusiaux}, C. and {Helmi}, A. and {Muraveva}, T. and {Reyl{\'e}}, C. and {Spoto}, F. and {Vallenari}, A. and {Antoja}, T. and {Balbinot}, E. and {Barache}, C. and {Bauchet}, N. and {Bragaglia}, A. and {Busonero}, D. and {Cantat-Gaudin}, T. and {Carrasco}, J.~M. and {Diakit{\'e}}, S. and {Fabrizio}, M. and {Figueras}, F. and {Garcia-Gutierrez}, A. and {Garofalo}, A. and {Jordi}, C. and {Kervella}, P. and {Khanna}, S. and {Leclerc}, N. and {Licata}, E. and {Lambert}, S. and {Marrese}, P.~M. and {Masip}, A. and {Ramos}, P. and {Robichon}, N. and {Robin}, A.~C. and {Romero-G{\'o}mez}, M. and {Rubele}, S. and {Weiler}, M.},
        title = "{Gaia Early Data Release 3. Catalogue validation}",
      journal = {\aap},
         year = 2021,
        month = may,
       volume = {649},
          eid = {A5},
        pages = {A5},
          doi = {10.1051/0004-6361/202039834},
archivePrefix = {arXiv},
       eprint = {2012.06242},
 primaryClass = {astro-ph.GA},
       adsurl = {https://ui.adsabs.harvard.edu/abs/2021A&A...649A...5F}
}

@ARTICLE{Holl_2023,
       author = {{Holl}, B. and {Fabricius}, C. and {Portell}, J. and {Lindegren}, L. and {Panuzzo}, P. and {Bernet}, M. and {Casta{\~n}eda}, J. and {Jevardat de Fombelle}, G. and {Audard}, M. and {Ducourant}, C. and {Harrison}, D.~L. and {Evans}, D.~W. and {Busso}, G. and {Sozzetti}, A. and {Gosset}, E. and {Arenou}, F. and {De Angeli}, F. and {Riello}, M. and {Eyer}, L. and {Rimoldini}, L. and {Gavras}, P. and {Mowlavi}, N. and {Nienartowicz}, K. and {Lecoeur-Ta{\"\i}bi}, I. and {Garc{\'\i}a-Lario}, P. and {Pourbaix}, D.},
        title = "{Gaia Data Release 3. Gaia scan-angle-dependent signals and spurious periods}",
      journal = {\aap},
         year = 2023,
        month = jun,
       volume = {674},
          eid = {A25},
        pages = {A25},
          doi = {10.1051/0004-6361/202245353},
archivePrefix = {arXiv},
       eprint = {2212.11971},
 primaryClass = {astro-ph.IM},
       adsurl = {https://ui.adsabs.harvard.edu/abs/2023A&A...674A..25H}
}

@ARTICLE{Oguri_2006,
       author = {{Oguri}, Masamune and {Inada}, Naohisa and {Pindor}, Bartosz and {Strauss}, Michael A. and {Richards}, Gordon T. and {Hennawi}, Joseph F. and {Turner}, Edwin L. and {Lupton}, Robert H. and {Schneider}, Donald P. and {Fukugita}, Masataka and et al.},
        title = "{The Sloan Digital Sky Survey Quasar Lens Search. I. Candidate Selection Algorithm}",
      journal = {\aj},
         year = 2006,
        month = sep,
       volume = {132},
       number = {3},
        pages = {999-1013},
          doi = {10.1086/506019},
archivePrefix = {arXiv},
       eprint = {astro-ph/0605571},
 primaryClass = {astro-ph},
       adsurl = {https://ui.adsabs.harvard.edu/abs/2006AJ....132..999O}
}

@ARTICLE{Lemon_2018,
       author = {{Lemon}, Cameron A. and {Auger}, Matthew W. and {McMahon}, Richard G. and {Ostrovski}, Fernanda},
        title = "{Gravitationally lensed quasars in Gaia - II. Discovery of 24 lensed quasars}",
      journal = {\mnras},
         year = 2018,
        month = oct,
       volume = {479},
       number = {4},
        pages = {5060-5074},
          doi = {10.1093/mnras/sty911},
archivePrefix = {arXiv},
       eprint = {1803.07601},
 primaryClass = {astro-ph.GA},
       adsurl = {https://ui.adsabs.harvard.edu/abs/2018MNRAS.479.5060L}
}

@ARTICLE{Lemon_2017,
       author = {{Lemon}, Cameron A. and {Auger}, Matthew W. and {McMahon}, Richard G. and {Koposov}, Sergey E.},
        title = "{Gravitationally lensed quasars in Gaia: I. Resolving small-separation lenses}",
      journal = {\mnras},
         year = 2017,
        month = dec,
       volume = {472},
       number = {4},
        pages = {5023-5032},
          doi = {10.1093/mnras/stx2094},
archivePrefix = {arXiv},
       eprint = {1709.08976},
 primaryClass = {astro-ph.GA},
       adsurl = {https://ui.adsabs.harvard.edu/abs/2017MNRAS.472.5023L}
}

@ARTICLE{Geiger_1996,
       author = {{Geiger}, Bernhard and {Schneider}, Peter},
        title = "{The light-curve reconstruction method for measuring the time delay of gravitational lens systems}",
      journal = {\mnras},
         year = 1996,
        month = sep,
       volume = {282},
       number = {2},
        pages = {530-546},
          doi = {10.1093/mnras/282.2.530},
archivePrefix = {arXiv},
       eprint = {astro-ph/9601164},
 primaryClass = {astro-ph},
       adsurl = {https://ui.adsabs.harvard.edu/abs/1996MNRAS.282..530G}
}

@ARTICLE{KroneMartins_2019,
       author = {{Krone-Martins}, A. and {Graham}, M.~J. and {Stern}, D. and {Djorgovski}, S.~G. and {Delchambre}, L. and {Ducourant}, C. and {Teixeira}, R. and {Drake}, A.~J. and {Scarano}, Jr., S. and {Surdej}, J. and {Galluccio}, L. and {Jalan}, P. and {Wertz}, O. and {Kl{\"u}ter}, J. and {Mignard}, F. and {Spindola-Duarte}, C. and {Dobie}, D. and {Slezak}, E. and {Sluse}, D. and {Murphy}, T. and {Boehm}, C. and {Nierenberg}, A.~M. and {Bastian}, U. and {Wambsganss}, J. and {LeCampion}, J.-F.},
        title = "{Gaia GraL: Gaia DR2 Gravitational Lens Systems. V. Doubly-imaged QSOs discovered from entropy and wavelets}",
      journal = {arXiv e-prints},
         year = 2019,
        month = dec,
          eid = {arXiv:1912.08977},
        pages = {arXiv:1912.08977},
          doi = {10.48550/arXiv.1912.08977},
archivePrefix = {arXiv},
       eprint = {1912.08977},
 primaryClass = {astro-ph.GA},
       adsurl = {https://ui.adsabs.harvard.edu/abs/2019arXiv191208977K}
}

@ARTICLE{Pindor_2005,
       author = {{Pindor}, Bart},
        title = "{Discovering Gravitational Lenses through Measurements of Their Time Delays}",
      journal = {\apj},
         year = 2005,
        month = jun,
       volume = {626},
       number = {2},
        pages = {649-656},
          doi = {10.1086/430048},
archivePrefix = {arXiv},
       eprint = {astro-ph/0501518},
 primaryClass = {astro-ph},
       adsurl = {https://ui.adsabs.harvard.edu/abs/2005ApJ...626..649P}
}

@ARTICLE{Laguna-Miralles_2026,
       author = {{Laguna-Miralles}, J{\'u}lia and {Belokurov}, Vasily and {Cranmer}, Miles},
        title = "{A hierarchical Bayesian framework for cosmology using Type 1 AGN variability}",
      journal = {arXiv e-prints},
         year = 2026,
        month = jun,
          eid = {arXiv:2606.28501},
        pages = {arXiv:2606.28501},
          doi = {10.48550/arXiv.2606.28501},
archivePrefix = {arXiv},
       eprint = {2606.28501},
 primaryClass = {astro-ph.CO},
       adsurl = {https://ui.adsabs.harvard.edu/abs/2026arXiv260628501L}
}

@ARTICLE{Brewer_2026,
       author = {{Brewer}, Brendon and {Lewis}, Geraint F. and {Yu}, Xiang and {Li}, Yuan},
        title = "{The Information Content of Quasar Variability Light Curves: How Well Can we Infer Stochastic Model Parameters?}",
      journal = {The Open Journal of Astrophysics},
         year = 2026,
        month = jul,
       volume = {9},
        pages = {64731},
          doi = {10.33232/001c.164731},
archivePrefix = {arXiv},
       eprint = {2606.01496},
 primaryClass = {astro-ph.GA},
       adsurl = {https://scixplorer.org/abs/2026OJAp....964731B}
}

@ARTICLE{Biggs_Browne_2018,
       author = {{Biggs}, A.~D. and {Browne}, I.~W.~A.},
        title = "{A revised lens time delay for JVAS B0218+357 from a reanalysis of VLA monitoring data}",
      journal = {\mnras},
         year = 2018,
        month = jun,
       volume = {476},
       number = {4},
        pages = {5393-5407},
          doi = {10.1093/mnras/sty565},
archivePrefix = {arXiv},
       eprint = {1802.10088},
 primaryClass = {astro-ph.CO},
       adsurl = {https://ui.adsabs.harvard.edu/abs/2018MNRAS.476.5393B}
}

@ARTICLE{Meyer_2023,
       author = {{Meyer}, Antoine D. and {van Dyk}, David A. and {Tak}, Hyungsuk and {Siemiginowska}, Aneta},
        title = "{TD-CARMA: Painless, Accurate, and Scalable Estimates of Gravitational Lens Time Delays with Flexible CARMA Processes}",
      journal = {\apj},
         year = 2023,
        month = jun,
       volume = {950},
       number = {1},
          eid = {37},
        pages = {37},
          doi = {10.3847/1538-4357/acbea1},
archivePrefix = {arXiv},
       eprint = {2207.09327},
 primaryClass = {astro-ph.IM},
       adsurl = {https://ui.adsabs.harvard.edu/abs/2023ApJ...950...37M}
}

@ARTICLE{Cheung_2014,
       author = {{Cheung}, C.~C. and {Larsson}, S. and {Scargle}, J.~D. and {Amin}, M.~A. and {Blandford}, R.~D. and {Bulmash}, D. and {Chiang}, J. and {Ciprini}, S. and {Corbet}, R.~H.~D. and {Falco}, E.~E. and et al.},
        title = "{Fermi Large Area Telescope Detection of Gravitational Lens Delayed {\ensuremath{\gamma}}-Ray Flares from Blazar B0218+357}",
      journal = {\apjl},
         year = 2014,
        month = feb,
       volume = {782},
       number = {2},
          eid = {L14},
        pages = {L14},
          doi = {10.1088/2041-8205/782/2/L14},
archivePrefix = {arXiv},
       eprint = {1401.0548},
 primaryClass = {astro-ph.HE},
       adsurl = {https://ui.adsabs.harvard.edu/abs/2014ApJ...782L..14C}
}

@ARTICLE{Falco_1999,
       author = {{Falco}, E.~E. and {Impey}, C.~D. and {Kochanek}, C.~S. and {Leh{\'a}r}, J. and {McLeod}, B.~A. and {Rix}, H.-W. and {Keeton}, C.~R. and {Mu{\~n}oz}, J.~A. and {Peng}, C.~Y.},
        title = "{Dust and Extinction Curves in Galaxies with z>0: The Interstellar Medium of Gravitational Lens Galaxies}",
      journal = {\apj},
         year = 1999,
        month = oct,
       volume = {523},
       number = {2},
        pages = {617-632},
          doi = {10.1086/307758},
archivePrefix = {arXiv},
       eprint = {astro-ph/9901037},
 primaryClass = {astro-ph},
       adsurl = {https://ui.adsabs.harvard.edu/abs/1999ApJ...523..617F}
}

@ARTICLE{Ducourant_2026,
       author = {{Ducourant}, C. and {Teixeira}, R. and {Vale-Cunha}, P.~H. and {Delchambre}, L. and {Krone-Martins}, A. and {Braine}, J. and {Galluccio}, L. and {Le Campion}, J.-F. and {Krinski-Moreira}, O.~S. and {Scarano}, Jr., S. and et al.},
        title = "{Gaia GraL: The GraL catalogue of gravitationally lensed quasars: X. Matched with Gaia data, redshifts, and time delays}",
      journal = {\aap},
         year = 2026,
        month = mar,
       volume = {707},
          eid = {A345},
        pages = {A345},
          doi = {10.1051/0004-6361/202558049},
archivePrefix = {arXiv},
       eprint = {2603.12810},
 primaryClass = {astro-ph.GA},
       adsurl = {https://ui.adsabs.harvard.edu/abs/2026A&A...707A.345D}
}

@ARTICLE{Mosquera_2011,
       author = {{Mosquera}, A.~M. and {Kochanek}, C.~S.},
        title = "{The Microlensing Properties of a Sample of 87 Lensed Quasars}",
      journal = {\apj},
         year = 2011,
        month = sep,
       volume = {738},
       number = {1},
          eid = {96},
        pages = {96},
          doi = {10.1088/0004-637X/738/1/96},
archivePrefix = {arXiv},
       eprint = {1104.2356},
 primaryClass = {astro-ph.CO},
       adsurl = {https://ui.adsabs.harvard.edu/abs/2011ApJ...738...96M}
}

@ARTICLE{Kozlowski_2010,
       author = {{Koz{\l}owski}, Szymon and {Kochanek}, Christopher S. and {Udalski}, A. and {Wyrzykowski}, {\L}. and {Soszy{\'n}ski}, I. and {Szyma{\'n}ski}, M.~K. and {Kubiak}, M. and {Pietrzy{\'n}ski}, G. and {Szewczyk}, O. and {Ulaczyk}, K. and et al.},
        title = "{Quantifying Quasar Variability as Part of a General Approach to Classifying Continuously Varying Sources}",
      journal = {\apj},
         year = 2010,
        month = jan,
       volume = {708},
       number = {2},
        pages = {927-945},
          doi = {10.1088/0004-637X/708/2/927},
archivePrefix = {arXiv},
       eprint = {0909.1326},
 primaryClass = {astro-ph.CO},
       adsurl = {https://ui.adsabs.harvard.edu/abs/2010ApJ...708..927K}
}

@article{Kelly_2014,
  author        = {{Kelly}, B. C. and {Becker}, A. C. and {Sobolewska}, M. and {Siemiginowska}, A. and {Uttley}, P.},
  title         = {{Flexible and Scalable Methods for Quantifying Stochastic Variability in the Era of Massive Time-domain Astronomical Data Sets}},
  journal       = {The Astrophysical Journal},
  volume        = {788},
  number        = {1},
  eid           = {33},
  pages         = {33},
  year          = {2014},
  doi           = {10.1088/0004-637X/788/1/33},
  archivePrefix = {arXiv},
  eprint        = {1402.5978},
  primaryClass  = {astro-ph.IM}
}

@ARTICLE{Yu_2022,
       author = {{Yu}, Weixiang and {Richards}, Gordon T. and {Vogeley}, Michael S. and {Moreno}, Jackeline and {Graham}, Matthew J.},
        title = "{Examining AGN UV/Optical Variability beyond the Simple Damped Random Walk}",
      journal = {\apj},
         year = 2022,
        month = sep,
       volume = {936},
       number = {2},
          eid = {132},
        pages = {132},
          doi = {10.3847/1538-4357/ac8351},
archivePrefix = {arXiv},
       eprint = {2201.08943},
 primaryClass = {astro-ph.GA},
       adsurl = {https://scixplorer.org/abs/2022ApJ...936..132Y}
}

@ARTICLE{Sesar_2007,
       author = {{Sesar}, Branimir and {Ivezi{\'c}}, {\v{Z}}eljko and {Lupton}, Robert H. and {Juri{\'c}}, Mario and {Gunn}, James E. and {Knapp}, Gillian R. and {DeLee}, Nathan and {Smith}, J. Allyn and {Miknaitis}, Gajus and {Lin}, Huan and et al.},
        title = "{Exploring the Variable Sky with the Sloan Digital Sky Survey}",
      journal = {\aj},
         year = 2007,
        month = dec,
       volume = {134},
       number = {6},
        pages = {2236-2251},
          doi = {10.1086/521819},
archivePrefix = {arXiv},
       eprint = {0704.0655},
 primaryClass = {astro-ph},
       adsurl = {https://ui.adsabs.harvard.edu/abs/2007AJ....134.2236S}
}

@ARTICLE{Kozlowski_2016,
       author = {{Koz{\l}owski}, Szymon},
        title = "{Revisiting Stochastic Variability of AGNs with Structure Functions}",
      journal = {\apj},
         year = 2016,
        month = aug,
       volume = {826},
       number = {2},
          eid = {118},
        pages = {118},
          doi = {10.3847/0004-637X/826/2/118},
archivePrefix = {arXiv},
       eprint = {1604.05858},
 primaryClass = {astro-ph.GA},
       adsurl = {https://ui.adsabs.harvard.edu/abs/2016ApJ...826..118K}
}

@ARTICLE{Zu_2013,
       author = {{Zu}, Ying and {Kochanek}, C.~S. and {Koz{\l}owski}, Szymon and {Udalski}, Andrzej},
        title = "{Is Quasar Optical Variability a Damped Random Walk?}",
      journal = {\apj},
         year = 2013,
        month = mar,
       volume = {765},
       number = {2},
          eid = {106},
        pages = {106},
          doi = {10.1088/0004-637X/765/2/106},
archivePrefix = {arXiv},
       eprint = {1202.3783},
 primaryClass = {astro-ph.CO},
       adsurl = {https://ui.adsabs.harvard.edu/abs/2013ApJ...765..106Z}
}

@ARTICLE{Suberlak_2021,
       author = {{Suberlak}, Krzysztof L. and {Ivezi{\'c}}, {\v{Z}}eljko and {MacLeod}, Chelsea},
        title = "{Improving Damped Random Walk Parameters for SDSS Stripe 82 Quasars with Pan-STARRS1}",
      journal = {\apj},
         year = 2021,
        month = feb,
       volume = {907},
       number = {2},
          eid = {96},
        pages = {96},
          doi = {10.3847/1538-4357/abc698},
archivePrefix = {arXiv},
       eprint = {2012.12907},
 primaryClass = {astro-ph.GA},
       adsurl = {https://ui.adsabs.harvard.edu/abs/2021ApJ...907...96S}
}

@ARTICLE{Lupton_1999,
       author = {{Lupton}, Robert H. and {Gunn}, James E. and {Szalay}, Alexander S.},
        title = "{A Modified Magnitude System that Produces Well-Behaved Magnitudes, Colors, and Errors Even for Low Signal-to-Noise Ratio Measurements}",
      journal = {\aj},
         year = 1999,
        month = sep,
       volume = {118},
       number = {3},
        pages = {1406-1410},
          doi = {10.1086/301004},
archivePrefix = {arXiv},
       eprint = {astro-ph/9903081},
 primaryClass = {astro-ph},
       adsurl = {https://ui.adsabs.harvard.edu/abs/1999AJ....118.1406L}
}

@ARTICLE{Uttley_2005,
       author = {{Uttley}, P. and {McHardy}, I.~M. and {Vaughan}, S.},
        title = "{Non-linear X-ray variability in X-ray binaries and active galaxies}",
      journal = {\mnras},
         year = 2005,
        month = may,
       volume = {359},
       number = {1},
        pages = {345-362},
          doi = {10.1111/j.1365-2966.2005.08886.x},
archivePrefix = {arXiv},
       eprint = {astro-ph/0502112},
 primaryClass = {astro-ph},
       adsurl = {https://ui.adsabs.harvard.edu/abs/2005MNRAS.359..345U}
}

@ARTICLE{Arevalo_2026,
       author = {{Ar{\'e}valo}, P. and {S{\'a}nchez-S{\'a}ez}, P. and {Sotomayor}, B. and {Lira}, P. and {Bauer}, F.~E. and {R{\'\i}os}, S.},
        title = "{Unlocking AGN variability with custom ZTF photometry for high-fidelity light curves and robust selection}",
      journal = {\aap},
         year = 2026,
        month = jan,
       volume = {705},
          eid = {A247},
        pages = {A247},
          doi = {10.1051/0004-6361/202556258},
archivePrefix = {arXiv},
       eprint = {2510.06898},
 primaryClass = {astro-ph.GA},
       adsurl = {https://ui.adsabs.harvard.edu/abs/2026A&A...705A.247A}
}

@ARTICLE{Guidry_2021,
       author = {{Guidry}, Joseph A. and {Vanderbosch}, Zachary P. and {Hermes}, J.~J. and {Barlow}, Brad N. and {Lopez}, Isaac D. and {Boudreaux}, Thomas M. and {Corcoran}, Kyle A. and {Bell}, Keaton J. and {Montgomery}, M.~H. and {Heintz}, Tyler M. and {Castanheira}, Barbara G. and {Reding}, Joshua S. and {Dunlap}, Bart H. and {Winget}, D.~E. and {Winget}, Karen I. and {Kuehne}, J.~W.},
        title = "{I Spy Transits and Pulsations: Empirical Variability in White Dwarfs Using Gaia and the Zwicky Transient Facility}",
      journal = {\apj},
         year = 2021,
        month = may,
       volume = {912},
       number = {2},
          eid = {125},
        pages = {125},
          doi = {10.3847/1538-4357/abee68},
archivePrefix = {arXiv},
       eprint = {2012.00035},
 primaryClass = {astro-ph.SR},
       adsurl = {https://ui.adsabs.harvard.edu/abs/2021ApJ...912..125G}
}

@manual{ZSDS_Supplement,
  author      = {{Masci}, F. J. and {Laher}, R. R. and {Rusholme}, B. and
                 {Shupe}, D. L. and {Groom}, S. and {Surace}, J. and others},
  title       = {{The ZTF Science Data System (ZSDS) Explanatory Supplement:
                  Pipelines, Definitions, Data Products, Access \& Usage}},
  organization = {IPAC, California Institute of Technology},
  year        = {2020},
  month       = jun,
  note = {Version 5.0},
  url  ={https://irsa.ipac.caltech.edu/data/ZTF/docs/ztf_explanatory_supplement.pdf}
}

@ARTICLE{Mushotzky_2011,
       author = {{Mushotzky}, R.~F. and {Edelson}, R. and {Baumgartner}, W. and {Gandhi}, P.},
        title = "{Kepler Observations of Rapid Optical Variability in Active Galactic Nuclei}",
      journal = {\apjl},
         year = 2011,
        month = dec,
       volume = {743},
       number = {1},
          eid = {L12},
        pages = {L12},
          doi = {10.1088/2041-8205/743/1/L12},
archivePrefix = {arXiv},
       eprint = {1111.0672},
 primaryClass = {astro-ph.GA},
       adsurl = {https://ui.adsabs.harvard.edu/abs/2011ApJ...743L..12M}
}

@ARTICLE{Kasliwal_2015,
       author = {{Kasliwal}, Vishal P. and {Vogeley}, Michael S. and {Richards}, Gordon T.},
        title = "{Are the variability properties of the Kepler AGN light curves consistent with a damped random walk?}",
      journal = {\mnras},
         year = 2015,
        month = aug,
       volume = {451},
       number = {4},
        pages = {4328-4345},
          doi = {10.1093/mnras/stv1230},
archivePrefix = {arXiv},
       eprint = {1505.00360},
 primaryClass = {astro-ph.GA},
       adsurl = {https://ui.adsabs.harvard.edu/abs/2015MNRAS.451.4328K}
}

@ARTICLE{Kasliwal_2017,
       author = {{Kasliwal}, Vishal P. and {Vogeley}, Michael S. and {Richards}, Gordon T.},
        title = "{Extracting information from AGN variability}",
      journal = {\mnras},
         year = 2017,
        month = sep,
       volume = {470},
       number = {3},
        pages = {3027-3048},
          doi = {10.1093/mnras/stx1420},
archivePrefix = {arXiv},
       eprint = {1607.04299},
 primaryClass = {astro-ph.GA},
       adsurl = {https://ui.adsabs.harvard.edu/abs/2017MNRAS.470.3027K}
}

@ARTICLE{Sluse_2012,
       author = {{Sluse}, D. and {Hutsem{\'e}kers}, D. and {Courbin}, F. and {Meylan}, G. and {Wambsganss}, J.},
        title = "{Microlensing of the broad line region in 17 lensed quasars}",
      journal = {\aap},
         year = 2012,
        month = aug,
       volume = {544},
          eid = {A62},
        pages = {A62},
          doi = {10.1051/0004-6361/201219125},
archivePrefix = {arXiv},
       eprint = {1206.0731},
 primaryClass = {astro-ph.CO},
       adsurl = {https://ui.adsabs.harvard.edu/abs/2012A&A...544A..62S}
}

@ARTICLE{Gilman_2020,
       author = {{Gilman}, Daniel and {Birrer}, Simon and {Nierenberg}, Anna and {Treu}, Tommaso and {Du}, Xiaolong and {Benson}, Andrew},
        title = "{Warm dark matter chills out: constraints on the halo mass function and the free-streaming length of dark matter with eight quadruple-image strong gravitational lenses}",
      journal = {\mnras},
         year = 2020,
        month = feb,
       volume = {491},
       number = {4},
        pages = {6077-6101},
          doi = {10.1093/mnras/stz3480},
archivePrefix = {arXiv},
       eprint = {1908.06983},
 primaryClass = {astro-ph.CO},
       adsurl = {https://ui.adsabs.harvard.edu/abs/2020MNRAS.491.6077G}
}

@ARTICLE{Oguri_2012,
       author = {{Oguri}, Masamune and {Inada}, Naohisa and {Strauss}, Michael A. and {Kochanek}, Christopher S. and {Kayo}, Issha and {Shin}, Min-Su and {Morokuma}, Tomoki and {Richards}, Gordon T. and {Rusu}, Cristian E. and {Frieman}, Joshua A. and et al.},
        title = "{The Sloan Digital Sky Survey Quasar Lens Search. VI. Constraints on Dark Energy and the Evolution of Massive Galaxies}",
      journal = {\aj},
         year = 2012,
        month = may,
       volume = {143},
       number = {5},
          eid = {120},
        pages = {120},
          doi = {10.1088/0004-6256/143/5/120},
archivePrefix = {arXiv},
       eprint = {1203.1088},
 primaryClass = {astro-ph.CO},
       adsurl = {https://ui.adsabs.harvard.edu/abs/2012AJ....143..120O}
}

@ARTICLE{Carnerero_2023,
       author = {{Carnerero}, Maria I. and {Raiteri}, Claudia M. and {Rimoldini}, Lorenzo and {Busonero}, Deborah and {Licata}, Enrico and {Mowlavi}, Nami and {Lecoeur-Ta{\"\i}bi}, Isabelle and {Audard}, Marc and {Holl}, Berry and {Gavras}, Panagiotis and {Nienartowicz}, Krzysztof and {Jevardat de Fombelle}, Gr{\'e}gory and {Carballo}, Ruth and {Clementini}, Gisella and {Delchambre}, Ludovic and {Klioner}, Sergei and {Lattanzi}, Mario G. and {Eyer}, Laurent},
        title = "{Gaia Data Release 3. The first Gaia catalogue of variable AGN}",
      journal = {\aap},
         year = 2023,
        month = jun,
       volume = {674},
          eid = {A24},
        pages = {A24},
          doi = {10.1051/0004-6361/202244035},
archivePrefix = {arXiv},
       eprint = {2207.06849},
 primaryClass = {astro-ph.HE},
       adsurl = {https://ui.adsabs.harvard.edu/abs/2023A&A...674A..24C}
}

@ARTICLE{Brown_2025,
       author = {{Brown}, Anthony G.~A.},
        title = "{Gaia: Ten Years of Surveying the Milky Way and Beyond}",
      journal = {arXiv e-prints},
         year = 2025,
        month = mar,
          eid = {arXiv:2503.01533},
        pages = {arXiv:2503.01533},
          doi = {10.48550/arXiv.2503.01533},
archivePrefix = {arXiv},
       eprint = {2503.01533},
 primaryClass = {astro-ph.GA},
       adsurl = {https://ui.adsabs.harvard.edu/abs/2025arXiv250301533B}
}

@ARTICLE{Moreno_2019,
       author = {{Moreno}, Jackeline and {Vogeley}, Michael S. and {Richards}, Gordon T. and {Yu}, Weixiang},
        title = "{Stochastic Modeling Handbook for Optical AGN Variability}",
      journal = {\pasp},
         year = 2019,
        month = mar,
       volume = {131},
       number = {1000},
        pages = {063001},
          doi = {10.1088/1538-3873/ab1597},
archivePrefix = {arXiv},
       eprint = {1811.00154},
 primaryClass = {astro-ph.IM},
       adsurl = {https://ui.adsabs.harvard.edu/abs/2019PASP..131f3001M}
}

@ARTICLE{Yu_2025,
       author = {{Yu}, Weixiang and {Richards}, Gordon T. and {Ruan}, John J. and {Vogeley}, Michael S. and {Bauer}, Franz E. and {Graham}, Matthew J.},
        title = "{Examining Active Galactic Nucleus UV/Optical Variability beyond the Simple Damped Random Walk. II. Insights from 22 yr Observations of SDSS, PS1, and ZTF}",
      journal = {\apj},
         year = 2025,
        month = oct,
       volume = {992},
       number = {1},
          eid = {130},
        pages = {130},
          doi = {10.3847/1538-4357/adfdd2},
archivePrefix = {arXiv},
       eprint = {2508.12076},
 primaryClass = {astro-ph.GA},
       adsurl = {https://ui.adsabs.harvard.edu/abs/2025ApJ...992..130Y}
}

@book{Howell_2006,
  author    = {{Howell}, Steve B.},
  title     = {Handbook of CCD Astronomy},
  edition   = {2nd},
  series    = {Cambridge Observing Handbooks for Research Astronomers},
  volume    = {5},
  publisher = {Cambridge University Press},
  address   = {Cambridge},
  year      = {2006}
}

@ARTICLE{Byrd_1995,
       author = {{Byrd}, Richard H. and {Lu}, Peihuang and {Nocedal}, Jorge and {Zhu}, Ciyou},
        title = "{A Limited Memory Algorithm for Bound Constrained Optimization}",
      journal = {SIAM Journal on Scientific Computing},
         year = 1995,
        month = sep,
       volume = {16},
       number = {5},
        pages = {1190-1208},
          doi = {10.1137/0916069},
       adsurl = {https://ui.adsabs.harvard.edu/abs/1995SJSC...16.1190B}
}

@article{Powell_1964,
    author = "Powell, M. J. D.",
    title = "{An efficient method for finding the minimum of a function of several variables without calculating derivatives}",
    doi = "10.1093/comjnl/7.2.155",
    journal = "Comput. J.",
    volume = "7",
    number = "2",
    pages = "155--162",
    year = "1964"
}

@techreport{Kraft_1988,
  author      = {{Kraft}, Dieter},
  title       = {A Software Package for Sequential Quadratic Programming},
  institution = {Deutsche Forschungs- und Versuchsanstalt f\"{u}r Luft- und Raumfahrt (DFVLR), Institut f\"{u}r Dynamik der Flugsysteme},
  type        = {Forschungsbericht},
  number      = {DFVLR-FB 88-28},
  address     = {Oberpfaffenhofen},
  month       = jul,
  year        = {1988}
}

@article{Wilks_1938,
    author = "Wilks, S. S.",
    title = "{The Large-Sample Distribution of the Likelihood Ratio for Testing Composite Hypotheses}",
    doi = "10.1214/aoms/1177732360",
    journal = "Annals Math. Statist.",
    volume = "9",
    number = "1",
    pages = "60--62",
    year = "1938"
}

@ARTICLE{Protassov_2002,
       author = {{Protassov}, Rostislav and {van Dyk}, David A. and {Connors}, Alanna and {Kashyap}, Vinay L. and {Siemiginowska}, Aneta},
        title = "{Statistics, Handle with Care: Detecting Multiple Model Components with the Likelihood Ratio Test}",
      journal = {\apj},
         year = 2002,
        month = may,
       volume = {571},
       number = {1},
        pages = {545-559},
          doi = {10.1086/339856},
archivePrefix = {arXiv},
       eprint = {astro-ph/0201547},
 primaryClass = {astro-ph},
       adsurl = {https://ui.adsabs.harvard.edu/abs/2002ApJ...571..545P}
}

@ARTICLE{Schwarz_1978,
       author = {{Schwarz}, Gideon},
        title = "{Estimating the Dimension of a Model}",
      journal = {Annals of Statistics},
         year = 1978,
        month = jul,
       volume = {6},
       number = {2},
        pages = {461-464},
       adsurl = {https://ui.adsabs.harvard.edu/abs/1978AnSta...6..461S}
}

@ARTICLE{Yu_2026,
       author = {{Yu}, Weixiang and {Ruan}, John J. and {Burke}, Colin J. and {Assef}, Roberto J. and {Ananna}, Tonima T. and {Bauer}, Franz E. and {De Cicco}, Demetra and {Horne}, Keith and {Hern{\'a}ndez-Garc{\'\i}a}, Lorena and {Ili{\'c}}, Dragana and et al.},
        title = "{Scalable and Robust Multiband Modeling of AGN Light Curves in Rubin-LSST}",
      journal = {\apj},
         year = 2026,
        month = feb,
       volume = {998},
       number = {1},
          eid = {144},
        pages = {144},
          doi = {10.3847/1538-4357/ae28d3},
archivePrefix = {arXiv},
       eprint = {2511.21479},
 primaryClass = {astro-ph.GA},
       adsurl = {https://ui.adsabs.harvard.edu/abs/2026ApJ...998..144Y}
}

@article{FritschButland_1984,
  author  = {Fritsch, F. N. and Butland, J.},
  title   = {A Method for Constructing Local Monotone Piecewise Cubic Interpolants},
  journal = {SIAM Journal on Scientific and Statistical Computing},
  year    = {1984},
  volume  = {5},
  number  = {2},
  pages   = {300--304},
  doi     = {10.1137/0905021}
}

@ARTICLE{Wenger_2000,
       author = {{Wenger}, M. and {Ochsenbein}, F. and {Egret}, D. and {Dubois}, P. and {Bonnarel}, F. and {Borde}, S. and {Genova}, F. and {Jasniewicz}, G. and {Lalo{\"e}}, S. and {Lesteven}, S. and {Monier}, R.},
        title = "{The SIMBAD astronomical database. The CDS reference database for astronomical objects}",
      journal = {\aaps},
         year = 2000,
        month = apr,
       volume = {143},
        pages = {9-22},
          doi = {10.1051/aas:2000332},
archivePrefix = {arXiv},
       eprint = {astro-ph/0002110},
 primaryClass = {astro-ph},
       adsurl = {https://ui.adsabs.harvard.edu/abs/2000A&AS..143....9W}
}

@ARTICLE{Massaro_2015,
       author = {{Massaro}, E. and {Maselli}, A. and {Leto}, C. and {Marchegiani}, P. and {Perri}, M. and {Giommi}, P. and {Piranomonte}, S.},
        title = "{The 5th edition of the Roma-BZCAT. A short presentation}",
      journal = {\apss},
         year = 2015,
        month = may,
       volume = {357},
       number = {1},
          eid = {75},
        pages = {75},
          doi = {10.1007/s10509-015-2254-2},
archivePrefix = {arXiv},
       eprint = {1502.07755},
 primaryClass = {astro-ph.HE},
       adsurl = {https://ui.adsabs.harvard.edu/abs/2015Ap&SS.357...75M}
}

@article{Varin_2011,
author  = {Varin, Cristiano and Reid, Nancy and Firth, David},
title   = {An overview of composite likelihood methods},
journal = {Statistica Sinica},
year    = {2011},
volume  = {21},
number  = {1},
pages   = {5--42}
}

@ARTICLE{Drton_2013,
       author = {{Drton}, Mathias and {Plummer}, Martyn},
        title = "{A Bayesian information criterion for singular models}",
      journal = {arXiv e-prints},
         year = 2013,
        month = sep,
          eid = {arXiv:1309.0911},
        pages = {arXiv:1309.0911},
          doi = {10.48550/arXiv.1309.0911},
archivePrefix = {arXiv},
       eprint = {1309.0911},
 primaryClass = {stat.ME},
       adsurl = {https://ui.adsabs.harvard.edu/abs/2013arXiv1309.0911D}
}

@ARTICLE{Patel_2025,
       author = {{Patel}, P. and {Lira}, P. and {Ar{\'e}valo}, P. and {Sun}, M. and {Bernal}, S. and {Mart{\'\i}nez-Aldama}, M.~L.},
        title = "{Probing the rest-frame wavelength dependence of quasar variability: Insights from the Zwicky Transient Facility Survey}",
      journal = {\aap},
         year = 2025,
        month = mar,
       volume = {695},
          eid = {A162},
        pages = {A162},
          doi = {10.1051/0004-6361/202449518},
archivePrefix = {arXiv},
       eprint = {2409.14999},
 primaryClass = {astro-ph.GA},
       adsurl = {https://ui.adsabs.harvard.edu/abs/2025A&A...695A.162P}
}

@ARTICLE{TDCOSMO_2026,
       author = {{TDCOSMO Collaboration} and {Birrer}, Simon and {Cappellari}, Michele and {Courbin}, Fr{\'e}d{\'e}ric and {Fassnacht}, Christopher D. and {Fleury}, Pierre and {Frieman}, Joshua A. and {Galan}, Aymeric and {Huang}, Xiang-Yu and {Johnson}, Daniel P. and {Khanolkar}, Sanjana and {Knabel}, Shawn and {Krizic}, Hrvoje and {Millon}, Martin and {Morishita}, Takahiro and {Motta}, Veronica and {Mozumdar}, Pritom and {Paic}, Eric and {Paugnat}, Hadrien and {Schweinfurth}, Allan G. and {Shajib}, Anowar J. and {Sheu}, William and {Sluse}, Dominique and {Stiavelli}, Massimo and {Suyu}, Sherry H. and {Treu}, Tommaso and {Urcelay}, Felipe and {Williams}, Devon M. and {Wong}, Kenneth C.},
        title = "{TDCOSMO 2026: new insights on the Hubble tension from strong lensing time delays}",
      journal = {arXiv e-prints},
         year = 2026,
        month = oct,
          eid = {arXiv:2610.14881},
        pages = {arXiv:2610.14881},
          doi = {10.48550/arXiv.2610.14881},
archivePrefix = {arXiv},
       eprint = {2610.14881},
 primaryClass = {astro-ph.CO},
       adsurl = {https://ui.adsabs.harvard.edu/abs/2026arXiv261014881T}
}

@ARTICLE{Tie_2018,
       author = {{Tie}, S.~S. and {Kochanek}, C.~S.},
        title = "{Microlensing makes lensed quasar time delays significantly time variable}",
      journal = {\mnras},
         year = 2018,
        month = jan,
       volume = {473},
       number = {1},
        pages = {80-90},
          doi = {10.1093/mnras/stx2348},
archivePrefix = {arXiv},
       eprint = {1707.01908},
 primaryClass = {astro-ph.GA},
       adsurl = {https://ui.adsabs.harvard.edu/abs/2018MNRAS.473...80T}
}

@ARTICLE{Tewes_2013,
       author = {{Tewes}, M. and {Courbin}, F. and {Meylan}, G.},
        title = "{COSMOGRAIL: the COSmological MOnitoring of GRAvItational Lenses. XI. Techniques for time delay measurement in presence of microlensing}",
      journal = {\aap},
         year = 2013,
        month = may,
       volume = {553},
          eid = {A120},
        pages = {A120},
          doi = {10.1051/0004-6361/201220123},
archivePrefix = {arXiv},
       eprint = {1208.5598},
 primaryClass = {astro-ph.CO},
       adsurl = {https://ui.adsabs.harvard.edu/abs/2013A&A...553A.120T}
}

@ARTICLE{Euclidlenses_2026,
       author = {{Euclid Collaboration} and {Lines}, N.~E.~P. and {Holloway}, P. and {Li}, T. and {Collett}, T.~E. and {Schuldt}, S. and {Vincken}, S.~H. and {Nightingale}, J.~W. and {Nersesian}, A. and {Fogliardi}, M. and et al.},
        title = "{Euclid Data Release 1 (DR1). Euclid Lens Identification Pipeline and Scientific Exploration (ECLIPSE) A: Strong lens catalogue overview}",
      journal = {arXiv e-prints},
         year = 2026,
        month = nov,
          eid = {arXiv:2611.15883},
        pages = {arXiv:2611.15883},
          doi = {10.48550/arXiv.2611.15883},
archivePrefix = {arXiv},
       eprint = {2611.15883},
 primaryClass = {astro-ph.GA},
       adsurl = {https://ui.adsabs.harvard.edu/abs/2026arXiv261115883E}
}

@ARTICLE{Inada_2012,
       author = {{Inada}, Naohisa and {Oguri}, Masamune and {Shin}, Min-Su and {Kayo}, Issha and {Strauss}, Michael A. and {Morokuma}, Tomoki and {Rusu}, Cristian E. and {Fukugita}, Masataka and {Kochanek}, Christopher S. and {Richards}, Gordon T. and et al.},
        title = "{The Sloan Digital Sky Survey Quasar Lens Search. V. Final Catalog from the Seventh Data Release}",
      journal = {\aj},
         year = 2012,
        month = may,
       volume = {143},
       number = {5},
          eid = {119},
        pages = {119},
          doi = {10.1088/0004-6256/143/5/119},
archivePrefix = {arXiv},
       eprint = {1203.1087},
 primaryClass = {astro-ph.CO},
       adsurl = {https://ui.adsabs.harvard.edu/abs/2012AJ....143..119I}
}

@ARTICLE{More_2016,
       author = {{More}, Anupreeta and {Oguri}, Masamune and {Kayo}, Issha and {Zinn}, Joel and {Strauss}, Michael A. and {Santiago}, Basilio X. and {Mosquera}, Ana M. and {Inada}, Naohisa and {Kochanek}, Christopher S. and {Rusu}, Cristian E. and et al.},
        title = "{The SDSS-III BOSS quasar lens survey: discovery of 13 gravitationally lensed quasars}",
      journal = {\mnras},
         year = 2016,
        month = feb,
       volume = {456},
       number = {2},
        pages = {1595-1606},
          doi = {10.1093/mnras/stv2813},
archivePrefix = {arXiv},
       eprint = {1509.07917},
 primaryClass = {astro-ph.GA},
       adsurl = {https://ui.adsabs.harvard.edu/abs/2016MNRAS.456.1595M}
}

@ARTICLE{EuclidDR1_2026,
       author = {{Euclid Collaboration} and {Aussel}, Herve},
        title = "{Euclid First Data Release 1 (DR1) - Data release overview}",
      journal = {arXiv e-prints},
         year = 2026,
        month = nov,
          eid = {arXiv:2611.16510},
        pages = {arXiv:2611.16510},
          doi = {10.48550/arXiv.2611.16510},
archivePrefix = {arXiv},
       eprint = {2611.16510},
 primaryClass = {astro-ph.CO},
       adsurl = {https://ui.adsabs.harvard.edu/abs/2026arXiv261116510E}
}

@ARTICLE{Turner_1984,
       author = {{Turner}, E.~L. and {Ostriker}, J.~P. and {Gott}, III, J.~R.},
        title = "{The statistics of gravitational lenses : the distributions of image angular separations and lens redshifts.}",
      journal = {\apj},
         year = 1984,
        month = sep,
       volume = {284},
        pages = {1-22},
          doi = {10.1086/162379},
       adsurl = {https://ui.adsabs.harvard.edu/abs/1984ApJ...284....1T}
}

@ARTICLE{Choi_2007,
       author = {{Choi}, Yun-Young and {Park}, Changbom and {Vogeley}, Michael S.},
        title = "{Internal and Collective Properties of Galaxies in the Sloan Digital Sky Survey}",
      journal = {\apj},
         year = 2007,
        month = apr,
       volume = {658},
       number = {2},
        pages = {884-897},
          doi = {10.1086/511060},
archivePrefix = {arXiv},
       eprint = {astro-ph/0611607},
 primaryClass = {astro-ph},
       adsurl = {https://ui.adsabs.harvard.edu/abs/2007ApJ...658..884C}
}

@ARTICLE{Lelli_2016,
       author = {{Lelli}, Federico and {McGaugh}, Stacy S. and {Schombert}, James M.},
        title = "{SPARC: Mass Models for 175 Disk Galaxies with Spitzer Photometry and Accurate Rotation Curves}",
      journal = {\aj},
         year = 2016,
        month = dec,
       volume = {152},
       number = {6},
          eid = {157},
        pages = {157},
          doi = {10.3847/0004-6256/152/6/157},
archivePrefix = {arXiv},
       eprint = {1606.09251},
 primaryClass = {astro-ph.GA},
       adsurl = {https://ui.adsabs.harvard.edu/abs/2016AJ....152..157L}
}

@ARTICLE{Zwaan_2010,
       author = {{Zwaan}, M.~A. and {Meyer}, M.~J. and {Staveley-Smith}, L.},
        title = "{The velocity function of gas-rich galaxies}",
      journal = {\mnras},
         year = 2010,
        month = apr,
       volume = {403},
       number = {4},
        pages = {1969-1977},
          doi = {10.1111/j.1365-2966.2009.16188.x},
archivePrefix = {arXiv},
       eprint = {0912.1754},
 primaryClass = {astro-ph.CO},
       adsurl = {https://ui.adsabs.harvard.edu/abs/2010MNRAS.403.1969Z}
}

@ARTICLE{Shahaf_2026,
       author = {{Shahaf}, Sahar and {Hogg}, David W. and {El-Badry}, Kareem and {Rix}, Hans-Walter and {Andrae}, Ren{\'e}},
        title = "{Gaia as a 0.1 arcsec point-source survey across the sky: dynamical scene reconstruction from one-dimensional scans}",
      journal = {arXiv e-prints},
         year = 2026,
        month = nov,
          eid = {arXiv:2611.21510},
        pages = {arXiv:2611.21510},
archivePrefix = {arXiv},
       eprint = {2611.21510},
 primaryClass = {astro-ph.IM},
       adsurl = {https://ui.adsabs.harvard.edu/abs/2026arXiv261121510S}
}

@ARTICLE{Ding_2021,
       author = {{Ding}, Xuheng and {Treu}, Tommaso and {Birrer}, Simon and {Agnello}, Adriano and {Sluse}, Dominique and {Fassnacht}, Chris and {Auger}, Matthew W. and {Wong}, Kenneth C. and {Suyu}, Sherry H. and {Morishita}, Takahiro and et al.},
        title = "{Testing the evolution of correlations between supermassive black holes and their host galaxies using eight strongly lensed quasars}",
      journal = {\mnras},
         year = 2021,
        month = feb,
       volume = {501},
       number = {1},
        pages = {269-280},
          doi = {10.1093/mnras/staa2992},
archivePrefix = {arXiv},
       eprint = {2005.13550},
 primaryClass = {astro-ph.GA},
       adsurl = {https://scixplorer.org/abs/2021MNRAS.501..269D}
}

@ARTICLE{Harris_1978,
       author = {{Harris}, Fredric J.},
        title = "{On the Use of Windows for Harmonic Analysis with the Discrete Fourier Transform}",
      journal = {IEEE Proceedings},
         year = 1978,
        month = jan,
       volume = {66},
        pages = {51-83},
          doi = {10.1109/PROC.1978.10837},
       adsurl = {https://scixplorer.org/abs/1978IEEEP..66...51H}
}

@ARTICLE{Zu_2011,
       author = {{Zu}, Ying and {Kochanek}, C.~S. and {Peterson}, Bradley M.},
        title = "{An Alternative Approach to Measuring Reverberation Lags in Active Galactic Nuclei}",
      journal = {\apj},
         year = 2011,
       volume = {735},
       number = {2},
          eid = {80},
        pages = {80},
          doi = {10.1088/0004-637X/735/2/80},
       eprint = {1008.0641},
       adsurl = {https://scixplorer.org/abs/2011ApJ...735...80Z}
}

@ARTICLE{Zu_2016,
       author = {{Zu}, Ying and {Kochanek}, C.~S. and {Koz{\l}owski}, Szymon and {Peterson}, B.~M.},
        title = "{Application of Stochastic Modeling to Analysis of Photometric Reverberation Mapping Data}",
      journal = {\apj},
         year = 2016,
       volume = {819},
       number = {2},
          eid = {122},
        pages = {122},
          doi = {10.3847/0004-637X/819/2/122},
       eprint = {1310.6774},
       adsurl = {https://scixplorer.org/abs/2016ApJ...819..122Z}
}

@ARTICLE{Chelouche_2013,
       author = {{Chelouche}, Doron and {Zucker}, Shay},
        title = "{Quasar Cartography: From Black Hole to Broad-line Region Scales}",
      journal = {\apj},
         year = 2013,
       volume = {769},
       number = {2},
          eid = {124},
        pages = {124},
          doi = {10.1088/0004-637X/769/2/124},
       eprint = {1305.6499},
       adsurl = {https://scixplorer.org/abs/2013ApJ...769..124C}
}

@ARTICLE{Barnacka_2011,
       author = {{Barnacka}, A. and {Glicenstein}, J.-F. and {Moudden}, Y.},
        title = "{First evidence of a gravitational lensing-induced echo in gamma rays with Fermi LAT}",
      journal = {\aap},
         year = 2011,
       volume = {528},
          eid = {L3},
        pages = {L3},
          doi = {10.1051/0004-6361/201016175},
       eprint = {1011.4498},
       adsurl = {https://scixplorer.org/abs/2011A&A...528L...3B}
}

@ARTICLE{Kroupa_2026a,
       author = {{Kroupa}, Namu and {Handley}, Will},
        title = "{Global structure of the time delay likelihood}",
      journal = {Physical Review Research},
         year = 2026,
       volume = {8},
       number = {2},
          eid = {023175},
        pages = {023175},
          doi = {10.1103/kj4s-nnyx},
       adsurl = {https://scixplorer.org/abs/2026PhRvR...8b3175K}
}

@ARTICLE{Kroupa_2026b,
       author = {{Kroupa}, Namu and {Yallup}, David and {Handley}, Will},
        title = "{Time delays and stationarity in quasar light curves}",
      journal = {\prd},
         year = 2026,
       volume = {113},
       number = {6},
          eid = {063549},
        pages = {063549},
          doi = {10.1103/87nn-klh9},
       eprint = {2602.11264},
       adsurl = {https://scixplorer.org/abs/2026PhRvD.113f3549K}
}

@ARTICLE{Rybicki_1995,
       author = {{Rybicki}, George B. and {Press}, William H.},
        title = "{Class of Fast Methods for Processing Irregularly Sampled or Otherwise Inhomogeneous One-Dimensional Data}",
      journal = {\prl},
         year = 1995,
       volume = {74},
       number = {7},
        pages = {1060-1063},
          doi = {10.1103/PhysRevLett.74.1060},
       adsurl = {https://scixplorer.org/abs/1995PhRvL..74.1060R}
}

@ARTICLE{Foreman-Mackey_2017,
       author = {{Foreman-Mackey}, Daniel and {Agol}, Eric and {Ambikasaran}, Sivaram and {Angus}, Ruth},
        title = "{Fast and Scalable Gaussian Process Modeling with Applications to Astronomical Time Series}",
      journal = {\aj},
         year = 2017,
       volume = {154},
       number = {6},
          eid = {220},
        pages = {220},
          doi = {10.3847/1538-3881/aa9332},
       eprint = {1703.09710},
       adsurl = {https://scixplorer.org/abs/2017AJ....154..220F}
}

@ARTICLE{Fian_2024,
       author = {{Fian}, C. and {Mu{\~n}oz}, J.~A. and {Jim{\'e}nez-Vicente}, J. and {Mediavilla}, E. and {Chelouche}, D. and {Kaspi}, S. and {For{\'e}s-Toribio}, R.},
        title = "{Revealing the inner workings of the lensed quasar SDSS J1339+1310: Insights from microlensing analysis}",
      journal = {\aap},
         year = 2024,
        month = sep,
       volume = {689},
          eid = {A129},
        pages = {A129},
          doi = {10.1051/0004-6361/202450151},
archivePrefix = {arXiv},
       eprint = {2407.05942},
 primaryClass = {astro-ph.GA},
       adsurl = {https://scixplorer.org/abs/2024A&A...689A.129F}
}

\appendix

\section{Lens-population calculation}
\label{app:lens-population-motivation}

For Figure~\ref{fig:lens-population-motivation}, we assume
spherical mass profiles and non-evolving comoving galaxy
abundances in a flat cosmology with a cosmological constant
and cold dark matter
(\(H_0=70\,{\rm km\,s^{-1}\,Mpc^{-1}}\),
\(\Omega_{\rm m}=0.3\)).

The velocity-dispersion function of \citet{Choi_2007}
sets the abundance weights of the E/S0 populations, with
SIS velocity scales taken equal to the line-of-sight stellar
velocity dispersion \(\sigma_*\). Disc profiles use 150 late-type galaxies from Spitzer Photometry
and Accurate Rotation Curves \citep[SPARC;][]{Lelli_2016},
selected for quality flags 1 or 2 and inclinations of at least
\(30^\circ\), with weights from the rotation-velocity function
of the H\,I Parkes All Sky Survey
\citep[HIPASS;][]{Zwaan_2010}.
The number density in each rotation-velocity bin is shared
equally among its templates, whose rotation curves provide
spherical-equivalent mass profiles with linearly rising
central rotation and flat outer extrapolation up to a finite
mass truncation.

After projecting each mass profile, we solve the spherical
lens equation for the two outermost images and sample source
positions uniformly in area, obtaining \(q\) from their
magnifications and \(\Delta_{\rm obs}\) from their geometric
and gravitational arrival-time differences.
Weighting these solutions by galaxy abundance,
multiple-imaging cross-section and comoving volume over lens-redshift range \(0.05\leq z_{\rm l}\leq1.95\) yields the separation and
joint separation--delay optical-depth histograms.
The low-dispersion SIS profiles and extrapolated disc centres
remain assumptions that affect the compact-lens predictions.

For the catalogue comparison \citep{Ducourant_2026}, we exclude
eleven identified group/cluster-scale lenses.
Panel (a) counts 336 systems using their maximum image
separations, whereas panel (b) contains 66 systems with
measured delays between individually identified images.
For each system in (b), we select the widest pair with
a measured delay and use its separation and absolute delay
together; for quads, this pair need not span the full
configuration.
We do not model magnification bias or the selection functions
of existing lens searches and delay measurements, so the
comparison does not quantify the fraction of lenses missing
from the catalogue.

\section{Notation summary}
\label{app:notation_summary}

Table~\ref{tab:notation_summary} lists the principal quantities,
their domains and where they are defined. The index $i\in \mathbb{N}_{>0}$ denotes objects or templates, $j\in\mathbb{N}_{>0}$
denotes epochs, and
$b\in\{g,r,G\}$ denotes the band. Hats indicate fitted estimates,
the superscripts $\mathrm{null}$ and $\mathrm{lens}$ identify the
fitted hypothesis, and the subscript $F$ denotes dimensionless relative
linear flux. Source-process times $t$, $\tau$ and $\Delta$ are
rest-frame quantities, while their observer-frame counterparts carry $\mathrm{obs}$.

 \begin{table*}
  \centering
  \footnotesize
  \setlength{\tabcolsep}{3pt}
  \renewcommand{\arraystretch}{1.0}
  \begin{tabularx}{\textwidth}{@{}
    >{\raggedright\arraybackslash}p{0.23\textwidth}
    >{\raggedright\arraybackslash}p{0.16\textwidth}
    >{\raggedright\arraybackslash}X@{}}
  \toprule
  Symbol & Domain & Description \\
  \midrule

  $z$
  & $\mathbb R_{\geq0}$
  & Source redshift; independently assigned in mocks
    (Sections~\ref{sec:preprocessing}, \ref{sec:calibration-grid}). \\

  $t_j$, $t_{j,{\rm obs}}$
  & $\mathbb R_{\geq0}$, $\mathbb R$
  & Rest-frame epoch relative to the adopted origin, and observer-frame
    epoch (Eq.~\eqref{eq:rest-frame-time}). \\

  $m_j$, $s_{m,j}$, $m_{\rm ref}$
  & $\mathbb R$, $\mathbb R_{>0}$, $\mathbb R$
  & Magnitude, its uncertainty and reference magnitude, in mag
    (Section~\ref{sec:preprocessing}). \\

  $y_j$, $s_j$
  & $\mathbb R_{>0}$ each
  & Observed relative flux and its uncertainty
    (Eq.~\eqref{eq:magnitude-to-flux}). \\

  $({\rm S/N})_{\rm epoch}$
  & $\mathbb R_{>0}$
  & Typical per-epoch signal-to-noise ratio
    (Eq.~\eqref{eq:epoch-snr}). \\

  $c_{\rm obs}$, $B_{\rm obs}$, $n_b$
  & $\mathbb R_{>0}$, $\mathbb R_{>0}$, $\mathbb N_{>0}$
  & Median intra-season night cadence, baseline and fitted epoch count
    (Fig.~\ref{fig:multiband_survey_distributions}). \\

  $\sigma_F$, $\tau$
  & $\mathbb R_{>0}$ each
  & Stationary DRW rms amplitude and damping time
    (Eq.~\eqref{eq:drw-kernel}). \\

  $\sigma_{{\rm bright},F}^{\rm lens}$
  & $\mathbb R_{>0}$
  & Brighter-image DRW amplitude
    (Section~\ref{sec:lens-model}). \\

  $\mu_F$, $\sigma_{{\rm add},F}$
  & $\mathbb R$, $\mathbb R_{\geq0}$
  & Mean flux and additional white-noise rms
    (Section~\ref{sec:null-model}). \\

  $D$, $D_{\rm obs}$
  & $\mathbb R_{>0}$ each
  & Rest-frame variability rate $\sigma_F^2/\tau$ and its
    observer-frame value $D/(1+z)$
    (Eq.~\eqref{eq:D-definition}). \\

    $\widehat D_{b,{\rm obs}}$
    & $\mathbb R_{>0}$
    & Fitted null-model observer-frame variability rate in band
      \(b\in\{g,r,G\}\)
      (Eq.~\eqref{eq:fitted-observer-D}). \\
$\Delta$, $\Delta_{\rm obs}$; $q$
  & $\mathbb R_{>0}$ each; $(0,1]$
  & Rest-frame and observer-frame delay; fainter-to-brighter flux ratio
    (Eqs.~\eqref{eq:rest-frame-time},
    \eqref{eq:lens-flux-ratio}). \\

  $S_{\rm sym}$, $\Omega_{b,{\rm obs}}$
  & $[0,1]$; subset of $\mathbb R_{>0}$
  & Temporal-support fraction and supported delay domain
    (Eq.~\eqref{eq:delay-support-fraction}). \\

  $\Lambda_b(\Delta_{\rm obs})$
  & $\mathbb R$
  & Twice the profiled log-likelihood improvement at a trial delay
    (Eq.~\eqref{eq:band-delay-statistic}). \\

  $T_b$, $T_{\rm coh}^{\mathcal B}$
  & $\mathbb R$ each
  & Maximum single-band and coherent likelihood-ratio statistics,
    with $\mathcal B=\{g,r\}$ or $\{g,r,G\}$
    (Eqs.~\eqref{eq:single-band-detection-statistic},
    \eqref{eq:coherent-statistic}). \\

  $\Delta\mathrm{BIC}$
  & $\mathbb R$
  & $\mathrm{BIC}_{\rm null}-\mathrm{BIC}_{\rm lens}$
    (Eq.~\eqref{eq:delta-bic-definition}). \\

  $\mathcal A$, $\alpha$,
  $T_{\mathcal A}^{\rm crit}(\alpha)$
  & Observable cell; $(0,1)$; $\mathbb R$
  & Calibration stratum, target false-positive rate and null cutoff
    (Eq.~\eqref{eq:empirical-detection-threshold}). \\

  $\tau_{\rm src}$, $\sigma_{{\rm src},m}$, $a_b$
  & $\mathbb R_{>0}$ each
  & Injected damping time, $r$-band magnitude rms and band amplitude
    multiplier (Section~\ref{sec:calibration-grid}). \\

  $D_m$ (injected)
  & $\mathbb R_{>0}$
  & Injected \(r\)-band magnitude-space rate
  $\sigma_{{\rm src},m}^{2}/\tau_{\rm src}$, in mag$^2$ per
  rest-frame day (Section~\ref{sec:calibration-grid}). \\

  $P_{\rm det}$
  & $[0,1]$
  & Conditional detection probability
    (Eq.~\eqref{eq:calibrated-selection-function}). \\

  $\tilde{y}_{b,i}(t_{\rm obs})$, $h$
  & Real-valued function; $\mathbb R_{>0}$
  & Empirical flux reconstruction and observer-frame smoothing width
    (Section~\ref{sec:empirical-injection}). \\

  \bottomrule
  \end{tabularx}
  \caption{Principal notation, with domains listed in symbol order.
  Times are in days. Domains refer to retained observations or
  scoreable fits; numerical search bounds are given in
  Section~\ref{sec:likelihood-fitting}.}
  \label{tab:notation_summary}
  \end{table*}

\section{Comparison of DRW formulations}
\label{sec:four-model-DRW-comparison}

We compare three single-source DRW formulations to separate intrinsic
variability from photometric-noise assumptions. All use the retained data
and rest-frame convention of Section~\ref{sec:framework}, with
$m_{\rm ref}$ and the reported uncertainties held fixed. We omit hypothesis superscripts in this single-source comparison.

\subsection{Model 1: linear-flux DRW}

This is the single-source null model of
Eqs.~\eqref{eq:null-generative-distribution}--\eqref{eq:null-covariance},
with $x_F\sim\mathrm{GP}(0,k_{\DRW,F})$ and
$\epsilon_{F,j}\sim\Normal(0,s_j^2+\sigma_{{\rm add},F}^2)$.
Its Gaussian likelihood is analytic, but intrinsic fluxes are not
constrained to be positive.

\subsection{Model 2: log-flux DRW}

As in Model 1, the measurement noise is independent, additive and
Gaussian in linear relative flux:
\[
  \epsilon_{F,j}\sim
  \Normal\!\left(0,s_j^2+\sigma_{{\rm add},F}^2\right).
\]
Both models use the same reported uncertainties $s_j$, with
$\sigma_{{\rm add},F}$ fitted separately in each, but the intrinsic DRW is instead defined in logarithmic flux
\citep{Kelly_2009}:
\begin{align}
  x_{\ln F}
  &\sim\mathrm{GP}\!\left(0,
  \sigma_{\ln F}^2 e^{-|t-t'|/\tau}\right),
  \nonumber\\
  y_j
  &=\mu_F\exp\!\left[
    x_{\ln F}(t_j)-\frac{\sigma_{\ln F}^2}{2}
  \right]+\epsilon_{F,j}.
  \label{eq:model-two}
\end{align}
Here $x_{\ln F}$ describes zero-mean log-flux fluctuations.
The variance correction ensures that $\mu_F>0$ is the stationary
mean intrinsic relative flux. Models 1 and 2 therefore differ in
their intrinsic variability, not their photometric-noise assumption.

\subsection{Model 3: magnitude DRW}

The magnitude-space formulation \citep{Kelly_2009,MacLeod_2010} is
\begin{equation}
  m_j=m_0+x_m(t_j)+\epsilon_{m,j},
  \label{eq:model-three}
\end{equation}
where \(x_m\sim\mathrm{GP}\!\left(0,
\sigma_m^2e^{-|t-t'|/\tau}\right)\) and
\(\epsilon_{m,j}\sim\Normal\!\left(
0,s_{m,j}^2+\sigma_{{\rm add},m}^2\right)\).
Models~2 and~3 are equivalent at the intrinsic-process level under
\begin{gather}
  \sigma_{\ln F}=\gamma\sigma_m,
  \label{eq:magnitude-logflux-amplitude}\\
  m_0=m_{\rm ref}-\frac{\ln\mu_F}{\gamma}
      +\frac{\sigma_{\ln F}^2}{2\gamma},
  \label{eq:magnitude-logflux-mean}
\end{gather}
where \(\gamma=0.4\ln10\). However, their observation models differ:
Gaussian errors and constant additional scatter in magnitude do not
become Gaussian errors and constant additive scatter in flux.

\subsection{Noise description}

Models 2 and 3 share the same intrinsic process and differ only in
the observation model. Model~2 adds Gaussian noise in linear relative
flux, with variance $s_j^2+\sigma_{{\rm add},F}^2$. Model~3 adds
Gaussian noise in magnitude, with variance
$s_{m,j}^2+\sigma_{{\rm add},m}^2$. The detector (Model~1), the calibration mocks (Section~\ref{sec:calibration-grid}) and the empirical injections (Section~\ref{sec:empirical-injection}) all use Gaussian noise additive in flux, with an additional-noise amplitude that is constant in flux. Model~2 combines this same noise prescription
with an intrinsic DRW in log flux. We explain now what that choice assumes and where it could fail.

The reported uncertainty $s_j$ is the
propagated photon-counting noise of the source, sky and detector for
that exposure, which is additive in counts and therefore in flux, and
approximately Gaussian at the counts of interest. The catalogue
magnitude error is derived from it, and Gaussian noise in magnitude is
an approximation that holds only when the fractional error is small
\citep{Lupton_1999}. For the reported term, then, Model~2 is the
description that matches the measurement and Model~3 is the
approximation. However, at high per-epoch S/N, the flux-to-magnitude conversion is nearly linear over the measurement uncertainties, making Gaussian magnitude errors a close approximation to Gaussian flux errors.

The two prescriptions differ in how excess scatter scales with
brightness: a constant \(\sigma_{{\rm add},F}\) gives the same additional flux uncertainty at every epoch, whereas a constant
\(\sigma_{{\rm add},m}\) gives approximately constant fractional
uncertainty. The latter therefore allows flux scatter to increase
with brightness, making it closer to flux-rescaling calibration errors
and the 0.01--0.02~mag precision floor of bright ZTF sources
\citep{Masci_2019}. Although strongly variable, well-sampled light curves help distinguish these scalings, our flux-constant term remains a
phenomenological description of the average excess scatter whose
fitted amplitude alone does not identify its physical origin.

The calibration mocks use the same flux-constant excess as the
detector, keeping this noise prescription fixed when testing the
intrinsic-variability approximation. In real data, an incorrect
brightness dependence can affect both the fitted excess and the
DRW parameters. The contrast $\Delta\ell_{\rm errors}$ below
compares the two prescriptions on observed light curves, while
Section~\ref{sec:empirical-injection} assesses whether the
mock-calibrated thresholds transfer to real data.

\subsection{Likelihood comparisons}

Let $\widehat\ell_k$ denote the maximised log likelihood of Model $k$
after integrating out the latent process, expressed in the same
relative-flux coordinates with $m_{\rm ref}$ fixed. For Model 3,
this requires adding the Jacobian correction
$-\sum_j\ln(\gamma y_j)$ to its magnitude-space log likelihood.
For Model 2, the latent integration uses a Laplace approximation.

We compare intrinsic variability through
$\Delta\ell_{\rm intrinsic}\coloneq\widehat\ell_2-\widehat\ell_1$,
the observation model through
$\Delta\ell_{\rm errors}\coloneq\widehat\ell_3-\widehat\ell_2$,
and their combined effect through
$\Delta\ell_{\rm natural}\coloneq\widehat\ell_3-\widehat\ell_1$.
All three models use the same observations and have four fitted
parameters, giving identical BIC penalties. These contrasts
measure relative fit, not posterior model probabilities or calibrated
significance. Interpreting small differences requires comparable physical parameter ranges, with errors from optimisation and the Laplace approximation smaller than the contrasts being interpreted.

\subsection{Results and interpretation}
\label{sec:four-model-preliminary-results}

We compare 311 ZTF-$r$ AGN light curves, stratified by variability amplitude, photometric S/N, cadence and epoch count, and weighted to the eligible parent population.

Model 2 has the higher likelihood for $77.4$ per cent of the weighted
population, but the intrinsic-coordinate contrast is generally
small: $\langle\Delta\ell_{\rm intrinsic}\rangle=0.97$, with
$84.6$ per cent having $|\Delta\ell_{\rm intrinsic}|<2$. The mean contrast
increases from $0.47$ at low variability amplitude to $3.32$ at
high amplitude, with weaker dependence on observing conditions. This closer agreement at low amplitude is consistent with the linear-flux approximation introduced in Section~\ref{sec:drw-source-model}, since exponentiation is approximately linear for \(\sigma_{\ln F}\ll1\) but changes both the flux distribution and its covariance at larger amplitudes.

Changing the noise prescription has a larger population-averaged
effect. The mean contrast is
$\langle\Delta\ell_{\rm errors}\rangle=-4.28$, with Model 2 having
the higher likelihood than Model 3 for $75.5$ per cent of the weighted
population. Including both changes gives
$\langle\Delta\ell_{\rm natural}\rangle=-3.31$, with Model 1
having the higher likelihood than Model 3 for $71.4$ per cent.

These results support generating the calibration mocks with
Model 2. It preserves the lognormal intrinsic variability shared
with Model 3, while adding measurement noise in flux after combining
the images. Sharing this noise prescription with the detector
allows the mocks to test the intrinsic-variability approximation
without simultaneously changing the observation model.

Together with its analytic delayed-mixture likelihood
(Section~\ref{sec:likelihood-fitting}), these findings support
Model 1 as a practical approximation for candidate ranking and
delay searches, without establishing equivalence between the
intrinsic models. The larger noise-dependent differences show that the fits are sensitive to how measurement uncertainties and excess scatter are modelled \citep{Zu_2013,Arevalo_2026}, although the comparison does not identify the physical origin of that scatter. The additional white-noise term can absorb uncorrelated excess scatter but cannot represent
time-correlated systematics.

\section{Frequency-domain signature and the effect of sampling}
\label{sec:fourier-sampling}

\begin{figure}
\centering
\includegraphics[width=\columnwidth]{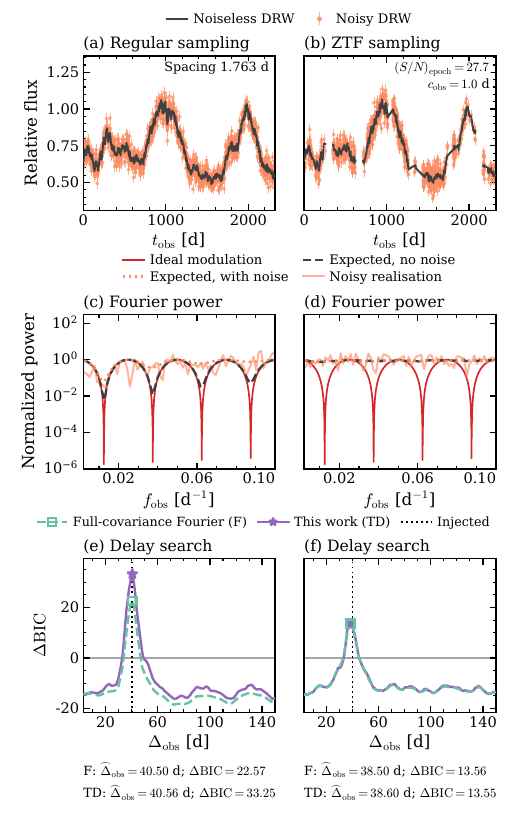}
\caption{Sampling and delay recovery for the same simulated lensed light curve sampled regularly (left) and at ZTF-\(r\) epochs (right),
using SDSS~J011908.94+350322.8 as the observing template.
The injected parameters are \(z=0.5\),
\(\tau_{\rm src}=300\,{\rm d}\),
\(\sigma_{{\rm src},m}=0.28\,{\rm mag}\) in \(r\),
\(\Delta_{\rm obs}=40\,{\rm d}\), \(q=1\) and
\(\sigma_{{\rm add},F}=0.02220\).
Top: noiseless flux and noisy measurements, with error bars showing the reported photometric uncertainties.
Middle: realised power, expectations with and without noise,
and ideal modulation, normalised by the matching unlensed
expectation and averaged over three bins.
Bottom: full-covariance Fourier and time-domain searches,
with fitted maxima marked and the injected delay dotted.
Each search uses an independently fitted null model.
\keyidea modelling the joint Fourier covariance recovers
delays consistent with the time-domain search, even when
sampling blurs the power-spectrum comb.}
\label{fig:fourier-time-domain-appendix}
\end{figure}

QUELLS infers lensing delays from the covariance of an
unresolved light curve in the time domain, but the same
delayed-copy signal produces the periodic modulation
illustrated in Figure~\ref{fig:drw-lensed-dho-psd}, with
frequency spacing \(1/\Delta_{\rm obs}\)
(Eq.~\eqref{eq:psd-lens}).
This motivates a complementary check: fitting the same
covariance model in Fourier space and comparing the recovered
delay with the time-domain result.
Figure~\ref{fig:fourier-time-domain-appendix} compares the Fourier-
and time-domain methods under regular and ZTF sampling, showing
that the delay can also be recovered by modelling the full Fourier
covariance when the observing window blurs the spectral comb.

For this comparison, we sample the same log-flux realisation
at regular and actual ZTF-\(r\) epochs, each with 1320
measurements over \(2325.85\,{\rm d}\), using the same
chronological uncertainty sequence and independent noise draws.
Within each cadence, both searches fit identical observations
with the Gaussian linear-flux lens model, varying the
amplitude, damping time, flux ratio and additional noise.

We compute the Fourier coefficients directly from the
observer-frame epochs, without interpolation or gap filling.
We apply a Hann taper, a raised-cosine weight that smoothly tapers the light-curve ends to reduce power leakage between frequencies  \citep{Harris_1978}, and subtract the mean calculated with the same weights. At each of 228 frequencies between \(0.002\) and
\(0.1\,{\rm d}^{-1}\), weighted sine and cosine sums give
the real and imaginary coefficients; the sum of their squares
gives the power plotted in the middle panels, normalised
by the matching unlensed expectation and averaged over
three adjacent frequency bins.
The expected curves give the average powers over repeated
realisations at fixed simulation parameters and observing
conditions, calculated from the covariance of the simulated lensed signal under the same sampling, mean subtraction and taper. The noisy expectation includes measurement and additional
noise, whereas the noise-free expectation omits both.
With the fixed-total-mean normalisation of
Eq.~\eqref{eq:calibration-lens-mock}, the ideal modulation
of Eq.~\eqref{eq:psd-lens} becomes
\([1+q^2+2q\cos(2\pi f_{\rm obs}\Delta_{\rm obs})]/(1+q)^2\),
which reduces to \(\cos^2(\pi f_{\rm obs}\Delta_{\rm obs})\)
at \(q=1\), but sampling mixes frequencies and fills
the troughs even without noise.
The Fourier likelihood therefore fits the real and imaginary
coefficients jointly, using the transformed Gaussian-flux
covariance of Eqs~\eqref{eq:null-covariance}
and~\eqref{eq:lens-covariance}.
After removing nearly dependent coefficient combinations,
this retains the correlations between the remaining
components, rather than treating the displayed powers
as independent measurements.

The Fourier search scans \(3\)--\(150\,{\rm d}\) in the
observer frame at \(1\,{\rm d}\) spacing, with
\(0.5\,{\rm d}\) refinement around the three leading modes;
the time-domain search uses its supported \(1\,{\rm d}\)
grid and local refinement.
Both report the descriptive \(\Delta\mathrm{BIC}\) of
Eq.~\eqref{eq:delta-bic-definition}, using the same epoch-count penalty as in the time-domain fit. The bottom panels show closely agreeing recovered delays near the injected value under both cadences, despite the weak power-spectrum comb under ZTF sampling.

The same covariance model can therefore be fitted in either domain,
while QUELLS uses time-domain inference to exploit the DRW Markov
structure directly, avoiding the additional Fourier projection.

\bsp	
\label{lastpage}
\end{document}